\documentclass[prl,aps,showpacs,twocolumn,preprintnumbers,
amsmath,amssymb,superscriptaddress,longbibliography]{revtex4-1}
\usepackage[english]{babel}
\usepackage{amsmath,amssymb,amsfonts}
\usepackage{mathrsfs}
\usepackage{graphicx}
\usepackage[colorlinks=True,linkcolor=red,citecolor=blue,urlcolor=blue]{hyperref}
\usepackage{bookmark}
\usepackage{xcolor}
\usepackage{chngcntr}
\usepackage{braket}
\usepackage{nicefrac}
\usepackage{bm}
\usepackage{bbm}
\usepackage{braket}
\usepackage{slashed}
\usepackage[normalem]{ulem}
\usepackage{array}
\newcolumntype{C}{>{$}c<{$}}

\newcommand{\diff}{\mathrm{d}}

\def\id{\mathbbm{1}}

\newcommand{\tr}{\mathop{\mathrm{tr}}}
\newcommand{\Z}{\mathbb{Z}}

\renewcommand{\Re}{\mathop{\mathrm{Re}}}
\renewcommand{\Im}{\mathop{\mathrm{Im}}}
\renewcommand{\i}{\mathop{\mathrm{i}}}

\newcommand{\bolds}[1]{\boldsymbol #1}
\allowdisplaybreaks
\DeclareMathAlphabet{\zc}{OT1}{pzc}{m}{it}

\begin{document}

\title{Non-Hermitian chiral anomalies}

 \author{Sharareh Sayyad}
 \email{sharareh.sayyad@neel.cnrs.fr}
\affiliation{Univ. Grenoble Alpes, CNRS, Grenoble INP, Institut N\'eel, 38000 Grenoble, France}
\author{Julia D. Hannukainen}
 \affiliation{Department of Physics, KTH Royal Institute of Technology, 106 91 Stockholm, Sweden}
  \author{Adolfo G. Grushin}
 \affiliation{Univ. Grenoble Alpes, CNRS, Grenoble INP, Institut N\'eel, 38000 Grenoble, France}
%

\begin{abstract}
The chiral anomaly underlies a broad number of phenomena, from enhanced electronic transport in topological metals to anomalous currents in the quark-gluon plasma.
The discovery of topological states of matter in non-Hermitian systems raises the question of whether there are anomalous conservation laws that remain unaccounted for.
To answer this question, we consider both two and four space-time dimensions, presenting a unified formulation to calculate anomalous responses in Hermitianized, anti-Hermitianized and non-Hermitian systems of 
massless electrons with complex Fermi velocities coupled to non-Hermitian gauge fields. 
Our results indicate that the quantum conservation laws of chiral currents of non-Hermitian systems are not related to those in Hermitianized and anti-Hermitianized systems, as would be expected classically, due to novel anomalous terms that we derive. 
We further present some physical consequences of our non-Hermitian anomaly that may have implications for a broad class of emerging experimental systems that realize non-Hermitian Hamiltonians.
\end{abstract}
\date{\today}
\maketitle

\paragraph*{\bf Introduction-.}
Quantum anomalies explain transport phenomena in many branches of physics, including high-energy physics~\cite{Adler1969, Bell1969, Kharzeev2008,Fukushima2008}, astrophysics~\cite{Kaminski2014,Basilakos2020}, and condensed matter physics~\cite{Nielsen1983, Frohlich2018, Burkov2015}.
Anomalies account for the fact that, upon quantization, the conservation law associated with a classical symmetry can be broken, resulting in observable anomalous transport currents~\cite{Landsteiner2016}.
While anomalous currents can be corrected by interactions among particles, the chiral anomaly~\cite{Adler2004}, the imbalance between left and right movers due to quantum fluctuations~\cite{Bertlmann1996}, remains universal.
The universal chiral anomaly coefficient appears in dissipationless transport currents~\cite{Landsteiner2016} due to electromagnetic~\cite{Xiong2015, Huang2015, Reis2016, Marsh2017, Liang2018, Yuan2020} and strain fields~\cite{Ilan2020} in condensed matter systems, and in anomalous transport in the quark-gluon plasma~\cite{Kharzeev2008, Landsteiner2016}.

The universality of the chiral anomaly, i.e., its robustness against local perturbations, is intimately related to topological properties of the Hamiltonian~\cite{Armitage2018}.
While most of these Hamiltonians respect the Hermiticity condition, non-Hermitian Hamiltonians, which are effective descriptions of systems coupled to an environment~\cite{Bernard2002, Chen2017,  Lieu2018, Carlstrom2018, Bergholts2019}, introduce new classes of topological systems which do not have any Hermitian analogue~\cite{ Magnea2008, Mussardo2010, Xu2017, Zhou2019, Kawabata2019}.
This motivates the question we address in this paper:~are chiral quantum anomalies in non-Hermitian systems different in any way from those of (anti-)Hermitian systems? 

\begin{figure}
    \centering
    \includegraphics[width=\columnwidth]{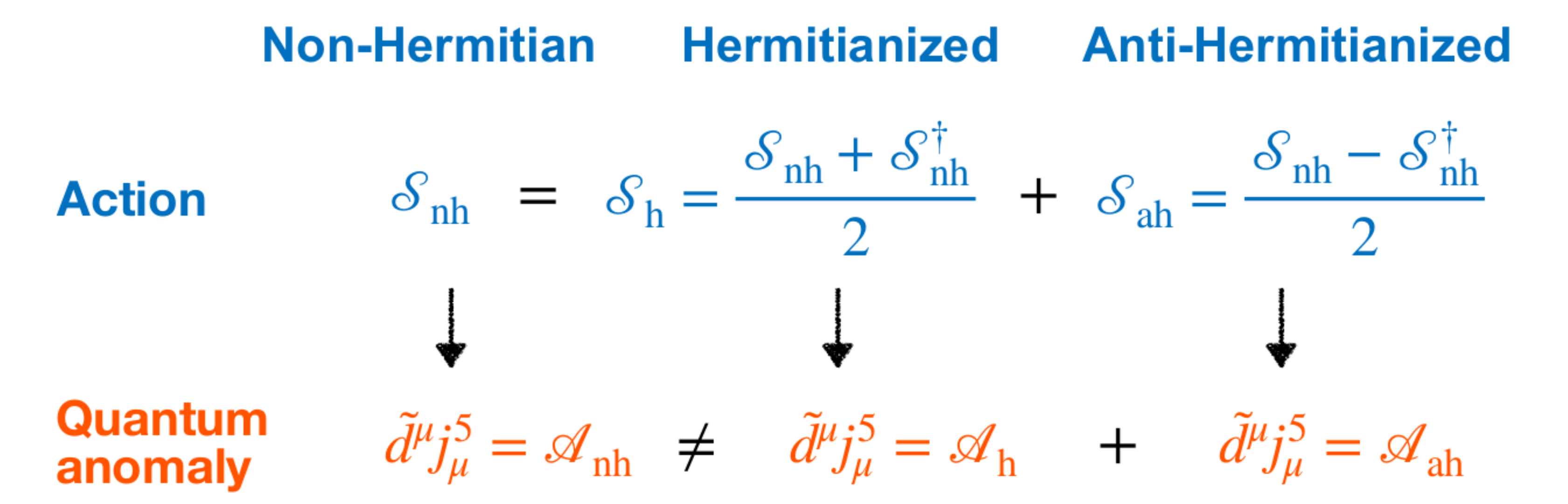}
    \caption{Anomalies in Hermitianized, anti-Hermitianized and non-Hermitian actions. For the non-Hermitian Hamiltonian~${\cal S}_{\rm nh}$, as well as its Hermitianized~(${\cal S}_{\rm h}$) and anti-Hermitianized~(${\cal S}_{\rm ah}$) forms, the conserved classical chiral current will be anomalous upon including quantum fluctuations. Note that the definition of current remains the same in all systems and $\tilde{d}_{\mu}=f_{\mu}^{\nu} \partial_{\nu}$ where $f_{\mu}^{\nu}$ is introduced in Table~\ref{tab:convertI}. While 
    ${\cal S}_{\rm nh}={\cal S}_{\rm h} +{\cal S}_{\rm ah}$, for the anomalous currents we get ${\cal A}_{\rm nh}\neq {\cal A}_{\rm h} + {\cal A}_{\rm ah}$.
    \label{fig:schematic}}
\end{figure}
So far, one of the main links between anomalous field theories and non-Hermitian systems are the extension of index theorem to non-Hermitian systems~\cite{Xi2020} and the reinterpretation of the non-Hermitian skin effect as a consequence of an anomaly~\cite{Lee2019II, Kawabata2020}.
The skin effect~\cite{Martinez2018b, Xiong_2018, Kunst2018, Martinez2018, Yao2018, Lee2019} is one of the central differences between topological non-Hermitian and Hermitian systems and results in a macroscopic number of states accumulating at the boundary of the system.
The action describing a non-Hermitian system in $d+0$ dimensions is mathematically equivalent to that of a $(d-1)+1$ dimensional Hermitian system~\cite{Kawabata2020}. 
This establishes a link between the non-Hermitian skin effect of a $d$ dimensional system to a $(d-1)+1$ dimensional anomalous Hermitian theory. 
Additionally, Ref.~\cite{Xi2020} showed that a lattice version of a ${\cal P T}$ symmetric continuum model, i.e., non-Hermitian systems with real spectrum, 
can display quantum anomalies similar to those in Hermitian systems.
In ${\cal PT}$-symmetric systems the chiral magnetic effect~\cite{Fukushima2008}, an anomalous transport current parallel to a magnetic field, can exist in equilibrium in non-Hermitian systems~\cite{Chernodub2020}.

In this paper, we present a generic $d+1$ dimensional non-Hermitian formulation that consolidates anomalous chiral responses in $d+1$ dimensional Hermitian, anti-Hermitian, and non-Hermitian systems.
More specifically, we study the chiral anomaly in two and four space-time dimensions for non-Hermitian massless Weyl fermions, with complex Fermi velocities, coupled to complex gauge fields. 
We also (anti-)symmetrize the non-Hermitian action and introduce an (anti-)Hermitianized action. 
All these models lack ${\cal P T}$ and Lorentz symmetry.
By establishing a unified notation for the Hermitianized, anti-Hermitianized, and non-Hermitian systems, we show that all of these systems classically conserve both the vector and chiral currents but exhibit anomalous responses when quantized~(Fig.~\ref{fig:schematic}). 
Furthermore, the anomalous currents in non-Hermitian systems are not given by the simple addition of currents associated with the Hermitianized and anti-Hermitianized actions, as would be expected classically~(Fig.~\ref{fig:schematic}).

In the main text, we follow Fujikawa's path integral approach~\cite{Fujikawa1979, Fujikawa1980, Fujikawa1981, Fujikawa2004}, which proves to be a controlled method to evaluate the chiral anomalies emerging from the non-Hermitian system. 
Based on these results, we present a Chern-Simons description of our non-Hermitian models and briefly discuss currents associated with the non-Hermitian anomalous Hall effect and non-Hermitian chiral magnetic effect.
We also evaluate the $1+1$ dimensional anomalies using the diagrammatic method, for which we discuss some subtleties arising from the non-Hermicity of the action.
Detailed calculations can be found in the supplemental material~(SM)~\cite{SuppMat}, which also includes an effective bosonic theory for the $1+1$ dimensional non-Hermitian systems.
%


\paragraph*{\bf Non-Hermitian chiral anomaly from Fujikawa's method-.} 
Within Fujikawa's method~\cite{Fujikawa1979, Fujikawa1980, Fujikawa1981, Fujikawa2004}, the covariant form of the chiral anomaly~\cite{Landsteiner2016} is evaluated by the change of the measure of the path integral after applying both vector and chiral transformations.

We apply Fujikawa's method to the three different actions presented in Fig.~\ref{fig:schematic}. 
These three systems, in the language of the path integral, correspond to a non-Hermitian action and its (anti-)symmetrized form, generating the (anti-)Hermitian action.

We consider a non-Hermitian, non-${\cal P T}$ symmetric system consisting of massless fermions, with complex Fermi velocities, coupled to non-Hermitian gauge fields $(V, W)$ described by the non-Hermitian Euclidean-space action ${\cal S}_{\rm{nh}}$ and the partition function ${\cal Z}_{\rm{nh}}$,
\begin{align}
{\cal Z}_{\rm{nh}} &\propto \int {\cal D} \Psi {\cal D} \bar{\Psi} 
e^{ 
 {\cal S}_{\rm{nh}}},
\label{eq:Z_nh}\\
{\cal S}_{\rm{nh}} &= \i \int {\rm d}^{d} x 
\Big[
\bar{\Psi} \gamma^{\mu} ({\zc D}_{\rm{nh}, \mu} \Psi)
\Big]
,
\label{eq:action_nh} \\
\slashed{\zc D}_{\rm{nh}} &=\gamma^{\mu} {\zc D}_{\rm{nh}, \mu} = \gamma^{\mu}  M_{\mu}^{\nu}
\partial_{\nu}
-\i \gamma^{\mu} M_{\mu}^{\nu}
\left(
V_{\nu}
+\gamma^{5} W_{\nu}
\right),
\end{align}
in units where $c=\hbar=1$.
The dimension $d$ is even, and the Greek indices take values between $1$ and $d$, where repeated indices are summed over.
The gamma matrices $\gamma^{\mu}$ obey $\{\gamma^\mu,\gamma^\nu\}=2g^{\mu\nu}$, where $g^{\mu\nu}=-\delta^{\mu \nu}$ is the Euclidean metric, and the fifth gamma-matrix is $\gamma^5 = - \prod_{\mu} \gamma^{\mu}$.
The $M$ is a rank $d$ diagonal matrix such that $M={\rm diag}[v_{1}, \ldots , v_{d}]$ with $v_{d}=1$ and spatial elements $v_{i \neq d}$ be complex-valued Fermi velocities.
%
%
Here ${\zc D}_{\rm nh}$ is the Dirac operator, and $\overline{\Psi} =\Psi^{\dagger} \gamma^{0} $ denotes the Dirac adjoint with $\gamma^{0}=\i \gamma^{4}$.

In the SM~\cite{SuppMat}, we propose a microscopic model based on a nonreciprocal anisotropic heterostructure as a platform to experimentally realize the introduced linear band system with complex Fermi velocities.

To show the differences between non-Hermitian and (anti-)Hermitian systems, we also explore the chiral anomaly in the (anti-)Hermitianized form of Eq.~\eqref{eq:action_nh}. By symmetrizing/anti-symmetrizing ${\cal S}_{\rm nh}$ in Eq.~\eqref{eq:action_nh}, the Hermitian/anti-Hermitian action~${\cal S}_{\rm{h}/\rm{ah}}$ yields
 \begin{align}
{\cal S}_{\rm{h}/\rm{ah}} &= \frac{\i}{2} \int {\rm d}^{d} x 
\Big[
\bar{\Psi} \gamma^{\mu} ({\zc D}_{\rm{nh},\mu} \Psi)
\mp 
\overline{({\zc D}_{\rm{nh},\mu} \Psi)} \gamma^{\mu} \Psi
\Big]
, \\
 &= \i 
\int {\rm d}^{d} x 
\bar{\Psi} \gamma^{\mu}
{\zc D}_{\rm{h}/\rm{ah}, \mu} \Psi,\label{eq:action_hah}
\end{align}
 where $\overline{({\zc D}_{\rm{nh}, \mu} \Psi)} = ({\zc D}_{\rm{nh}, \mu} \Psi)^{\dagger} \gamma^{0}$. Here, $ {\zc D}_{\rm h}, {\zc D}_{\rm ah} $ are the modified Dirac operators which are given by
 \begin{align}
     {\zc D}_{\rm{h}, \mu}
 &=
\Re[M_{\mu}^{\nu}] \partial_{\nu}
 -\i \Re[M_{\mu}^{\nu}  V_{\nu} ]
 -\i \gamma^{5}  \Re[M_{\mu}^{\nu}  W_{\nu}]
 ,\label{eq:Dd_h}\\
{\zc D}_{\rm{ah}, \mu}
 &=
\i \Im[M_{\mu}^{\nu}] \partial_{\nu}
 + \Im[M_{\mu}^{\nu}  V_{\nu} ]
 + \gamma^{5}  \Im[M_{\mu}^{\nu}  W_{\nu}]
 .\label{eq:Dd_ah}
 \end{align}
 Note that our Hermitianized system with $M\in\mathbb{R}$ represents the Lorentz preserving Hermitian model when $M= \id_{d \times d}$ and $\{V, W \} \in \mathbb{R}$. 

Our calculations are simplified by defining a unified representation that incorporates all cases 
  \begin{align}
   \tilde{\cal Z} &\propto \int {\cal D} \Psi {\cal D} \bar{\Psi} 
e^{  \tilde{{\cal S}}},
\,\text{ with }
\tilde{{\cal S} }
 = \i 
\int {\rm d}^{d} x 
\bar{\Psi} \gamma^{\mu}
\tilde{\zc D}_{ \mu} \Psi,\label{eq:action} \\
\tilde{{\zc D}}_{ \mu}
 &=
 \tilde{d}_{\mu} -\i \tilde{V}_{\mu} - \i \gamma^{5} \tilde{W}_{\mu} .\label{eq:Dd}
\end{align}
Here $\tilde{d}_{\mu} = f_{\mu}^{\nu} \partial_{\nu}$.
The matrix $f$ and the gauge field $\tilde{A} \in \{ \tilde{V}, \tilde{W} \}$, with their associated elements given in Table~\ref{tab:convertI}, map our generic action~$\tilde{S}$ into the Hermitianized, anti-Hermitianized, or non-Hermitian actions.

\begin{table}
\begin{center}
\begin{tabular}{c|c|c|c}
&  $\mathcal{S}_{\rm{h}}$& $\mathcal{S}_{\rm{ah}}$&  $\mathcal{S}_{\rm{nh}}$
    \\[-0.05em]
     \hline
    &&\\[-0.05em]
    $f_{\mu}^{\nu}$ & $\, \Re[M_{\mu}^{\nu}] \, $  & $\, \i \Im[M_{\mu}^{\nu}] \,$ &$\, M_{\mu}^{\nu} \,$
    \\[-0.05em]
    \hline
    &&\\[-0.05em]
    $\tilde{A}_{\mu}$& $\, \quad \Re[M_{\mu}^{\nu} A_{\nu}]\, $ &$\, \quad \i \Im[M_{\mu}^{\nu} A_{\nu}]\, $ &$\, M_{\mu}^{\nu} A_{\nu}\, $ 
      \\[-0.05em]
     \hline
    &&\\[-0.05em]
   $\tilde{\cal F}_{2}$  &  $\, 4 \pi |\Re[v_{\rm f} ]| \,$ &$\, 4 \pi |\Im[v_{\rm f} ]| \,$& $\, 4 \pi \sqrt{\det[B]} \,$
     \\[-0.05em]
    \hline
    &&\\[-0.05em]
    $\tilde{\cal F}_{4}$&$\, 32 \pi^2 |\det[\Re[M]]| $ \, & $\, 32 \pi^2 |\det[\Im[M]]|$ \, & $\, 32 \pi^2 \sqrt{\det[B]}$
    \\[-0.05em]
\end{tabular}
\end{center}
\caption{\label{tab:convertI} $f$, $\tilde{A}$  and $\tilde{\cal F}$ for Hermitianized, $\mathcal{S}_{\rm{h}}$, anti-Hermitianized, $\mathcal{S}_{\rm{ah}}$,  and non-Hermitian, $\mathcal{S}_{\rm{nh}}$, systems. $A$ stands for gauge fields $V$ and $W$. $\tilde{\cal F}$ is presented for $2$ and $4$ dimensions. The matrix $B$ for non-Hermitian systems is introduced in Eq.~\eqref{eq:constB}.
}
\end{table}

Classically, $\tilde{{\cal S} }$, or equivalently, ${\cal S}_{\rm{h}} $ , ${\cal S}_{\rm{ah}} $  or ${\cal S}_{\rm{nh}} $, carries a local $U_{\rm{A}}(1) \times U_{\rm{V}}(1)$ symmetry, where $U_{\rm{A}(\rm{V})}(1)$ is the chiral~(vector) $U(1)$ symmetry; see the SM~\cite{SuppMat}.
Quantum mechanical fluctuations reduce this underlying classical symmetry in all cases,  
as we will show in the following.

 While the quantum action in Eq.~\eqref{eq:action} remains invariant under a chiral transformation of the spinor,
 \begin{equation}
 \Psi_{\rm rot} = e^{- \i \gamma^{5} \beta} \Psi, \label{eq:rotPsibeta}
 \end{equation}
the rotated partition function acquires a phase given by the Jacobian of its measure.
Hence, we continue by calculating the Jacobian of the rotated path-integral measure.
For this purpose, one may decompose the rotated spinors into the eigenbasis of the non-Hermitian Dirac operator, ($\tilde{\slashed{\zc D}} \neq \tilde{\slashed{\zc D}}^{\dagger}$), which is biorthogonal with right and left eigenvectors.
 However, rather than working with the eigenbasis of $\tilde{\slashed{\zc D}}$, which introduces unnecessary complications due to the biorthogonal basis of a non-Hermitian $\tilde{\slashed{\zc D}}$, we employ an equally correct but more straightforward approach which is using the eigenbases of the Hermitian Laplacian operators $\slashed{\zc D} \slashed{\zc D}^{\dagger}$ and $ \slashed{\zc D}^{\dagger} \slashed{\zc D}$, given by
\begin{align}
   \tilde{\slashed{\zc D}} \tilde{\slashed{\zc D}}^{\dagger} |\eta_{n} \rangle &= |\lambda|^{2}_{n} | \eta_{n} \rangle , \quad \tilde{\slashed{\zc D}}^{\dagger} \tilde{\slashed{\zc D}} | \xi_{n} \rangle = |\lambda|^{2}_{n} | \xi_{n} \rangle,\\
    \tilde{\slashed{\zc D}}^{\dagger} |\eta_{n} \rangle &= \lambda^{*}_{n} |\xi_{n} \rangle ,
    \quad \tilde{\slashed{\zc D}} | \xi_{n} \rangle = \lambda_{n} |\eta_{n} \rangle.
\end{align}
Here $\{ \lambda_{n} \}$ are complex eigenvalues and $\{| \xi \rangle \}$ and $\{ | \eta \rangle \}$ are the corresponding eigenvectors, see the SM~\cite{SuppMat}. 
Using these eigenbases and following the standard Fujikawa method, allows us to write
\begin{align}
 {\cal D} \Psi_{\rm rot} {\cal D}\bar{\Psi}_{\rm rot}= &\tilde{{\cal J}}^{5}[\beta] {\cal D} \Psi {\cal D}\bar{\Psi}= e^{\tilde{{S}}^{5}[\beta]} {\cal D} \Psi {\cal D}\bar{\Psi},\\
 \tilde{{ S}}^{5}[\beta] =& \i \int {\rm d}^{d} x \beta(x) \tilde{{\cal A}}^{5}(x),\label{eq:S5beta_main}
   \end{align}
    where $\tilde{{\cal J}}^{5}$ is the path integral Jacobian due to an infinitesimal chiral transformation. The exponent of this Jacobian is regularized by the heat-kernel regularization method~\cite{Nakahara1990, regFuji}, see details in the SM~\cite{SuppMat}. 
    Up to the first-order in the fields, $ \tilde{{\cal A}}^{5} $ in $d=2$ dimensions reads
 \begin{align}
 \tilde{{\cal A}}^{5}  =&\frac{ -\varepsilon^{\mu \nu} }{\tilde{\cal F}_{2}} 
       \left[
\i (
  \tilde{F}_{\mu \nu}[ \tilde{V}^{\dagger}] 
         -
  \tilde{F}^{\dagger}_{\mu \nu}[ \tilde{V}] 
  )
 \right]
, \label{eq:A5_1p1_mt} 
 \end{align}
 and up to the second-order in the fields, $\tilde{\cal A}^{5}$ in $d=4$ dimensions casts 
    \begin{align}
\tilde{\cal A}^{5}
 =&
   \frac{   \varepsilon^{\mu \nu \eta \zeta}}{\tilde{\cal F}_{4}} 
     \left[
\tilde{F}_{\mu \nu}[ \tilde{V}^{\dagger}] 
\tilde{F}_{\eta \zeta}[ \tilde{V}^{\dagger}] 
+
 \tilde{F}^{\dagger}_{\mu \nu}[ \tilde{V}] 
\tilde{F}^{\dagger}_{\eta \zeta}[ \tilde{V}] 
  \right.
    \nonumber \\
  &
  \left.
+
\tilde{F}^{\dagger}_{\mu \nu} [\tilde{W}]
  \tilde{F}^{\dagger}_{\eta\zeta} [\tilde{W}]
  +
 \tilde{F}_{\mu \nu} [\tilde{W}^{\dagger}]
 \tilde{F}_{\eta \zeta} [\tilde{W}^{\dagger}]
 \right]
.\label{eq:A5_3p1}
\end{align}

Here, $\tilde{F}_{\mu \nu} [\tilde{A}]= \tilde{d}_{\mu} \tilde{A}_{\nu}- \tilde{d}_{\nu} \tilde{A}_{\mu} $, and $\tilde{F}^{\dagger}_{\mu \nu} [\tilde{A}]= \tilde{d}^{\dagger}_{\mu} \tilde{A}_{\nu}- \tilde{d}^{\dagger}_{\nu} \tilde{A}_{\mu} $ with $\tilde{d}_{\mu}^{\dagger} = - f^{ * \nu }_{\mu} \partial_{\nu}$.  The explicit form of $\tilde{{\cal F}}_{d+1}$ for all cases is presented in Table.~\ref{tab:convertI}, where it is formulated in terms of the determinant of a matrix $B$, with matrix elements:
  \begin{equation}
 B^{\alpha \beta}= \delta^{\mu \nu}
 f_{\mu}^{*\alpha} f_{\nu}^{\beta}
 -
  \frac{1}{2} [ \gamma^{ \mu},  \gamma^{ \nu}]
\frac{f_{\mu}^{*\alpha} f_{\nu}^{\beta} 
- f_{\mu}^{\alpha} f_{\nu}^{*\beta}   
}{2}. \label{eq:constB}
 \end{equation}
For real gauge-fields $V$ and $W$, and for $M=\id_{d \times d }$, $ \tilde{{\cal A}}^{5}$ reproduces the Lorentz invariant Hermitian results~\cite{Kim1988, Bertlmann1996, SuppMat}.
 
We evaluate the change of the path-integral measure under a local vector transformation using the same method as for the chiral rotations
 \begin{align}
  \Psi_{\rm rot} = e^{-\i \kappa(x)} \Psi, \label{eq:rotPsikappa}
 \end{align}
 where $\kappa$ is the rotation angle.
 For an infinitesimal $\kappa$, 
 \begin{equation}
  {\cal D} \Psi_{\rm rot} {\cal D}\bar{\Psi}_{\rm rot}
   =e^{\tilde{ S}[\kappa]}{\cal D} \Psi {\cal D}\bar{\Psi}
 =e^{\i \int {\rm d}^{d} x \kappa(x) \tilde{\cal A}} {\cal D} \Psi {\cal D}\bar{\Psi}
 .
  \end{equation}
Up to the first-order in the fields,  $\tilde{\cal A}$ in $d=2$ dimensions is
  \begin{align}
 \tilde{\cal A}= & \frac{-\varepsilon^{\mu \nu} }{\tilde{\cal F}_{2}} 
     \left[ 
 \i \big( 
 \tilde{F}^{\dagger}_{\mu \nu} [\tilde{W}]
  -  \tilde{F}_{\mu \nu} [\tilde{W}^{\dagger}]
      \big)
 \right]
    ,\label{eq:A_1p1_mt}
 \end{align}
 and up to the second-order in the fields, $\tilde{\cal A}$ in $d=4$ dimensions casts~\cite{NoteAA5}
\begin{align}
\tilde{\cal A} 
 &=
  \frac{- \varepsilon^{\mu \nu \eta \zeta} }{\tilde{\cal F}_{4}} 
   \left[ 
 \tilde{F}_{\mu \nu}[ \tilde{V}^{\dagger}] 
 \tilde{F}_{\eta \zeta} [\tilde{W}^{\dagger}]
+
\tilde{F}^{\dagger}_{\mu \nu}[ \tilde{V}]  
 \tilde{F}^{\dagger}_{\eta \zeta} [\tilde{W}] 
 \right],\label{eq:A_3p1}
 \end{align}
In the limit where $W=0$, $V\in \mathbb{R}$ and $M=\id_{d \times d}$, $\tilde{\cal A} $ reduces to the well-known Lorentz preserving Hermitian result ${\cal A}=0$~\cite{ Fujikawa1979, Bertlmann1996}.

The rotated action in Eq.~\eqref{eq:action} is modified under the chiral and vector transformations in Eqs.~(\ref{eq:rotPsibeta},~\ref{eq:rotPsikappa}), such that
\begin{align}
\label{eq:currentrot}
    \tilde{\cal S}_{\rm rot} -\tilde{\cal S} =& -  \int {\rm d}^{2} x   \Big[\beta(x)  \tilde{d}_{\mu}  j^{5,\mu} 
    - \kappa(x) 
      \tilde{d}_{\mu}  j^{\mu} \Big],
\end{align}
where the chiral and vector currents are $j^{5,\mu}= \bar{\Psi} \gamma^{\mu} \gamma^{5} \Psi$ and $j^{\mu} = \bar{\Psi} \gamma^{\mu} \Psi$, respectively. To enforce the invariance of $\tilde{\cal Z}$ under the chiral and vector rotations, we differentiate the partition function with respect to $\beta$ and $\kappa$ and obtain the anomaly equations $\tilde{\cal A}^{5}= \i \tilde{d}_{\mu}  j^{5,\mu} $ and $\tilde{\cal A}= -\i \tilde{d}_{\mu}  j^{\mu}$.

After analytical continuation $\tau \rightarrow \i t$, we obtain the divergence of currents in the Minkowski space.  Using the elements of $M$ and $V$, we rewrite the divergence of chiral currents as
\begin{equation}
\tilde{d}_{\mu} j^{5,\mu} \propto  v_{1} \tilde{E}_{1}^{\dagger} + v^{*}_{1} \tilde{E}_{1} 
\qquad 
\text{
 in $d=2$
} 
,\label{eq:dJ5_1p1}
\end{equation}
\begin{align}
\tilde{d}_{\mu} j^{5,\mu} \propto& v_{1}v_{2}v_{3} (\tilde{E}^{\dagger}.\tilde{B}^{\dagger} + \tilde{E}^{5\dagger}.\tilde{B}^{5\dagger}) 
\nonumber \\
&
+v^{*}_{1}v^{*}_{2}v^{*}_{3} (\tilde{E}.\tilde{B}+ \tilde{E}^{5}.\tilde{B}^{5}) 
\quad 
\text{ in $d=4$}.
\label{eq:dJ5_3p1}
\end{align}
Here, the complex electric fields read $\tilde{E}_{j}= (\exp[2 \i \phi_{j} ] \partial_{t} V_{j} -  \partial_{j} V_{0})$ and $\tilde{E}^{5}_{j}= (\exp[2 \i \phi_{j} ] \partial_{t} W_{j} -  \partial_{j} W_{0})$, where $\exp[\i \phi_{j}]=v_{j}/ |v_{j}|$ with $i,j,k \neq t$ being a spatial index. Similarly, the complex magnetic field casts $\tilde{B}^{i} =\varepsilon^{ijk}\tilde{B}_{jk}$ and  $\tilde{B}^{5,i} =\varepsilon^{ijk} \tilde{B}_{jk}^{5}$ with $\tilde{B}_{jk}= \exp[2 \i \phi_{k}] \partial_{j} V_{k} -\exp[2 \i \phi_{j}]  \partial_{k} V_{j} $ and $\tilde{B}_{jk}^{5}= \exp[2 \i \phi_{k}] \partial_{j} W_{k} -\exp[2 \i \phi_{j}]  \partial_{k} W_{j}$. It is notable that $\tilde{d}_{\mu} j^{5,\mu}$ in Eqs.~(\ref{eq:dJ5_1p1}, \ref{eq:dJ5_3p1}) obey the common form of the chiral anomaly reported in Hermitian systems in the absence~\cite{Fujikawa2004} or presence~\cite{Rylands2021} of interactions.

The non-Hermitian anomaly $[\tilde{d}_{\mu} j^{5,\mu}]_{\rm nh}$ is different from the naive summation of $[\tilde{d}_{\mu} j^{5,\mu}]_{\rm h}+[\tilde{d}_{\mu} j^{5,\mu}]_{\rm ah}$. One can see this difference by noticing, for example, that the prefactor $\tilde{\cal F}_{2}$ in \eqref{eq:A5_1p1_mt} (or $\tilde{\cal F}_{4}$ in Eq.~(B73)) is different for all three cases. This is the  main result of this work (see Fig.~\ref{fig:schematic}).

\paragraph*{\bf A Chern-Simons description of non-Hermitian Weyl semimetals-.} To explore some physical consequences of non-Hermitian anomalies, we explore a situation in which the chiral gauge field $W$ is absent. In this case, the anomaly-induced action, given in Eq.~\eqref{eq:S5beta_main}, in real-time representation casts a Chern-Simons action as
\begin{align}
\tilde{S}^{5}[\beta]=&
\int {\rm d}t {\rm d}^3 x 
 \frac{   4 \varepsilon^{\mu \nu \eta \zeta}}{\tilde{\cal F}_{4}} 
\tilde{d}_{\mu}\beta(x)  \tilde{V}_{\nu}^{\dagger}
\tilde{d}_{\eta}  \tilde{V}_{\zeta}^{\dagger}
\nonumber \\
&
+
\int {\rm d}t {\rm d}^3 x 
 \frac{   4 \varepsilon^{\mu \nu \eta \zeta}}{\tilde{\cal F}_{4}} 
 \tilde{d}^{\dagger}_{\mu} \beta(x) \tilde{V}_{\nu} 
\tilde{d}^{\dagger}_{\eta}  \tilde{V}_{\zeta} ,
\label{eq:CS}
\end{align}
where we have performed an integration by parts and dropped a total derivative term. We then obtain the associated current as a sum of the functional derivatives of $\tilde{S}^{5}[\beta]$ with respect to both $V$ and $V^\dagger$, which results in
\begin{align}
M_{\nu}^{\alpha} j^{\nu}
=&\frac{8 \varepsilon^{\mu \nu \eta \zeta} \partial_{\delta} \beta }{\tilde{\cal F}_{4}}
\Re[M^{\alpha *}_{\nu} M^{\delta}_{\mu} M^{\rho}_{\zeta} \tilde{E}^{\dagger}_{\rho}],
\, \mu =1,2,3,
\label{eq:hall}
\\
M_{\nu}^{\alpha} j^{\nu}
=&\frac{8 \varepsilon^{0 \nu \eta \zeta} \partial_{t} \beta }{\tilde{\cal F}_{4}}
\Re[M^{\alpha *}_{\nu} M^{\iota}_{\eta} M^{\rho}_{\zeta} \tilde{B}^{\dagger}_{\iota \rho}],
\label{eq:cme}
\end{align}
where the complex electric~($\tilde{E}$) and magnetic fields~($ \tilde{B}$) are defined in the paragraph below Eq.~\eqref{eq:dJ5_3p1}. In the limit where $V \in \mathbb{R}$ and $M= \id_{4\times 4}$, Eq.~\eqref{eq:hall} represents the Hermitian anomalous Hall effect and Eq.~\eqref{eq:cme} coincides with the Hermitian chiral magnetic effect~\cite{Zyuzin2012}. $\partial_{0} \beta $ and $\partial_{\delta} \beta$ can be associated to the energy and spatial separation of Weyl nodes, respectively, which for a non-Hermitian Weyl semimetal can be complex-valued. As a result, one can view the complex currents in Eqs.~(\ref{eq:hall},\ref{eq:cme}) as a representation of the non-Hermitian anomalous Hall effect and non-Hermitian chiral magnetic effect.

Assigning $\bolds{j}$ as the polarization current as $\bolds{j} =\partial_{t} \bolds{P}$ where $\bolds{P}$ denotes the electric polarization, Eq.~\eqref{eq:cme} simplifies to
\begin{align}
M_{\nu}^{\alpha} P^{\nu}
=&\frac{8 \varepsilon^{0 \nu \eta \zeta} \beta }{{\cal F}_{4}}
\Re[M^{\alpha *}_{\nu} M^{\iota}_{\eta} M^{\rho}_{\zeta} \tilde{B}^{\dagger}_{\iota \rho}].
\end{align}
When $\beta=\pi$, which can occur in time-reversal-invariant  topological insulators~\cite{Qi2008}, the  quantized magnetoelectric effect of Hermitian systems, written as $\bolds{P} = e^2 \bolds{B}/4\pi$, is modified to
\begin{align}
\label{eq:pol}
M_{\nu}^{\alpha} P^{\nu}
=&\frac{ \varepsilon^{0 \nu \eta \zeta}  }{4 \pi \sqrt{\det[B]} }
\Re[M^{\alpha *}_{\nu} M^{\iota}_{\eta} M^{\rho}_{\zeta} \tilde{B}^{\dagger}_{\iota \rho}],
\end{align}
in non-Hermitian systems. Such a modified magneto-electric polarization will modify the electromagnetic responses of non-Hermitian topological insulators, including image monopole charges~\cite{Qi2009} or the Casimir effect~\cite{Grushin2011}. In the SM, we study the Witten effect as an example. The polarization in Eq.~\eqref{eq:pol} will change the induced electric charge created by a magnetic monopole from $e/2$ of the Hermitian system \cite{Rosenberg2010}, to an arbitrary value (see SM) that depends on the phase of the complex Fermi velocity.

\paragraph*{\bf One-loop diagrammatic calculation of the $1+1$ dimensional anomaly-.}
We also derive the chiral anomaly for the (anti)-Hermitianized, and non-Hermitian actions in $1+1$ dimensions using the diagrammatic method, see the SM~\cite{SuppMat}.
Within this approach, we integrate out the fermionic degrees of freedom from the underlying partition function, which results in an effective action for the gauge fields.
The functional integration of an Hermitian action, with a Lagrangian density $\mathcal{L}=\Psi^\dagger(\i\gamma^0\slashed{\zc D})\Psi$, requires an orthogonal eigenbasis of the self-adjoint operator $(\i\gamma^0\slashed{\zc D})$ to yield the effective action $\Gamma[V,W]=-\i\ln[\det(\i\gamma^0\slashed{\zc D})]$.
By using the product rules of determinants and logarithms, and omitting any constant vacuum contributions $\Gamma[V,W]$ is rewritten as $\Gamma[V,W]=-\i\ln[\det(\i\slashed{\zc D})]$.
When the operator $(\i\gamma^0\slashed{\zc D})$ is not self-adjoint, one should instead use the corresponding biorthogonal eigenbasis to evaluate the functional integral.
To take advantage of the well-developed Hermitian field theory and to avoid the complexities of finding functional determinants from a biorthogonal basis, we construct the self-adjoint operator $\slashed{\zc D}\slashed{\zc D}_\mathcal{L}^\dagger$, where $\slashed{\zc D}_\mathcal{L}^\dagger \equiv \gamma^\mu{\zc D}_\mu^\dagger$.
Formally, this is done by considering the sum of the effective actions with respect to the operators $\i\gamma^0\slashed{\zc D}$ and $\i\gamma^0\slashed{\zc D}_\mathcal{L}^\dagger$ and using properties of the determinant and the logarithm to obtain $\Gamma[V,W]=-\frac{\i}{2}\ln[\det(\slashed{\zc D}\slashed{\zc D}_{\mathcal{L}}^\dagger)]$~(see SM~\cite{SuppMat}).
For the various actions given in Eqs.~(\ref{eq:action_nh}, \ref{eq:action_hah}), we expand  
this effective action up to the second-order in the gauge fields. These calculations give rise to divergent momentum integrals which are treated with gauge-invariant regularization methods~\cite{Peskin1995}.

The vector (chiral) currents are defined as the sum of the functional derivatives of the effective action with respect to $V$ and $V^\dagger$ ($W$ and $W^\dagger$), from which the expressions for divergence of the vector and chiral currents~(terms first-order in the gauge fields), e.g., in Eq.~(\ref{eq:dJ5_1p1}), follow.


\paragraph*{\bf Summary-.} 
We have found non-Hermitian anomalies in massless Dirac fermions with complex velocities coupled to non-Hermitian gauge fields. We have presented a unified non-Hermitian formulation to bring the Hermitianized, anti-Hermitianized, and non-Hermitian cases under one umbrella.
Our results show that non-Hermiticity allows new anomalous terms in the conservation laws for the chiral current in both in $2$ and $4$ dimensions. Interestingly, these anomalous terms could not be inferred by simply adding the Hermitianized and anti-Hermitianized results, as would be expected classically, see Fig.~\ref{fig:schematic}. In this sense, these anomalies are different and richer than those that occur in Hermitian systems. 
We further demonstrate this point by presenting non-Hermitian anomalous Hall effect and non-Hermitian chiral magnetic effect.

Finally, we note that novel parity-like anomalies may exist in non-Hermitian systems. Exploring them might give a different interpretation to the non-universality of the Hall response in non-Hermitian Chern insulators~\cite{Hirsbrunner2019}.

\paragraph*{\bf Acknowledgements-.}
We are grateful to M. Arminjon, J.~H. Bardarson, M. Chernodub, A. Cortijo, K. Landsteiner, T. Neupert, M.~A.~H. Vozmediano, and S.~B. Zhang for fruitful discussions, and to S. Aich for discussions and input at early stages of this project. 
A.~G.~G. and Sh.~S. acknowledge financial support by
 the ANR under the grant ANR-18-CE30-0001-01~(TOPODRIVE).
  A.~G.~G. is also supported by the European Union 
  Horizon 2020 research and innovation programme
   under grant agreement No. 829044~(SCHINES). This project has further received funding from the European Research Council~(ERC) under the European Union’s Horizon 2020 research and innovation programme~(grant agreement No. 679722) and by the Swedish Research Council through grants number 2019-04736 and 2020-00214.

\paragraph*{\bf Author contributions-.}  
Sh.~S. carried out the path integral and bosonization calculations and proposed the microscopic model with input from all authors.
J.~D.~H. carried out the diagrammatic calculations with input from all authors.
Sh.~S. drafted the manuscript, to which all authors contributed.
A.~G.~G. conceived and supervised the project.

\bibliography{nHanomaly.bib}

\begin{thebibliography}{65}
\bibitem [{regFuji()}]{regFuji}%
  \BibitemOpen
  \bibinfo {note} {To be precise, our employed heat-Kernel regularization
  suppresses states with large eigenvalues in the Euclidean space.}\BibitemShut
  {Stop}%
\bibitem [{NR_AN()}]{NR_AN}%
  \BibitemOpen
  \bibinfo {note} {The suggested model is anisotropic as there is a preferred
  growth direction of multilayers. Our system is nonreciprocal due to the
  imbalance between tunneling to the bottom and top surfaces of the nearest
  neighboring TI layers.}\BibitemShut {Stop}%
\end{thebibliography}

\newpage

\onecolumngrid

\setcounter{secnumdepth}{5}
\renewcommand{\theparagraph}{\bf \thesubsubsection.\arabic{paragraph}}

\renewcommand{\thefigure}{S\arabic{figure}}
\setcounter{figure}{0}

\appendix

\section*{Supplemental Material}


\section{Weyl semimetals with complex energy dispersion}

\begin{figure}[b]
    \centering
    \includegraphics[width=0.4\columnwidth]{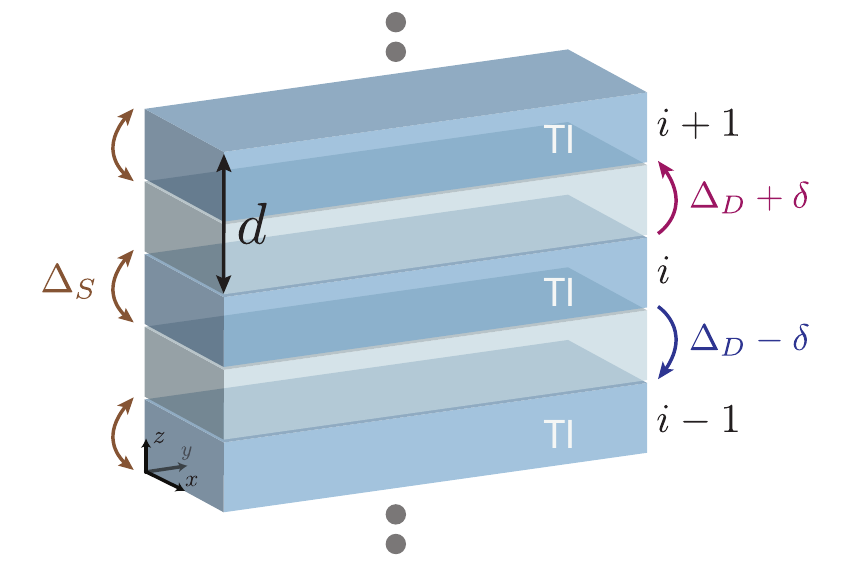}
    \caption{
    Schematic drawing of the nonreciprocal multilayer structure. The blue~(grey) layers are TI~(NI) layers. The magnetization direction in all TI layer are assumed to be along the $z$ direction. Tunneling between bottom~(top) surface state of TI layer to the top~(bottom) surface state of the nearest neighboring TI layer has the amplitude $\Delta_{D}-\delta$~($\Delta_{D}+\delta$). Tunneling within the same TI layer is with amplitude $\Delta_{S}$. The distance between two neighboring TI layers is $d$. TI layers are labeled by the symbol $i$. 
    \label{fig:schematic_app}}
\end{figure}

To experimentally realize our target system which is a linear band model with complex Fermi velocities, we propose a nonreciprocal anisotropic heterostructure~\cite{NR_AN} where layers of topological insulators~(TI) are separated by layers of nonreciprocal spacer material, e.g., nonreciprocal normal insulator~(NI), as shown in Fig.~\ref{fig:schematic_app}. This model is a generalization of the superlattice heterostucture with a real band structure introduced in Refs.~\cite{Burkov2011, Zyuzin2012}, which capture the anomalous Axion insulator and Weyl phases of MnBi$_2$Te$_4$/(Bi$_2$Te$_3$)$_n$ heterostructures~\cite{Gu2021}. The Hamiltonian of our non-Hermitian system reads
\begin{align}
H=\sum_{\bolds{k}_{\bot}, ij}\Big[ &
v_{\rm f} 
\tau^{z} (\hat{z} \times \bolds{\sigma}).\bolds{k}_{\bot} \delta_{ij}
+ m \sigma^{z} \delta_{ij}
+\Delta_{S} \tau^{x} \delta_{ij}
+\lambda \tau^{y} \sigma^{z} \delta_{ij}
+V \tau^{z} \delta_{ij}
\nonumber \\
&
+\frac{1}{2} (\Delta_{d} -\delta) \tau^{+} \delta_{j,i+1}
+\frac{1}{2} (\Delta_{d} +\delta) \tau^{-} \delta_{j,i-1}
\Big]c^{\dagger}_{\bolds{k}_{\bot,i} } c_{\bolds{k}_{\bot,j}}
,
\label{eq:Ham_TINI}
\end{align}
where $c^{\dagger}_{\bolds{k}_{\bot},i}~(c_{\bolds{k}_{\bot},i} )$ creates~(annihilates) an electron on the surface states of the $i$th TI layer with perpendicular momentum $\bolds{k}_{\bot}=(k_{x},k_{y})$. 

The first term in the Hamiltonian describes the top and bottom surface states of a single TI. We assume that the surface Dirac fermions in the 2D Brillouin zone have the same Fermi velocities, set by $v_{\rm f}$, along the $k_x$ and $k_y$ directions. $\bolds{\sigma}~(\bolds{\tau})$ are Pauli matrices acting on the real spin (surface pseudospin) degrees of freedom. Exchange spin splitting, induced by doped magnetic impurities, is captured by the second term in Eq.~\eqref{eq:Ham_TINI}. Tunneling between the top and bottom surfaces within the same TI layer, given by the third term of the above Hamiltonian, is adjusted by $\Delta_{S}$. Hopping between the bottom~(top) surface state of the TI layer to the top~(bottom) surface state of the nearest neighboring TI layer has the amplitude $\Delta_{D}-\delta$~($\Delta_{D}+\delta$) given in the fourth~(fifth) term in Eq.~\eqref{eq:Ham_TINI}, see also Fig.~\ref{fig:schematic_app}, with $\tau^{\pm}=(\tau_{x} \pm \i \tau_{y})/2$. 
Here $\delta \ll \Delta_{D}$ is a real-valued parameter that encodes the possible non-reciprocity between upward and downward inter-layer hoppings, along the stacking direction. Because of this asymmetry, $\delta$ leads to a non-Hermitian Hamiltonian, as occurs in a 1D model with asymmetric hoppings, such as the celebrated Hatano-Nelson model~\cite{Hatano1996}. The spin-orbit interaction term, with coupling $\lambda$, and the term proportional to $V$ are included to break the inversion symmetry of the Hamiltonian.

Performing the Fourier transformation by $c_{\bolds{k}} = \sum_{j} c_{\bolds{k}_{\bot}, j} e^{-\i k_{z} d j }$ on Eq.~\eqref{eq:Ham_TINI} gives
\begin{align}
H
=
\sum_{\bolds{k}_{\bot}, k_{z}}\Big[ &
v_{\rm f} 
\tau^{z} (\hat{z} \times \bolds{\sigma}).\bolds{k}_{\bot} 
+ m \sigma^{z}
+\Delta_{S} \tau^{x}
+\lambda \tau^{y} \sigma^{z} 
+V \tau^{z}
\nonumber \\
&
+\Delta_{d} \left[  \tau^{x} \cos( k_{z} d) - \tau^{y} \sin(k_{z} d) \right]
+\i \delta  \left[  \tau^{x} \sin( k_{z} d) + \tau^{y} \cos(k_{z} d) \right]
\Big]c^{\dagger}_{\bolds{k} } c_{\bolds{k}}.
\label{eq:Hk}
\end{align}
Here it is evident that nonzero nonreciprocal paramater $\delta$ makes $H$ a non-Hermitian Hamiltonian with a complex energy dispersion.

To see how this model can realize the non-Hermitian Weyl systems with complex parameters that we consider in the main text, e.g., complex Fermi velocities, we present a few special cases in the following. In all cases, real/imaginary Weyl nodes in the low-energy complex band structure can be seen in associated figures, but presenting an analytical formulation for linear Hamiltonian close to Weyl nodes may depend on parameters.

\begin{itemize}
\item When $V=\lambda=0 $, Eq.~\eqref{eq:Ham_TINI} describes a three-dimensional Weyl semimetals~\cite{Burkov2011} with energies
\begin{align}
\epsilon^{2}_{\bolds{k}, \pm}
=&
-\delta ^2-2 i \delta  \Delta_{S} \sin (k_{z} d)
+\Delta(k_{z})^2
+v_{\rm f }^2 \left(k_{x}^2+k_{y}^2\right)+m^2
\pm 2 \left| m\right|  \sqrt{-\delta ^2-2 i \delta  \Delta_{S} \sin (k_{z} d)+\Delta(k_{z})^2 },
\label{eq:EVL0}
\end{align}
see also Fig.~\ref{fig:EVL0}.
Close to $k_{z}=0, 2n \pi$ with $n \in \Z$, the linear imaginary component of the $\epsilon^{2}_{\bolds{k}, \pm}$ touch zero with the effective Hamiltonian $H=\i v k_{z} +\text{other real terms}$ with $v= 2 \delta \Delta_{s} d$. 
The real part of $\epsilon^{2}_{\bolds{k}, \pm}$ also exhibits two Weyl nodes separated along the $k_{z}$ axis located at $k_{z}=\Re[k_{0}]$ with
\begin{align}
& k_{0}
=
\pm \frac{1}{d}
\arccos
\Big(
\frac{\pm \left| \delta \right|  \left| \Delta_{S}\right|  \sqrt{\left(\delta ^2-\Delta_{D}^2+\Delta_{S}^2\right)^2+m^4+2 m^2 \left(\delta ^2-\Delta_{D}^2-\Delta_{S}^2\right)}+\Delta_{D} \Delta_{S} \left(-\delta ^2+\Delta_{D}^2+\Delta_{S}^2-m^2\right)}{2 \Delta_{S}^2 \left(\delta ^2-\Delta_{D}^2\right)}
\Big),
\end{align}
which is evaluated on the intersection between $k_{x}=0$ and $k_{y}=0$ planes.

\begin{figure}
\includegraphics[width=5.47cm,height=5.47cm,keepaspectratio]{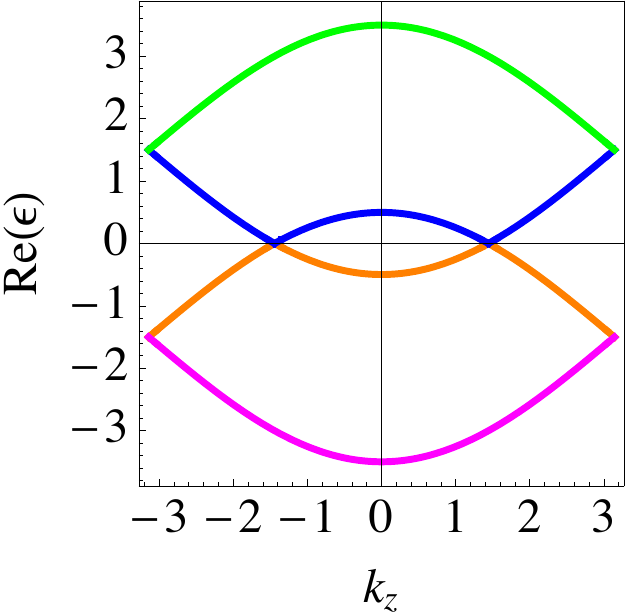}
\includegraphics[width=6cm,height=6cm,keepaspectratio]{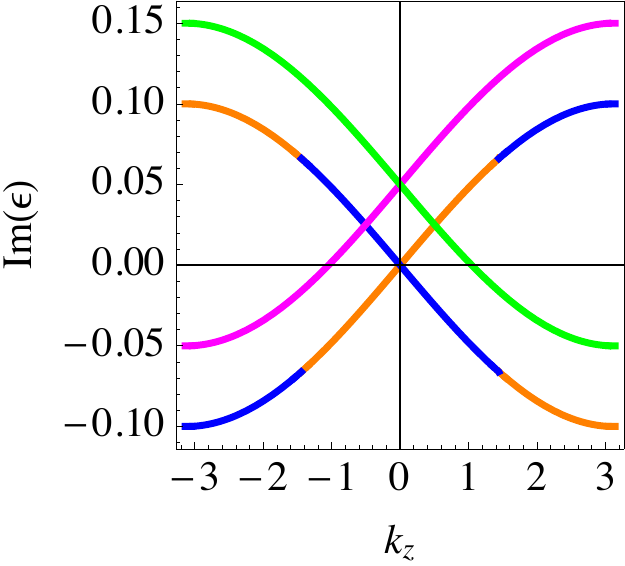}
\caption{Real~(left panel) and imaginary~(right) part of energy dispersion given in Eq.~\eqref{eq:EVL0} for $d=1$, $\delta=0.1$, $\lambda=0$, $m=1.5$, $V=0.0$, $\Delta_{S}=1.0$, and $\Delta_{D}=1.0$. For visibility purpose, green and pink lines in the right panels are shifted along the energy axis by $0.05$. 
\label{fig:EVL0}
}
\end{figure}

Expanding $\epsilon_{\bolds{k},\pm}^{2}$ around $k_{z}=\pi/d$ and assuming that $\Delta_{D}=\Delta_{S}$ gives 
\begin{align}
\epsilon^{2}_{\bolds{k}, \pm}
=&
v_{\rm f}^{2} (k_{x}^2 + k_{y}^{2}) 
+ 
\left[
m \pm  \sqrt{k_{z}^{2} \Delta_{D}^{2} - \delta^2}
\right]^2.
\end{align}
We can approximately rewrite it as
\begin{align}
\epsilon^{2}_{\bolds{k}, \pm}
=&
v_{\rm f}^{2} (k_{x}^2 + k_{y}^{2}) 
+ 
k_{z}^{2} \Delta_{D}^{2}
\left[
\frac{m d}{\pi \Delta_{D} } \pm  \sqrt{1 -\frac{ \delta^2 d^2}{\pi^{2} \Delta_{D}^{2}}}
\right]^2,
\\
=&
v_{\rm f}^{2} (k_{x}^2 + k_{y}^{2}) 
+ 
\tilde{k}_{z}^{2} v_{z}^2,
\end{align}

Similarly, expanding $\epsilon_{\bolds{k},\pm}^{2}$ around $k_{z}=0$ and assuming that $\Delta_{D}=\Delta_{S}$ gives 
\begin{align}
\epsilon^{2}_{\bolds{k}, \pm}
=&
v_{\rm f}^{2} (k_{x}^2 + k_{y}^{2}) 
+ 
\left[
m \pm \i \sqrt{
\Big(
k_{z}^{2} \Delta_{D}^{2} + \delta^2
-4 \Delta_{D}^2 + 2 \i k_{z} \delta \Delta_{D}
\Big)
}
\right]^2,
\end{align}
which can also be rewritten as
\begin{align}
\epsilon^{2}_{\bolds{k}, \pm}
=&
v_{\rm f}^{2} (k_{x}^2 + k_{y}^{2}) 
+ 
(k_{z} + \i \frac{\delta}{\Delta_{D} })^{2} \Delta_{D}^2
\left[
-\i \frac{m}{\delta} \pm \i \sqrt{
1- \frac{ 
\delta^2
-4 \Delta_{D}^2 
}{\delta^2}
}
\right]^2,
\\
=&
v_{\rm f}^{2} (k_{x}^2 + k_{y}^{2}) 
+ 
\tilde{k}_{z}^{2} v_{z}^2,
\end{align}
where $\tilde{k}_{z}= k_{z} + \i \frac{\delta}{\Delta_{D} } $ and  
$v_{z}^2= \Delta_{D}^2
\left[
-\i \frac{m}{\delta} \pm \i \sqrt{
1- \frac{ 
\delta^2
-4 \Delta_{D}^2 
}{\delta^2}
}
\right]^2$.

\item When $m=0$, the dispersion relation reads
\begin{align}
&\epsilon^{2}_{\bolds{k}, \pm}
=
-2 i \delta  \Delta_{S} \sin (k_{z} d)
+ \Delta(k_{z} )^2
+\lambda ^2+v_{\rm f}^2 \left(k_{x}^2+k_{y}^2\right)+V^2
-\delta ^2
\nonumber 
\\&
-\sqrt{2 \lambda ^2 \left(-\left(\delta ^2+\Delta_{D}^2\right) \cos (2 k_{z} d)+2 i \delta  \Delta_{D} \sin (2 k_{z} d)\right)+4 V^2 v_{\rm f}^2 \left(k_{x}^2+k_{y}^2\right)+2 \lambda ^2 \left(-\delta ^2+\Delta_{D}^2+2 v_{\rm f}^2 \left(k_{x}^2+k_{y}^2\right)\right)},
\label{eq:Em0}
\end{align}
see also Fig.~\ref{fig:Em0}.
The real part of $\epsilon^{2}_{\bolds{k}, \pm}$ also exhibits two Weyl nodes separated along the $k_{z}$ axis located at $k_{z}=\Re[k_{0}]$ with
\begin{align}
k_{0}=
\frac{\arctan\left(\frac{-\delta ^2-\Delta_{D}^2}{\delta ^2-\Delta_{D}^2},\frac{2 i \delta  \Delta_{D}}{\delta ^2-\Delta_{D}^2}\right)}{2 d},
\end{align}
which is evaluated on the intersection between $k_{x}=0$ and $k_{y}=0$ planes.
The imaginary part of the band structure also exhibit linear band dispersions with two Weyl nodes, but evaluating their separation seems not analytically feasible.

\begin{figure}
\includegraphics[width=5.47cm,height=5.47cm,keepaspectratio]{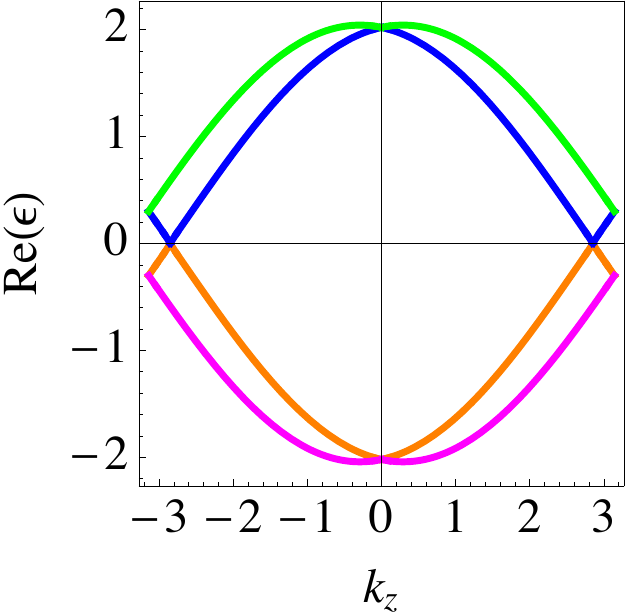}
\includegraphics[width=6cm,height=6cm,keepaspectratio]{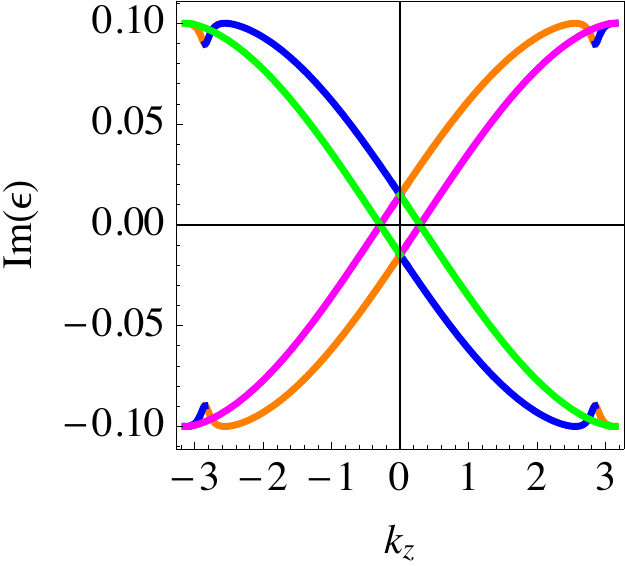}
\caption{Real~(left panel) and imaginary~(right) part of energy dispersion given in Eq.~\eqref{eq:Em0} for $d=1$, $\delta=0.1$, $\lambda=0.3$, $m=0$, $V=0.0$, $\Delta_{S}=1.0$, and $\Delta_{D}=1.0$. 
\label{fig:Em0}
}
\end{figure}

\item When $\lambda=0$,   eigenvalues of $H$ cast
\begin{align}
\epsilon^{2}_{\bolds{k}, \pm}
=&
-\delta ^2-2 i \delta  \Delta_{S} \sin (k_{z} d)
+\Delta(k_{z} )^2
+v_{\rm f}^2 \left(k_{x}^2+k_{y}^2\right)+m^2+V^2
\nonumber 
\\ &
\pm 
2 \sqrt{2 \Delta_{S} m^2 (\Delta_{D} \cos (k_{z} d)-i \delta  \sin (k_{z} d))+V^2 v_{\rm f}^2 \left(k_{x}^2+k_{y}^2\right)+m^2 \left(-\delta ^2+\Delta_{D}^2+\Delta_{S}^2+V^2\right)},
\label{eq:EL0}
\end{align}
see also Fig.~\ref{fig:EL0}.
The real part of $\epsilon^{2}_{\bolds{k}, \pm}$ also exhibit two Weyl nodes separated along the $k_{z}$ axis located at $k_{z}=\Re[k_{0}]$ with
\begin{align}
k_{0}=
\pm \frac{1}{d}\arccos\Big( &
\frac{
-\delta ^2 \Delta_{D} \Delta_{S}+\Delta_{D}^3 \Delta_{S}+\Delta_{D} \Delta_{S}^3-\Delta_{D} \Delta_{S} m^2+\Delta_{D} \Delta_{S} V^2
}{2 \Delta_{S}^2 \left(\delta ^2-\Delta_{D}^2\right)}
\nonumber \\
&
\pm
\frac{
\left| \delta \right|  \left| \Delta_{S}\right|  \sqrt{\left(\delta ^2-\Delta_{D}^2+\Delta_{S}^2\right)^2+m^4-2 m^2 \left(-\delta ^2+\Delta_{D}^2+\Delta_{S}^2+V^2\right)+V^4+2 V^2 \left(-\delta ^2+\Delta_{D}^2+\Delta_{S}^2\right)}
}{2 \Delta_{S}^2 \left(\delta ^2-\Delta_{D}^2\right)}
\Big),
\end{align}
where $k_{0}$ is evaluated when $k_{x},k_{y}$ are set to zero in $\epsilon^{2}_{\bolds{k}, \pm}$.

\begin{figure}
\includegraphics[width=5.47cm,height=5.47cm,keepaspectratio]{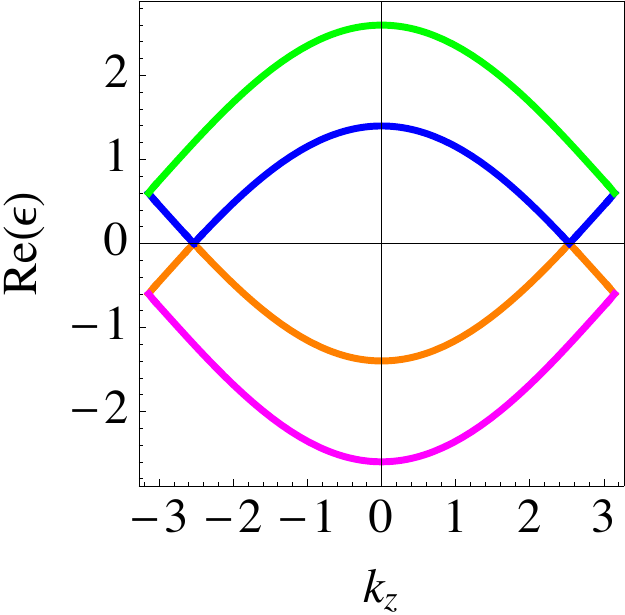}
\includegraphics[width=6cm,height=6cm,keepaspectratio]{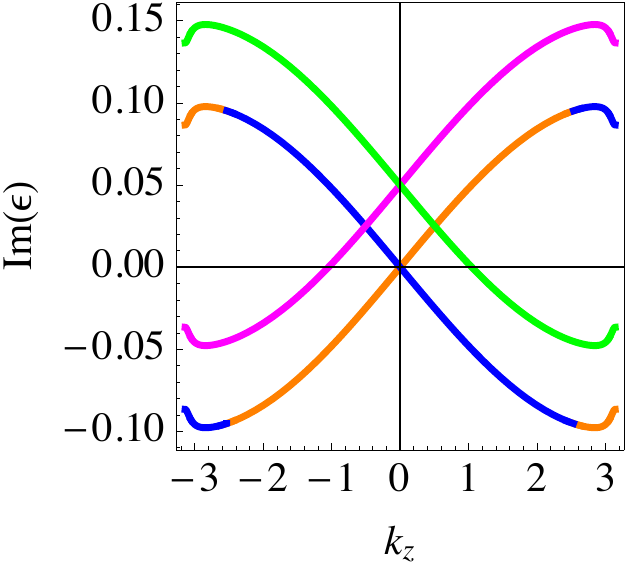}
\caption{Real~(left panel) and imaginary~(right) part of energy dispersion given in Eq.~\ref{eq:EL0} for $d=1$, $\delta=0.1$, $\lambda=0$, $m=0.6$, $V=0.05$, $\Delta_{S}=1.0$, and $\Delta_{D}=1.0$. For visibility purpose, green and pink lines in the right panels are shifted along the energy axis by $0.05$. 
\label{fig:EL0}
}
\end{figure}

\item When $V=0$, we can analytically find energy dispersion at $k_{x}=k_{y}=0$ which reads
\begin{align}
\epsilon_{\bolds{k},s, \pm}
=
\pm m \pm \sqrt{-\delta ^2+\sin (k_{z} d) (-2 \Delta_{D} \lambda -2 i \delta  \Delta_{S})+2 \cos (k_{z} d) (\Delta_{D} \Delta_{S}-i \delta  \lambda )+\Delta_{D}^2+\Delta_{S}^2+\lambda ^2},
\label{eq:EV0}
\end{align}
see also Fig.~\ref{fig:EV0}. 
The separations of two real Weyl nodes, evident in Fig.~\ref{fig:EV0}(right panel), are given by
\begin{align}
k_{0}=
\pm 
\frac{1}{d} \arccos\left(
\frac{\sqrt{-\frac{\Delta_{D}^2 \left(m^2-\lambda ^2\right) \left(\lambda ^2 \left(\Delta_{D}^2-\delta ^2\right)-m^2 \left(\Delta_{S}^2+\lambda ^2\right)\right)}{\left(\delta ^2-\Delta_{D}^2\right)^2}}}{\lambda ^2}
+
\frac{\Delta_{D} \Delta_{S} m^2}{\lambda ^2 \left(\delta ^2-\Delta_{D}^2\right)}
+
\frac{m \left| \delta \right| 
 \sqrt{\frac{A}{\Delta_{D} \left(\delta ^2-\Delta_{D}^2\right)^2} } }{\lambda ^2}
\right),
\end{align}
\begin{align}
k_{0}=\pm 
\frac{1}{d}
\arccos \Big(
\frac{1}{\lambda ^2 \left(\delta ^2-\Delta_{D}^2\right)}
\Big[ &
\left(\delta ^2-\Delta_{D}^2\right) \sqrt{-\frac{\Delta_{D}^2 \left(m^2-\lambda ^2\right) \left(\lambda ^2 \left(\Delta_{D}^2-\delta ^2\right)-m^2 \left(\Delta_{S}^2+\lambda ^2\right)\right)}{\left(\delta ^2-\Delta_{D}^2\right)^2}}+\Delta_{D} \Delta_{S} m^2
\nonumber
\\&
+
m \left| \delta \right|  \left(\Delta_{D}^2-\delta ^2\right) \sqrt{\frac{B}{\Delta_{D} \left(\delta ^2-\Delta_{D}^2\right)^2}}
\Big]
\Big),
\end{align}
where $A,B$ read
\begin{align}
A=&
\Delta_{D} m^2 \left(2 \Delta_{S}^2+\lambda ^2\right)
+
\delta ^2 \left(\Delta_{D} \lambda ^2+2 \Delta_{S} \sqrt{-\frac{\Delta_{D}^2 \left(m^2-\lambda ^2\right) \left(\lambda ^2 \left(\Delta_{D}^2-\delta ^2\right)-m^2 \left(\Delta_{S}^2+\lambda ^2\right)\right)}{\left(\delta ^2-\Delta_{D}^2\right)^2}}\right)
\nonumber \\
&
-\Delta_{D} \left(\Delta_{D}^2 \lambda ^2+\Delta_{S}^2 \lambda ^2+\lambda ^4+2 \Delta_{D} \Delta_{S} \sqrt{-\frac{\Delta_{D}^2 \left(m^2-\lambda ^2\right) \left(\lambda ^2 \left(\Delta_{D}^2-\delta ^2\right)-m^2 \left(\Delta_{S}^2+\lambda ^2\right)\right)}{\left(\delta ^2-\Delta_{D}^2\right)^2}}\right),
\end{align}
\begin{align}
B=&
\delta ^2 \left(\Delta_{D} \lambda ^2+2 \Delta_{S} \sqrt{-\frac{\Delta_{D}^2 \left(m^2-\lambda ^2\right) \left(\lambda ^2 \left(\Delta_{D}^2-\delta ^2\right)-m^2 \left(\Delta_{S}^2+\lambda ^2\right)\right)}{\left(\delta ^2-\Delta_{D}^2\right)^2}}\right)
\nonumber 
\\&
-\Delta_{D} \left(\Delta_{D}^2 \lambda ^2+\Delta_{S}^2 \lambda ^2+\lambda ^4+2 \Delta_{D} \Delta_{S} \sqrt{-\frac{\Delta_{D}^2 \left(m^2-\lambda ^2\right) \left(\lambda ^2 \left(\Delta_{D}^2-\delta ^2\right)-m^2 \left(\Delta_{S}^2+\lambda ^2\right)\right)}{\left(\delta ^2-\Delta_{D}^2\right)^2}}\right)+\Delta_{D} m^2 \left(2 \Delta_{S}^2+\lambda ^2\right).
\end{align}
Finding the separation of imaginary Weyl nodes, left panel in Fig.~\ref{fig:EV0}, seems not to be analytically feasible.

\end{itemize}

\begin{figure}
\includegraphics[width=5.47cm,height=5.47cm,keepaspectratio]{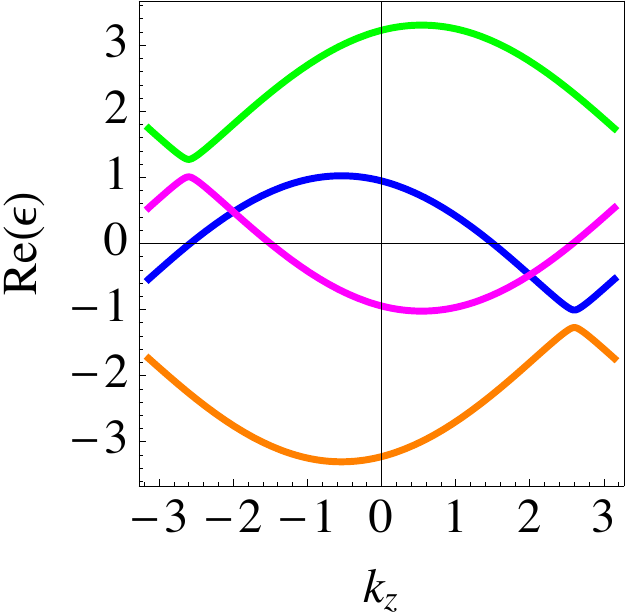}
\includegraphics[width=6cm,height=6cm,keepaspectratio]{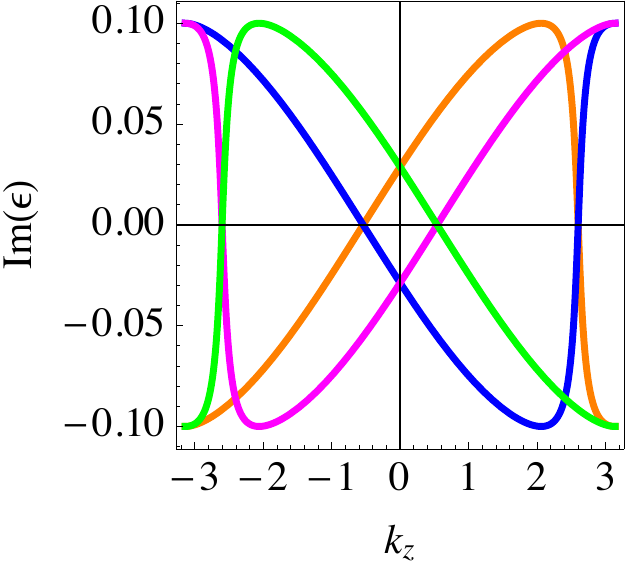}
\caption{Real~(left panel) and imaginary~(right) part of energy dispersion given in Eq.~\eqref{eq:EV0} for $d=1$, $\delta=0.1$, $\lambda=0.6$, $m=1.14$, $V=0.0$, $\Delta_{S}=1.0$, and $\Delta_{D}=1.0$.
\label{fig:EV0}
 }
\end{figure}

\section{Non-Hermitian chiral anomaly: Fujikawa's path integral formalism}\label{App:1p1}
In this section, we provide a detailed derivation of the chiral anomaly in $2$ and $4$ dimensions using Fujikawa's method.

The massless fermions with complex Fermi velocities coupled to non-Hermitian gauge fields $(V, W)$ in Euclidean space are described by the Dirac action ${\cal S}_{\rm{nh}}$ and partition function ${\cal Z}_{\rm{nh}}$,
\begin{align}
{\cal Z}_{\rm{nh}} &\propto \int {\cal D} \Psi {\cal D} \bar{\Psi} 
e^{ {\cal S}_{\rm{nh}}},
\label{eqapp:Z_nh}\\
{\cal S}_{\rm{nh}} &= \i \int {\rm d}^{d} x 
\Big[
\bar{\Psi} \gamma^{\mu} ({\zc D}_{\rm{nh}, \mu} \Psi)
\Big]
,
\label{eqapp:action_nh} \\
\slashed{\zc D}_{\rm{nh}} &=\gamma^{\mu} {\zc D}_{\rm{nh}, \mu} = \gamma^{\mu}  M_{\mu}^{\nu}
\partial_{\nu}
-\i \gamma^{\mu} M_{\mu}^{\nu}
\left(
V_{\nu}
+\gamma^{5} W_{\nu}
\right),\label{eqapp:dirac_nh}
\end{align}
in units where $c=\hbar=1$. For simplicity we assume that the gauge fields do not fluctuate, but the existence of the anomalies we derive does not rely on this assumption.
In $d=1+1$ dimensions, $\mu, \nu \in \{1,2 \}$, $\overline{\Psi}=\Psi^{\dagger} \gamma^{0}$ denotes the Dirac adjoint, and $\gamma$ stands for gamma matrices. The gamma matrices are given by $\gamma^{2}= -\i \sigma^{2}$, $\gamma^{1}=\i \sigma^{1}$ and $\gamma^{5}= \sigma^{3}$, where $\sigma$ stands for the Pauli matrices and our gamma matrices satisfy $\tr[\gamma^{5} \gamma^{\mu} \gamma^{\nu}]=-2 \varepsilon^{\mu \nu}$ with $\varepsilon^{21}=\i$. We assume the Fermi velocity matrix to be diagonal but complex in general. The metric reads $g^{\mu \nu} =-\delta^{\mu \nu}$.

In $d=3+1$ dimensions, $\mu, \nu \in \{1,2,3,4 \}$ and the metric reads $g^{\mu \nu} =-\delta^{\mu \nu}$. Here our convention for gamma matrices is $\gamma^{4}= -i\tau^{1} \otimes \sigma^{0}$, $\gamma^{j}=\i \tau^2 \otimes \sigma^{j}$, with $j \in \{1,2,3\}$, and $\gamma^{5}= \tau^{3} \otimes \sigma^{0}$, where $\sigma$ and $\tau$ denote Pauli matrices and $\tr[\gamma^{5} \gamma^{\mu } \gamma^{\nu} \gamma^{\alpha} \gamma^{\beta}]=-4 \varepsilon^{\mu \nu \alpha \beta}$ with $\varepsilon^{4321}=1$. 

When it is needed we transform Euclidean spacetime into the Minkowski spacetime by setting $\gamma^{0}=\i \gamma^{d}$ and rewriting the Minkowski action~(${\cal S}_{M} $) as ${\cal S}_{M} =\i {\cal S}_{E}$, where ${\cal S}_{E}$ is the action defined in the Euclidean spacetime.

To draw the connection between the anomalous currents generated from Eq.~\eqref{eqapp:action_nh} and their Hermitianized and anti-Hermitian counterparts, we also present the path integral calculations for the symmetrized and anti-symmetrized forms of  ${\cal S}_{ \rm{nh}}$.
The symmetrized form of ${\cal S}_{ \rm{nh}}$ defines the Hermitianized action~${\cal S}_{\rm{h}}$ which reads
 \begin{align}
 {\cal Z}_{\rm{h}} &\propto \int {\cal D} \Psi {\cal D} \bar{\Psi} 
e^{  {\cal S}_{\rm{h}}},
\label{eqapp:Z_h}\\
{\cal S}_{\rm{h}} &= \frac{\i}{2} \int {\rm d}^{d} x 
\Big[
\bar{\Psi} \gamma^{\mu} ({\zc D}_{\rm{nh}, \mu} \Psi)
-\overline{({\zc D}_{\rm{nh}, \mu} \Psi)} \gamma^{\mu} \Psi
\Big]
= \i 
\int {\rm d}^{2} x 
\bar{\Psi} \gamma^{\mu}
{\zc D}_{\rm{h}, \mu} \Psi,\label{eqapp:action_h} \\
{\zc D}_{\rm{h}, \mu}
 &=
  \Big[
\Re[M_{\mu}^{\nu}] \partial_{\nu}
 -\i \Re[M_{\mu}^{\nu}  V_{\nu} ]
 -\i \gamma^{5}  \Re[M_{\mu}^{\nu}  W_{\nu}]
 \Big] ,\label{eqapp:Dd_h}
\end{align}
and the anti-Hermitianized action ${\cal S}_{\rm{ah}}$ after anti-symmetrizing ${\cal S}_{\rm{nh}}$ reads
 \begin{align}
  {\cal Z}_{\rm{ah}} &\propto \int {\cal D} \Psi {\cal D} \bar{\Psi} 
e^{ {\cal S}_{\rm{ah}}},
\label{eqapp:Z_ah}\\
{\cal S}_{\rm{ah}} &= \frac{\i}{2} \int {\rm d}^{d} x 
\Big[
\bar{\Psi} \gamma^{\mu} ({\zc D}_{\rm{nh},\mu} \Psi)
+\overline{({\zc D}_{\rm{nh},\mu} \Psi)} \gamma^{\mu} \Psi
\Big]
= \i 
\int {\rm d}^{2} x 
\bar{\Psi} \gamma^{\mu}
{\zc D}_{\rm{ah}, \mu} \Psi,\label{eq:action_ah} \\
{\zc D}_{\rm{ah}, \mu}
 &=
  \Big[
\i \Im[M_{\mu}^{\nu}] \partial_{\nu}
 + \Im[M_{\mu}^{\nu}  V_{\nu} ]
 + \gamma^{5}  \Im[M_{\mu}^{\nu}  W_{\nu}]
 \Big] ,\label{eqapp:Dd_ah}
\end{align}
 where $\overline{({\zc D}_{\rm{nh}, \mu} \Psi)} = ({\zc D}_{\rm{nh}, \mu} \Psi)^{\dagger} \gamma^{0}$, and $ {\zc D}_{\rm{h}}, {\zc D}_{\rm{ah}} $ are modified Dirac operators for the Hermitianized and anti-Hermitianized systems, respectively. Here we have used ${\partial}^{\dagger}_{\mu}= - \partial_{\mu}$ to obtain ${\zc D}_{\rm{h}}$ and ${\zc D}_{\rm{ah}}$. To facilitate later comparison of the results and to avoid repetition, we formulate the Hermitian, anti-Hermitian and non-Hermitian actions in a unified language as
  \begin{align}
   \tilde{\cal Z} &\propto \int {\cal D} \Psi {\cal D} \bar{\Psi} 
e^{  \tilde{{\cal S}}},
\label{eq:Z_1p1}\\
\tilde{{\cal S} }
 &= \i 
\int {\rm d}^{d} x 
\bar{\Psi} 
( \tilde{\slashed{\zc D}} \Psi )
,\label{eq:action_1p1} \\
\tilde{{\zc D}}_{ \mu}
 &=
 \tilde{d}_{\mu} -\i \tilde{V}_{\mu} - \i \gamma^{5} \tilde{W}_{\mu} ,\label{eq:Dd_1p1}
\end{align}
where $\tilde{\slashed{\zc D}} = \gamma^{\mu} \tilde{\zc D}_{ \mu}  $, and functions and operators with tilde are defined in the unified notation with the mapping relations defined in Table~\ref{tabapp:convert}.
 \begin{table}
\begin{center}
\begin{tabular}{|c | c | c| c|c|}
\hline
     & $\tilde{d}_{\mu}$ & $\tilde{V}_{\mu}$ & $\tilde{W}_{\mu}$ & $f^{\nu}_{\mu}$ 
    \\
     \hline
    &&&&\\[-0.05em]
    Hermitianized & $\Re[M_{\mu}^{\nu} ] \partial_{\nu}$ & $\Re[M_{\mu}^{\nu}  V_{\nu} ]$& $\Re[M_{\mu}^{\nu}  W_{\nu} ]$ & $\Re[M_{\mu}^{\nu} ]$
    \\
    \hline
    &&&&\\[-0.05em]
    Anti-Hermitianized & $\i \Im[M_{\mu}^{\nu} ] \partial_{\nu}$ & $\i \Im[M_{\mu}^{\nu}  V_{\nu} ]$ & $\i \Im[M_{\mu}^{\nu}  W_{\nu} ]$
    & $\i \Im[M_{\mu}^{\nu} ]$
    \\
    \hline
    &&&&\\[-0.05em]
    Non-Hermitian & $M_{\mu}^{\nu}  \partial_{\nu}$ & $M_{\mu}^{\nu}  V_{\nu}$ & $M_{\mu}^{\nu}  W_{\nu}$  & $M_{\mu}^{\nu} $ 
    \\
    \hline
\end{tabular}
\end{center}
\caption{\label{tabapp:convert} Mapping relations between the unified notation of operators in $\tilde{\cal S}$ and operators in the Hermitianized~${\cal S}_{\rm{h}}$, anti-Hermitianized~${\cal S}_{\rm{ah}}$ and non-Hermitian~${\cal S}_{\rm{nh}}$ actions. The matrix B is introduced in Eq.~\eqref{eq:constA}.}
\end{table}

\subsection{Conserved classical currents}
Let us first present the classical vector and chiral currents for the Dirac action in Eq.~\eqref{eq:action_1p1} using the Dirac equations of motion for the spinor $\Psi$ and its adjoint. We calculate these equations of motion using the Euler-Lagrange equation and obtain
\begin{align}
   -  
\bar{\Psi} 
\gamma^{\mu}  (\tilde{V}_{\mu} +\gamma^{5} \tilde{W}_{\mu}) 
+\i
\tilde{d}_{\mu} \bar{\Psi} \gamma^{\mu} 
=0 ,\label{eq:EMpsibar}\\
\gamma^{0}\gamma^{\mu}  (\tilde{V}_{\mu} +\gamma^{5} \tilde{W}_{\mu})
\Psi
+\i
\gamma^{0} \gamma^{\mu} \tilde{d}_{\mu} \bar{\Psi}
=0.\label{eq:EMpsi}
\end{align}
Eqs.~(\ref{eq:EMpsibar},~\ref{eq:EMpsi}) then enable us to derive the classical chiral~($j^{5}$) and vector~($j$) currents as
\begin{align}
\tilde{d}_{\nu} j^{\mu} &=0 \text{, where } j^{\mu}=  \bar{\Psi} \gamma^{\mu} \Psi,\\
\tilde{d}_{\nu} j^{5\mu} &=0 \text{, where } j^{5\mu}=   \bar{\Psi} \gamma^{5}  \gamma^{\mu} \Psi.
\end{align}
As a result of the above equations, both vector and chiral currents are classically conserved.

\subsection{Non-conservation of quantum currents} \label{App:Fujikawa1p1}
 Performing chiral and vector rotations, with rotating angles $\beta$ and $\kappa$, respectively, on the Grassmann variables,
\begin{align}
\bar{\Psi}_{\rm rot} = \bar{\Psi} e^{-\i \kappa - \i \gamma^{5} \beta}, 
\qquad 
\Psi_{\rm rot} = e^{\i \kappa - \i \gamma^{5} \beta} {\Psi},\label{eq:barpsi_psirot}
\end{align}
keeps the rotated action ${\cal S}_{\rm rot}= \i \int {\rm d}^{d} x
 \bar{\Psi}_{\rm rot}  \tilde{\slashed{\zc D}} \Psi_{\rm rot}$ invariant under this transformation
 if one imposes 
 \begin{align}
 \gamma^{\mu} \tilde{V}_{\mu} + \gamma^{\mu} \gamma^{5} \tilde{W}_{\mu}
  = \gamma^{\mu} M_{\mu}^{\nu} \partial_{\nu}  \kappa
    -\gamma^{\mu} \gamma^{5}  M_{\mu}^{\nu} \partial_{\nu}  \beta.\label{eq:VW_kb}
\end{align}

While the above relation ensures the chiral symmetry of the action~($\tilde{\cal S}$), the measure in the generating function~($\tilde{\cal Z}$) 
exhibits the chiral anomaly after applying the chiral transformations in Eq.~\eqref{eq:barpsi_psirot}.
To evaluate the chiral anomaly, we employ Fujikawa's method~\cite{Fujikawa2004} and calculate the changes of the path-integral measure under the chiral and vector transformations.

To begin with, we note that $\tilde{\zc D}$ is non-Hermitian,
\begin{equation}
    \tilde{\slashed{\zc D}}^{\dagger}=
      ( \tilde{d}^{\dagger}_{\mu} + \i \tilde{V}^{\dagger}_{\mu} + \i \gamma^{5 \dagger} \tilde{W}_{\mu}^{\dagger} ) \gamma^{\mu \dagger}
     =- \gamma^{\mu} ( \tilde{d}^{\dagger}_{\mu} + \i \tilde{V}^{\dagger}_{\mu} - \i \gamma^{5} \tilde{W}_{\mu}^{\dagger} )=-\gamma^{\mu} \tilde{\zc D}_{\mu}^{\dagger} \neq \tilde{\slashed{\zc D}}
    ,\label{eq:ddag_fuji}
\end{equation}
which is obtained using the self-adjoint properties of gamma matrices~($\gamma^{\mu\dagger} = - \gamma^{\mu}, \gamma^{5 \dagger} = \gamma^{5}$). Note that by placing $ \gamma^{\mu}$ on the left hand side of $\tilde{\zc D}_{\mu}^{\dagger}$ in Eq.~\eqref{eq:ddag_fuji}, $\tilde{\zc D}_{\mu}^{\dagger}$ is no longer equal to the complex conjugate of $\tilde{\zc D}_{\mu}$, i.e., $\tilde{\zc D}_{\mu}^{\dagger} \neq [ \tilde{\zc D}_{\mu}]^{\dagger}$.
To avoid dealing with the unnecessary complexities of the non-Hermitian eigenvalue problem, we consider the Hermitian Laplacian operators $\tilde{\slashed{\zc D}} \tilde{\slashed{\zc D}}^{\dagger}$ and $\tilde{\slashed{\zc D}}^{\dagger}\tilde{\slashed{\zc D}}$, with eigenvalues $\{\lambda^{2}\}$, such that
\begin{align}
   \tilde{\slashed{\zc D}} \tilde{\slashed{\zc D}}^{\dagger} |\eta_{n} \rangle &= \lambda^{2}_{n} | \eta_{n} \rangle , \quad \tilde{\slashed{\zc D}}^{\dagger} \tilde{\slashed{\zc D}} | \xi_{n} \rangle = \lambda^{2}_{n} | \xi_{n} \rangle,\\
    \tilde{\slashed{\zc D}}^{\dagger} |\eta_{n} \rangle &= \lambda^{*}_{n} |\xi_{n} \rangle ,
    \quad \tilde{\slashed{\zc D}} | \xi_{n} \rangle = \lambda_{n} |\eta_{n} \rangle.
\end{align}
where $\{ | \xi_{n} \rangle \}$ and $\{ | \eta_{n} \rangle \}$ are eigenvectors. We further assume that these sets of eigenbases are normalizable and complete
\begin{align}
\forall \{| \phi \rangle \} \in \Big\{\{| \xi \rangle \}, \{| \eta \rangle \} \Big\}, \quad 
\int {\rm d}^{2} x \phi^{\dagger}_{n}(x) \phi_{m}(x) =\delta_{nm},\quad
\sum_{n} \phi^{\dagger}_{n}(y) \phi_{n}(x) = \delta(x-y), \label{eq:norm_comp_1p1}
\end{align}
where $x,y$ are $2$ space-time coordinates.
Using these complete eigenbases, we decompose the spinor fields into the eigenvectors of the Laplacian operators as
\begin{align}
\Psi(x) = \sum\limits_{n} a_{n} \xi_{n}(x) , \qquad \bar{\Psi}(x)= \sum\limits_{n}  \eta^{\dagger}_{n}(x) \bar{b}_{n},
\end{align}
where $a_{n}$ and $\bar{b}_{n}$ are new Grassmann variables. Subsequently, the path-integral measure in this representation reads ${\cal D}\Psi  {\cal D}\bar{\Psi}= \prod_{n} {\rm d}a_{n} {\rm d} \bar{b}_{n}$. 
In the following, we will evaluate the changes of this path-integral measure after performing infinitesimal chiral and vector transformations to calculate chiral anomaly and obtain quantum currents both in $2$ and $4$ dimensions.

\subsubsection{$2$ dimensions}\label{sec:1p1app}

\paragraph{\bf Chiral Transformation-.}

To evaluate the change of the path-integral measure under an infinitesimal chiral rotation, we first expand the spinor fields in the eigenbases of Laplacian operators as
\begin{align}
\Psi_{\rm rot} &= e^{-\i \gamma^{5} \beta} \Psi(x)=\left(1-\i \beta \gamma^{5} \right) \sum_{m}a_{m} \xi_{m}(x) = \sum\limits_{m}  \xi_{m} a^{\rm rot}_{m},\\
\bar{\Psi}_{\rm rot} &= \bar{\Psi}(x) e^{-\i \gamma^{5} \beta}= \sum_{m}\bar{b}_{m} \eta^{\dagger}_{m}(x)  \left(1-\i \beta \gamma^{5} \right)= \sum_{m}\bar{b}^{\rm rot}_{m} \eta^{\dagger}_{m}(x). 
\end{align}
Here the orthogonality of eigenfunctions imposes that
\begin{align}
a_{n}^{\rm rot} = \left(\delta_{nm} + C_{nm} \right)a_{m} =
\left[ \delta_{nm} - \i \beta \int {\rm d}^2 x \xi^{\dagger}_{n}(x) \gamma^{5} \xi_{m}(x) \right]a_{m}
,\\
 \bar{b}_{n}^{\rm rot} = \bar{b}_{m} \left(\delta_{nm} + D_{nm} \right) = \bar{b}_{m}
\left[ \delta_{mn} - \i \beta \int {\rm d}^2 x \eta^{\dagger}_{m}(x) \gamma^{5} \eta_{n}(x) \right].
\end{align}
The path-integral measure for the rotated spinors, and up to the first-order in $C$ and $D$, then yields
\begin{align}
{\cal D} \Psi_{\rm rot} {\cal D} \bar{\Psi}_{\rm rot} &=
\left[
{\rm det}(1+C) {\rm det}(1+D)
\right]^{-1}
{\cal D} \Psi {\cal D} \bar{\Psi} = e^{-\tr [D+C]} {\cal D} \Psi {\cal D} \bar{\Psi} ,\\
&=
e^{
\i \int {\rm d}^{2} x 
\beta(x)
\left(
\sum\limits_{n} \xi^{\dagger}_{n}(x) \gamma^{5} \xi_{n}(x)
+
 \sum\limits_{n} \eta^{\dagger}_{n}(x) \gamma^{5} \eta_{n}(x)
\right)
}
{\cal D} \Psi {\cal D} \bar{\Psi} 
= e^{\i \int {\rm d}^{2} x  \beta(x) {\cal A}_{5}(x)} {\cal D} \Psi {\cal D} \bar{\Psi} = e^{{\cal S}_{5}[\beta]}{\cal D} \Psi {\cal D} \bar{\Psi}. \label{eq:Sbeta_1p1}
\end{align}
Here ${\cal S}_{5}[\beta]$ is the Jacobian contribution to the action.
${\cal A}_{5}$ is ill-defined as using the completeness relation in Eq.~\eqref{eq:norm_comp_1p1} results in ${\cal A}_{5} \propto \tr[\gamma^{5}.\delta(0)]$.
To evaluate ${\cal A}_{5}$, Fujikawa proposed using a heat-kernel regulator~\cite{Nakahara1990}, with a Gaussian cut-off $m \rightarrow \infty$, such that
\begin{align}
{\cal A}_{5} (x) &= \lim_{m \rightarrow \infty} \sum\limits_{n} 
e^{\frac{-\lambda_{n}^{2}}{m^2}}
 \left[
  \xi^{\dagger}_{n}(x) \gamma^{5} \xi_{n}(x)
  +
    \eta^{\dagger}_{n}(x)
     \gamma^{5} \eta_{n}(x)
 \right],\\
 &=
  \lim_{m \rightarrow \infty} \sum\limits_{n} 
 \left[
  \xi^{\dagger}_{n}(x)   
 \gamma^{5}   
 e^{\frac{-\tilde{\slashed{\zc D}}^{\dagger} \tilde{\slashed{\zc D}} }{m^2}}
 \xi_{n}(x)
  +
    \eta^{\dagger}_{n}(x)
     \gamma^{5}
  e^{\frac{-\tilde{\slashed{\zc D}}  \tilde{\slashed{\zc D}}^{\dagger} }{m^2}}     
      \eta_{n}(x)
 \right]
 ,\\
 &=
   \lim_{m \rightarrow \infty} \sum\limits_{n} 
   \int \frac{{\rm d}^2 k}{(2\pi)^2}
   e^{-\i k x}
   \tr \left[
 \gamma^{5}   
 e^{\frac{-\tilde{\slashed{\zc D}}^{\dagger} \tilde{\slashed{\zc D}} }{m^2}}
  +
     \gamma^{5}
  e^{\frac{-\tilde{\slashed{\zc D}}  \tilde{\slashed{\zc D}}^{\dagger} }{m^2}}     
 \right]
 e^{\i k x},\label{eq:A5}
\end{align}
where we have substituted the Fourier transforms of the eigenbases and used the completeness relation of eigenvectors in the last line. Note that $\lambda$ is defined in the Minkowski space and upon mapping it to the Euclidean space, both exponents in the exponential functions in Eq.~\eqref{eq:A5} acquire minus signs which ensures cutting off the large Euclidean eigenvalues.
Using Eqs.~(\ref{eq:Dd_1p1}, \ref{eq:ddag_fuji}), $\tilde{\slashed{\zc D}}  \tilde{\slashed{\zc D}}^{\dagger}$ and $\tilde{\slashed{\zc D}}^{\dagger}  \tilde{\slashed{\zc D}}$ read
\begin{align}
\tilde{\slashed{\zc D}}  \tilde{\slashed{\zc D}}^{\dagger} &=
-\big[ \tilde{\zc D}^{\mu} \tilde{\zc D}_{\mu}^{\dagger}
+\frac{1}{4} [\gamma^{\mu}, \gamma^{\nu}] 
( 
  \tilde{\zc D}_{\mu} \tilde{\zc D}_{\nu}^{\dagger} 
- \tilde{\zc D}_{\nu} \tilde{\zc D}_{\mu}^{\dagger} )], \label{eq:Dddag}
\end{align}

    \begin{align}
  \tilde{\zc D}^{\mu}  \tilde{\zc D}^{\dagger}_{\mu} =&
\tilde{d}^{\mu} \tilde{d}^{\dagger}_{\mu}
+\i \tilde{d}^{\mu} ( \tilde{V}^{\dagger}_{\mu} ) +\i  \tilde{V}^{\dagger}_{\mu} \tilde{d}^{\mu}
-\i\gamma^{5} \tilde{d}^{\mu} ( \tilde{W}^{\dagger}_{\mu} ) -\i \gamma^{5} \tilde{W}^{\dagger}_{\mu}  \tilde{d}^{\mu}
\nonumber \\
&
-\i \tilde{V}^{\mu} \tilde{d}^{\dagger}_{\mu}
-\i \gamma^{5} \tilde{W}^{\mu} \tilde{d}^{\dagger}_{\mu}
+\tilde{V}^{\mu} \tilde{V}^{\dagger}_{\mu}
- \gamma^{5} \tilde{V}^{\mu} \tilde{W}^{\dagger}_{\mu}
+ \gamma^{5} \tilde{W}^{\mu} \tilde{V}^{\dagger}_{\mu}
-\tilde{W}^{\mu} \tilde{W}^{\dagger}_{\mu},
\\
       \tilde{\zc D}_{\mu}  \tilde{\zc D}^{\dagger}_{\nu} 
  -
   \tilde{\zc D}_{\nu}  \tilde{\zc D}^{\dagger}_{\mu} 
   =&
   (\tilde{d}_{\mu} \tilde{d}_{\nu}^{\dagger} - \tilde{d}_{\nu} \tilde{d}_{\mu}^{\dagger})
   +
  \i \tilde{F}_{\mu \nu}[ \tilde{V}^{\dagger}] 
   -\i  \gamma^{5} \tilde{F}_{\mu \nu} [\tilde{W}^{\dagger}]
+\big( \tilde{V}_{\mu} \tilde{V}^{\dagger}_{\nu} - \tilde{V}_{\nu} \tilde{V}^{\dagger}_{\mu}  \big)   
\nonumber \\
&
+\i \big( \tilde{V}^{\dagger}_{\nu} \tilde{d}_{\mu} - \tilde{V}^{\dagger}_{\mu} \tilde{d}_{\nu} - \tilde{V}_{\mu} \tilde{d}_{\nu}^{\dagger}  + \tilde{V}_{\nu} \tilde{d}^{\dagger}_{\mu}
 \big)
 +\i \gamma^{5}
 \big(
 -\tilde{W}^{\dagger}_{\nu} \tilde{d}_{\mu} 
 +\tilde{W}_{\mu}^{\dagger} \tilde{d}_{\nu}
 -\tilde{W}_{\mu} \tilde{d}_{\nu}^{\dagger}
 +\tilde{W}_{\nu} \tilde{d}_{\mu}^{\dagger}
 \big)
\nonumber \\
&
+ \gamma^{5} \big( 
-\tilde{V}_{\mu} \tilde{W}^{\dagger}_{\nu}
+\tilde{V}_{\nu} \tilde{W}^{\dagger}_{\mu}
+\tilde{W}_{\mu} \tilde{V}_{\nu}^{\dagger} 
- \tilde{W}_{\nu} \tilde{V}_{\mu}^{\dagger}
 \big)
 -\big( \tilde{W}_{\mu} \tilde{W}^{\dagger}_{\nu} - \tilde{W}_{\nu} \tilde{W}^{\dagger}_{\mu}  \big)  ,
    \end{align}
    and 
\begin{align}
\tilde{\slashed{\zc D}}^{\dagger}  \tilde{\slashed{\zc D}} &= 
-\big[
\tilde{\zc D}^{\mu\dagger} \tilde{\zc D}_{\mu}
+\frac{1}{4} [\gamma^{\mu}, \gamma^{\nu}]
(
 \tilde{\zc D}_{\mu }^{\dagger} \tilde{\zc D}_{\nu}
-\tilde{\zc D}_{\nu }^{\dagger} \tilde{\zc D}_{\mu}
)\big]
,\label{eq:DdagD}
\end{align}

      \begin{align}
  \tilde{\zc D}^{\mu \dagger}  \tilde{\zc D}_{\mu} =&
\tilde{d}^{\mu \dagger} \tilde{d}_{\mu}
-\i \tilde{d}^{\mu \dagger} ( \tilde{V}_{\mu} ) -\i  \tilde{V}_{\mu} \tilde{d}^{\mu \dagger}
-\i \gamma^{5} \tilde{d}^{\mu \dagger} ( \tilde{W}_{\mu} ) -\i \gamma^{5} \tilde{W}_{\mu}  \tilde{d}^{\mu \dagger}
\nonumber \\
&
+\i \tilde{V}^{\mu \dagger} \tilde{d}_{\mu}
-\i \gamma^{5} \tilde{W}^{\mu \dagger} \tilde{d}_{\mu}
+\tilde{V}^{\mu \dagger} \tilde{V}_{\mu}
+ \gamma^{5} ( \tilde{V}^{\mu \dagger} \tilde{W}_{\mu}
-  \tilde{W}^{\mu \dagger} \tilde{V}_{\mu} )
-\tilde{W}^{\mu \dagger} \tilde{W}_{\mu},
\\
       \tilde{\zc D}_{\mu}^{\dagger}  \tilde{\zc D}_{\nu} 
  -
   \tilde{\zc D}_{\nu}^{\dagger}  \tilde{\zc D}_{\mu} 
   =&
   (\tilde{d}_{\mu}^{\dagger} \tilde{d}_{\nu} - \tilde{d}^{\dagger}_{\nu} \tilde{d}_{\mu})
   -
  \i \tilde{F}^{\dagger}_{\mu \nu}[ \tilde{V}] 
   -\i  \gamma^{5} \tilde{F}^{\dagger}_{\mu \nu} [\tilde{W}]
+\big( \tilde{V}^{\dagger}_{\mu} \tilde{V}_{\nu} - \tilde{V}^{\dagger}_{\nu} \tilde{V}_{\mu}  \big)   
\nonumber \\
&
-\i \big( \tilde{V}_{\nu} \tilde{d}^{\dagger}_{\mu} - \tilde{V}_{\mu} \tilde{d}^{\dagger}_{\nu} - \tilde{V}^{\dagger}_{\mu} \tilde{d}_{\nu}  + \tilde{V}^{\dagger}_{\nu} \tilde{d}_{\mu}
 \big)
 -\i \gamma^{5}
 \big(
 \tilde{W}_{\nu} \tilde{d}^{\dagger}_{\mu} - \tilde{W}_{\mu} \tilde{d}^{\dagger}_{\nu}
 +\tilde{W}^{\dagger}_{\mu} \tilde{d}_{\nu}-\tilde{W}^{\dagger}_{\nu} \tilde{d}_{\mu}
 \big)
\nonumber \\
&
+ \gamma^{5} \big( 
\tilde{V}^{\dagger}_{\mu} \tilde{W}_{\nu}
-\tilde{V}^{\dagger}_{\nu} \tilde{W}_{\mu}
-\tilde{W}^{\dagger}_{\mu} \tilde{V}_{\nu} 
+ \tilde{W}^{\dagger}_{\nu} \tilde{V}_{\mu}
 \big)
 -\big( \tilde{W}^{\dagger}_{\mu} \tilde{W}_{\nu} - \tilde{W}^{\dagger}_{\nu} \tilde{W}_{\mu}  \big)   
 .
    \end{align}
Here, the field strengths are $F_{\mu \nu} [\tilde{A}] = \tilde{d}_{\mu} \tilde{A}_{\nu} - \tilde{d}_{\nu} \tilde{A}_{\mu}$ and $F^{\dagger}_{\mu \nu} [\tilde{A}] = \tilde{d}^{\dagger}_{\mu} \tilde{A}_{\nu} - \tilde{d}^{\dagger}_{\nu} \tilde{A}_{\mu}$.

Next, we insert Eqs.~(\ref{eq:Dddag},~\ref{eq:DdagD}) into Eq.~\eqref{eq:A5}  and apply variable changes $k_{\mu} \rightarrow p_{\mu}/m$ such that

\begin{align}
 &  e^{-\i k x}
   \tr \left[
 \gamma^{5}   
 (
 e^{\frac{-\tilde{\slashed{\zc D}}^{\dagger} \tilde{\slashed{\cal D}} }{m^2}}  
  )
 \right]
 e^{\i k x}
  =
   e^{-\i k x}
   \tr   \left[
 \gamma^{5}  
 e^{\frac{
 \tilde{\zc D}^{\mu \dagger}  \tilde{\zc D}_{\mu} 
+  \frac{1}{4} [\gamma^{\mu} , \gamma^{\nu}]
  \big[
  \tilde{\zc D}_{\mu}^{\dagger}  \tilde {\zc D}_{\nu} 
  -
   \tilde{\zc D}_{\nu}^{\dagger}   \tilde{\zc D}_{\mu} 
  \big]
  }{m^2}}
 \right]
 e^{\i k x},\\
 =
   \tr   &  \left[
 \gamma^{5}  
 \exp\Big[ 
 \frac{
(\tilde{d}^{\mu \dagger} -\i \tilde{k}^{\mu \dagger}) 
(\tilde{d}_{\mu} +\i \tilde{k}_{\mu} )
-\i \tilde{d}^{\mu \dagger} ( \tilde{V}_{\mu} ) 
-\i  \tilde{V}_{\mu} (\tilde{d}^{\mu \dagger} 
- \i \tilde{k}^{\mu \dagger} )
-\i \gamma^{5} \tilde{d}^{\mu \dagger} ( \tilde{W}_{\mu} )
 -\i \gamma^{5} \tilde{W}_{\mu} ( \tilde{d}^{\mu \dagger} -\i \tilde{k}^{\mu \dagger })
+\i \tilde{V}^{\mu \dagger} ( \tilde{d}_{\mu} + \i \tilde{k}_{\mu})
  }{m^2}
  \right.
  \nonumber \\
  & 
  +
  \frac{
-\i \gamma^{5} \tilde{W}^{\mu \dagger} (\tilde{d}_{\mu} + \i \tilde{k}_{\mu})
+\tilde{V}^{\mu \dagger} \tilde{V}_{\mu}
+ \gamma^{5} ( \tilde{V}^{\mu \dagger} \tilde{W}_{\mu}
-  \tilde{W}^{\mu \dagger} \tilde{V}_{\mu} )
-\tilde{W}^{\mu \dagger} \tilde{W}_{\mu}
  }{m^2}
  \nonumber \\
  & 
  + 
  \frac{
\frac{[\gamma^{\mu} , \gamma^{\nu}]}{4}
\Big(
   \big( (\tilde{d}^{\dagger}_{\mu} -\i \tilde{k}^{\dagger}_{\mu} ) 
    (\tilde{d}_{\nu} +\i \tilde{k}_{\nu} )
   - (\tilde{d}^{\dagger}_{\nu} -\i \tilde{k}^{\dagger}_{\nu})
   (\tilde{d}_{\mu} + \i \tilde{k}_{\mu})
    \big)
       -
  \i \tilde{F}^{\dagger}_{\mu \nu}[ \tilde{V}] 
   -\i  \gamma^{5} \tilde{F}^{\dagger}_{\mu \nu} [\tilde{W}]
+\big( \tilde{V}^{\dagger}_{\mu} \tilde{V}_{\nu} - \tilde{V}^{\dagger}_{\nu} \tilde{V}_{\mu}  \big)   
-\big( \tilde{W}^{\dagger}_{\mu} \tilde{W}_{\nu} - \tilde{W}^{\dagger}_{\nu} \tilde{W}_{\mu}  \big)   
  }{m^2}
    \nonumber \\
  & 
  \frac{ 
  -\i \big( \tilde{V}_{\nu} (\tilde{d}^{\dagger}_{\mu} -\i \tilde{k}^{\dagger}_{\mu})
   - \tilde{V}_{\mu} (\tilde{d}^{\dagger}_{\nu}-\i \tilde{k}^{\dagger}_{\nu})
    - \tilde{V}^{\dagger}_{\mu} (\tilde{d}_{\nu}+\i \tilde{k}_{\nu})
      + \tilde{V}^{\dagger}_{\nu} (\tilde{d}_{\mu} + \i \tilde{k}_{\mu} ) 
 \big)
 + \gamma^{5} \big( 
\tilde{V}^{\dagger}_{\mu} \tilde{W}_{\nu}
-\tilde{V}^{\dagger}_{\nu} \tilde{W}_{\mu}
-\tilde{W}^{\dagger}_{\mu} \tilde{V}_{\nu} 
+ \tilde{W}^{\dagger}_{\nu} \tilde{V}_{\mu}
 \big)
   }{m^2} 
     \nonumber \\
  & \left.
  \frac{
 -\i \gamma^{5}
 \big(
 \tilde{W}_{\nu} (\tilde{d}^{\dagger}_{\mu} -\i \tilde{k}^{\dagger}_{\mu})
  - \tilde{W}_{\mu} (\tilde{d}^{\dagger}_{\nu} -\i \tilde{k}^{\dagger}_{\nu} )
 +\tilde{W}^{\dagger}_{\mu} (\tilde{d}_{\nu} + \i \tilde{k}_{\nu} )
 -\tilde{W}^{\dagger}_{\nu} (\tilde{d}_{\mu} + \i \tilde{k}_{\mu} )
 \big)
\Big)
  }{m^2} \Big]
 \right],
  \label{eq:smpddagd2}
\end{align}
and
\begin{align}
 &  e^{-\i k x}
   \tr \left[
 \gamma^{5}   
 (
 e^{\frac{-\tilde{\slashed{\zc D}} \tilde{\slashed{\zc D}^{\dagger}} }{m^2}}  
  )
 \right]
 e^{\i k x}
  =
   e^{-\i k x}
   \tr   \left[
 \gamma^{5}  
 e^{\frac{
 \tilde{\zc D}^{\mu }  \tilde{\zc D}^{\dagger}_{\mu} 
+  \frac{1}{4} [\gamma^{\mu} , \gamma^{\nu}]
  \big[
  \tilde{\zc D}_{\mu}   \tilde{\zc D}^{\dagger}_{\nu} 
  -
   \tilde{\zc D}_{\nu}  \tilde{\zc D}^{\dagger}_{\mu} 
  \big]
  }{m^2}}
 \right]
 e^{\i k x},\\
 =
   \tr   &  \left[
 \gamma^{5}  
 \exp\Big[ 
 \frac{
(\tilde{d}^{\mu} + \i \tilde{k}^{\mu})
( \tilde{d}^{\dagger}_{\mu} - \i \tilde{k}^{\dagger}_{\mu})
+\i \tilde{d}^{\mu} ( \tilde{V}^{\dagger}_{\mu} ) 
+\i  \tilde{V}^{\dagger}_{\mu} (\tilde{d}^{\mu} + \i \tilde{k}^{\mu} )
-\i\gamma^{5} \tilde{d}^{\mu} ( \tilde{W}^{\dagger}_{\mu} ) 
-\i \gamma^{5} \tilde{W}^{\dagger}_{\mu}  (\tilde{d}^{\mu} + \i \tilde{k}^{\mu} )
-\i \tilde{V}^{\mu} ( \tilde{d}^{\dagger}_{\mu} -\i \tilde{k}^{\dagger}_{\mu} )
  }{m^2}
  \right.
  \nonumber \\
  & 
  +
  \frac{
-\i \gamma^{5} \tilde{W}^{\mu} ( \tilde{d}^{\dagger}_{\mu} -\i \tilde{k}^{\dagger}_{\mu} )
+\tilde{V}^{\mu} \tilde{V}^{\dagger}_{\mu}
- \gamma^{5} \tilde{V}^{\mu} \tilde{W}^{\dagger}_{\mu}
+ \gamma^{5} \tilde{W}^{\mu} \tilde{V}^{\dagger}_{\mu}
-\tilde{W}^{\mu} \tilde{W}^{\dagger}_{\mu}
  }{m^2}
  \nonumber \\
  & 
  + 
  \frac{
\frac{[\gamma^{\mu} , \gamma^{\nu}]}{4}
\Big(
   \big( 
   (\tilde{d}_{\mu} + \i \tilde{k}_{\mu} )
   ( \tilde{d}_{\nu}^{\dagger} -\i \tilde{k}_{\nu}^{\dagger} )
   -( \tilde{d}_{\nu} + \i \tilde{k}_{\nu})
   ( \tilde{d}_{\mu}^{\dagger} - \i \tilde{k}^{\dagger}_{\mu} )
   \big)
   +
  \i \tilde{F}_{\mu \nu}[ \tilde{V}^{\dagger}] 
   -\i  \gamma^{5} \tilde{F}_{\mu \nu} [\tilde{W}^{\dagger}]
+\big( \tilde{V}_{\mu} \tilde{V}^{\dagger}_{\nu} - \tilde{V}_{\nu} \tilde{V}^{\dagger}_{\mu}  \big)   
-\big( \tilde{W}_{\mu} \tilde{W}^{\dagger}_{\nu} - \tilde{W}_{\nu} \tilde{W}^{\dagger}_{\mu}  \big)   
  }{m^2}
    \nonumber \\
  & 
  \frac{ 
+\i \big( 
\tilde{V}^{\dagger}_{\nu} ( \tilde{d}_{\mu} +\i \tilde{k}_{\mu} )
- \tilde{V}^{\dagger}_{\mu} (\tilde{d}_{\nu}  +\i \tilde{k}_{\nu})
- \tilde{V}_{\mu} (\tilde{d}_{\nu}^{\dagger}  -\i \tilde{k}^{\dagger}_{\nu})
+ \tilde{V}_{\nu} (\tilde{d}^{\dagger}_{\mu} -\i \tilde{k}^{\dagger}_{\mu})
 \big)
+ \gamma^{5} \big( 
-\tilde{V}_{\mu} \tilde{W}^{\dagger}_{\nu}
+\tilde{V}_{\nu} \tilde{W}^{\dagger}_{\mu}
+\tilde{W}_{\mu} \tilde{V}_{\nu}^{\dagger} 
- \tilde{W}_{\nu} \tilde{V}_{\mu}^{\dagger}
 \big)
   }{m^2} 
     \nonumber \\
  & \left.
  \frac{
 +\i \gamma^{5}
 \big(
 -\tilde{W}^{\dagger}_{\nu} (\tilde{d}_{\mu} +\i \tilde{k}_{\mu} )
 +\tilde{W}_{\mu}^{\dagger} (\tilde{d}_{\nu} + \i \tilde{k}_{\nu})
 -\tilde{W}_{\mu} (\tilde{d}_{\nu}^{\dagger} -\i \tilde{k}^{\dagger}_{\nu} )
 +\tilde{W}_{\nu} (\tilde{d}_{\mu}^{\dagger} -\i \tilde{k}^{\dagger}_{\mu} )
 \big)
\Big)
  }{m^2} \Big]
 \right].
  \label{eq:smpdddag2}
\end{align}
Here, we use $e^{-\i k x} C(\tilde{d}_{\mu}) e^{\i k x}= C(\tilde{d}_{\mu} +\i \tilde{k}_{\mu})$ and  $e^{-\i k x} C(\tilde{d}^{\dagger}_{\mu}) e^{\i k x}= C(\tilde{d}^{\dagger}_{\mu} -\i \tilde{k}^{\dagger}_{\mu})$ where $\tilde{k}_{x}=f_{x}^{\nu} k_{\nu}$ with $f$ for the Hermitianized, anti-Hermitianized, and non-Hermitian systems are given in Table~\ref{tabapp:convert}.

We next re-scale the momentum $\tilde{k}_{\mu}=f_{\mu}^{\nu}k_{\nu} \rightarrow m f_{\mu}^{\nu}p_{\nu}= m \tilde{p}_{\mu}$.  Doing this results in canceling the factor $1/m^2$ in terms proportional with $\tilde{k}^{\dagger} \tilde{k}$.
Now, the challenge is evaluating the two Gaussian integrals with exponents $\gamma^{\mu} \gamma^{\nu}\tilde{p}_{\mu } \tilde{p}_{\nu}^{\dagger}$  and $\gamma^{\mu} \gamma^{\nu}\tilde{p}_{\mu }^{\dagger} \tilde{p}_{\nu} $ which are coming from terms like  $k^{\mu}k_{\mu}^{\dagger} +\frac{1}{4} [\gamma^{\mu}, \gamma^{\nu}] (k_{\mu} k^{\dagger}_{\nu}  - k_{\nu} k^{\dagger}_{\mu}) $. 
The subsequent Gaussian integrals read
\begin{align}
\int \frac{{\rm d}^{2} p }{(2 \pi)^{2}}  e^{p_{\alpha} f_{\mu}^{*\alpha} \gamma^{\mu} \gamma^{\nu} f_{\nu}^{\beta} p_{\beta}}
&= \int \frac{{\rm d}^{2} p }{(2 \pi)^{2}} 
 e^{p_{\alpha} f_{\mu}^{*\alpha}
  \big(
 \frac{1}{2} \{\gamma^{\mu} , \gamma^{\nu} \}
 +
 \frac{1}{2} [ \gamma^{\mu},  \gamma^{\nu}]
 \big)
  f_{\nu}^{\beta} p_{\beta}}
    ,\\
&= 
 \int \frac{{\rm d}^{2} p }{(2 \pi)^{2}} 
 e^{p_{\alpha} f_{\mu}^{*\alpha}
  \big(
g^{\mu \nu}
 +
 \frac{1}{2} [ \gamma^{\mu},  \gamma^{\nu}]
 \big)
  f_{\nu}^{\beta} p_{\beta}}
  ,\\
  &= 
   \int \frac{{\rm d}^{2} p }{(2 \pi)^{2}} 
 e^{p_{\alpha} f_{\mu}^{*\alpha}
  \big(
-\id \delta^{\mu \nu}
 +
 \frac{1}{2} [ \gamma^{\mu},  \gamma^{\nu}]
 \big)
  f_{\nu}^{\beta} p_{\beta}}
  ,\\
   &= 
     \int \frac{{\rm d}^{2} p }{(2 \pi)^{2}} 
 e^{ -p_{\alpha} 
 \Big(
\id  \delta^{\mu \nu}
 f_{\mu}^{*\alpha} f_{\nu}^{\beta}
 -
  \frac{1}{2} [ \gamma^{\mu},  \gamma^{\nu}]
f_{\mu}^{*\alpha} f_{\nu}^{\beta} 
 \Big)
 p_{\beta}} = \frac{\pi}{\sqrt{{\rm det}[B]}}, \label{eq:G0}
\end{align}
where the matrix $B$ is the symmetric part of $-f_{\mu}^{*\alpha} \gamma^{\mu} \gamma^{\nu} f_{\nu}^{\beta}$  

 \begin{equation}
 B^{\alpha \beta }= f_{\mu}^{*\alpha}f_{\nu}^{\beta}\delta^{\mu\nu} -\dfrac{i}{2}[ \gamma^{\mu},  \gamma^{\nu}]
\mathrm{Im}[f_{\mu}^{*\alpha}f_{\nu}^{\beta}],\label{eq:constA}
 \end{equation}
 which equals \eqref{eq:constB} in the main text.
For example, for $f=\mathrm{diag}(v_f e^{\i \phi},1)$ the determinant, $\det[B]$, reads 
\begin{align}
\det[B]= v^2_f\cos^2\phi
\label{eq:detconstA},
\end{align}

We next expand the exponential functions in Eqs.~(\ref{eq:smpddagd2},\ref{eq:smpdddag2}) as a power series, take the limit $m \rightarrow \infty$ and only keep terms which are are proportional to $1/m^2$ and with a nonzero trace of gamma matrices. After performing momentum integration, we finally obtain 
\begin{align}
{\cal S}_{5}[\beta] &= \i \int {\rm d}^{2} x  \beta(x) {\cal A}_{5}(x), \label{eq:Sbeta_1p1_2}\\
{\cal A}_{5}
& =
 \frac{1 }{ 4 \pi  \sqrt{\det[B]}}
       \left[
2\Big( 
-\i\tilde{d}^{\mu} ( \tilde{W}^{\dagger}_{\mu} ) 
-\i  \tilde{d}^{\mu \dagger} ( \tilde{W}_{\mu} )
+ ( \tilde{V}^{\mu \dagger} \tilde{W}_{\mu}
-  \tilde{W}^{\mu \dagger} \tilde{V}_{\mu} 
-  \tilde{V}^{\mu} \tilde{W}^{\dagger}_{\mu}
+ \tilde{W}^{\mu} \tilde{V}^{\dagger}_{\mu} )
\Big)
-\varepsilon^{\mu \nu}
\Big(
(
  \i \tilde{F}_{\mu \nu}[ \tilde{V}^{\dagger}] 
         -
  \i \tilde{F}^{\dagger}_{\mu \nu}[ \tilde{V}] 
  )
 \Big)
 \right]
.\label{eq:A5_1p1_final}
\end{align}
 In $d=2$ dimensions, the presented ${\cal A}_{5} $ in the Euclidean spacetime coincides with the associated function in the Minkowski space~\cite{Bertlmann1996}.
By setting $M=\id_{2 \times 2}$ and inserting $V=\Re[V]$ and $W=0$, we retrieve the Hermitian result
${\cal A}_{5}=-\i \varepsilon_{\mu \nu} F^{\mu \nu}/2 \pi $~\cite{Bertlmann1996}.

\paragraph{\bf Non-Chiral Transformation-.}

Under an infinitesimal non-chiral rotation, we again expand the spinor fields in the eigenbases of the Laplacian operators as
\begin{align}
\Psi_{\rm rot} &= e^{\i  \kappa} \Psi(x)=\left(1+\i \kappa \right) \sum_{m}a_{m} \xi_{m}(x) = \sum\limits_{m} a^{\rm rot}_{m} \xi_{m},\\
\bar{\Psi}_{\rm rot} &= \bar{\Psi}(x) e^{-\i  \kappa}= \sum_{m}\bar{b}_{m} \eta^{\dagger}_{m}(x)  \left(1-\i \kappa \right)= \sum_{m}\bar{b}^{\rm rot}_{m} \eta^{\dagger}_{m}(x). 
\end{align}
Here the orthogonality of eigenvectors imposes that
\begin{align}
a_{n}^{\rm rot} = \left(\delta_{nm} + C_{nm} \right)a_{m} =
\left[ \delta_{nm} + \i \kappa \int {\rm d}^2 x \xi^{\dagger}_{n}(x)  \xi_{m}(x) \right],\\ \bar{b}_{n}^{\rm rot} = \left(\delta_{nm} + D_{nm} \right)a_{m} =
\left[ \delta_{nm} - \i \kappa \int {\rm d}^2 x \eta^{\dagger}_{n}(x) \eta_{m}(x) \right].
\end{align}
The path-integral measure for the rotated spinors, and up to the first-order in $C$ and $D$, then yields
\begin{align}
{\cal D} \Psi_{\rm rot} {\cal D} \bar{\Psi}_{\rm rot} &=
\left[
{\rm det}(1+C) {\rm det}(1+D)
\right]^{-1}
{\cal D} \Psi {\cal D} \bar{\Psi}
=
e^{-\tr[D+C]}
{\cal D} \Psi {\cal D} \bar{\Psi}
,\\
&=
e^{
\i \int {\rm d}^{2} x 
\left(
-\kappa(x) \sum\limits_{n} \xi^{\dagger}_{n}(x)  \xi_{n}(x)
+
\kappa(x) \sum\limits_{n} \eta^{\dagger}_{n}(x)\eta_{n}(x)
\right)
}
{\cal D} \Psi {\cal D} \bar{\Psi} = e^{\i \int {\rm d}^{2} x  \kappa(x) {\cal A}(x)} {\cal D} \Psi {\cal D} \bar{\Psi}= e^{{\cal S}[\kappa]}{\cal D} \Psi {\cal D} \bar{\Psi}. \label{eq:Skappa_1p1}
\end{align}
Here ${\cal S}[\kappa]$ is the Jacobian contribution on the action.
To evaluate ${\cal A}$, we employ the Gaussian regulator~\cite{Nakahara1990}, with $m \rightarrow \infty$, such that
\begin{align}
{\cal A} (x) &= \lim_{m \rightarrow \infty} \sum\limits_{n} 
e^{\frac{-\lambda_{n}^{2}}{m^2}}
 \left[
 - \xi^{\dagger}_{n}(x) \xi_{n}(x)
  +
    \eta^{\dagger}_{n}(x)
      \eta_{n}(x)
 \right],\\
 &=
  \lim_{m \rightarrow \infty} \sum\limits_{n} 
 \left[
 - \xi^{\dagger}_{n}(x)   
 e^{\frac{-\tilde{\slashed{\zc D}}^{\dagger} \tilde{\slashed{\zc D}} }{m^2}}
 \xi_{n}(x)
  +
    \eta^{\dagger}_{n}(x)
  e^{\frac{-\tilde{\slashed{\zc D}}  \tilde{\slashed{\zc D}}^{\dagger} }{m^2}}     
      \eta_{n}(x)
 \right]
 ,\\
 &=
   \lim_{m \rightarrow \infty} \sum\limits_{n} 
   \int \frac{{\rm d}^2 k}{(2\pi)^2}
   e^{-\i k x}
   \tr \left[
 -
 e^{\frac{-\tilde{\slashed{\zc D}}^{\dagger} \tilde{\slashed{\zc D}} }{m^2}}
  +
  e^{\frac{-\tilde{\slashed{\zc D}}  \tilde{\slashed{\zc D}}^{\dagger} }{m^2}}     
 \right]
 e^{\i k x},\label{eq:A}
\end{align}
where we have substituted the Fourier transforms of the eigenbases and used the completeness relation of eigenvectors in the last line.
Here $\tilde{\slashed{\zc D}}  \tilde{\slashed{\zc D}}^{\dagger}$ and $\tilde{\slashed{\zc D}}^{\dagger}  \tilde{\slashed{\zc D}}$ are given in Eqs.~(\ref{eq:Dddag},~\ref{eq:DdagD}).
To finally evaluate~\eqref{eq:A}, we apply the variable changes $k_{\mu} \rightarrow p_{\mu}/m$, and for $m \rightarrow \infty$ only keep terms proportional to $1/m^2$ and with nonzero trace of gamma matrices. After performing the momentum integration, we obtain
\begin{align}
{\cal S}[\kappa] &= \i \int {\rm d}^{2} x  \kappa(x) {\cal A}(x), \label{eq:Skappa_1p1_2}\\
{\cal A} 
 &=
  \frac{1 }{ 4 \pi \sqrt{\det[B]}}
     \left[ 2
\Big(
\i \tilde{d}^{\mu} ( \tilde{V}^{\dagger}_{\mu} ) 
+\i \tilde{d}^{\mu \dagger} ( \tilde{V}_{\mu} ) 
\Big)
-\varepsilon^{\mu \nu}
\Big(
   -\i   \tilde{F}_{\mu \nu} [\tilde{W}^{\dagger}]
      +\i  \tilde{F}^{\dagger}_{\mu \nu} [\tilde{W}]
+4  \big( 
\tilde{V}_{\nu} \tilde{W}^{\dagger}_{\mu}
+\tilde{W}_{\mu} \tilde{V}_{\nu}^{\dagger} 
 \big)
\Big)
 \right]
.\label{eq:A_1p1_final}
\end{align}
where $B$ is given in Eq.~\eqref{eq:constA}. In $d=2$ dimensions, the presented ${\cal A} $ in the Euclidean spacetime coincides with the associated function in the Minkowski space~\cite{Bertlmann1996}.
Note that by setting $M=\id_{2\times 2}$ and inserting $V=\Re[V]$ and $W=0$, we retrieve ${\cal A}=0 $~\cite{Bertlmann1996}.

\paragraph{\bf Anomalous chiral and vector currents-.}

Under the chiral and vector transformations in Eqs.~\eqref{eq:barpsi_psirot}, the rotated action in Eq.~\eqref{eq:action_1p1} shifts such that
\begin{align}
    \tilde{\cal S}_{\rm rot} -\tilde{\cal S} =& - \int {\rm d}^{2} x \Big[\beta(x) \tilde{d}_{\nu}  j^{5,\mu}
    - \kappa(x) 
    \tilde{d}_{\nu}  j^{\mu} \Big],
\end{align}
where the chiral and vector currents are $j^{5,\mu}=\bar{\Psi} \gamma^{\mu} \gamma^{5} \Psi$ and $j^{\mu} = \bar{\Psi} \gamma^{\mu} \Psi$, respectively. Finally, the variation of the partition function with respect to $\beta$ and $\kappa$ enforces the invariance of $\tilde{\cal Z}$ under the chiral and vector rotations by satisfying ${\cal A}_{5}= \i \tilde{d}_{\mu}  j^{5,\mu} $, and ${\cal A}= -\i \tilde{d}_{\mu}  j^{\mu}$. As a result we get
\begin{align}
    \i \tilde{d}_{\mu}  j^{\mu} &= 
  \frac{-1 }{ 4 \pi \sqrt{\det[B]}}
     \left[ 2
\Big(
\i \tilde{d}^{\mu} ( \tilde{V}^{\dagger}_{\mu} ) 
+\i \tilde{d}^{\mu \dagger} ( \tilde{V}_{\mu} ) 
\Big)
-\varepsilon^{\mu \nu}
\Big(
\i ( \tilde{F}^{\dagger}_{\mu \nu} [\tilde{W}]
   -  \tilde{F}_{\mu \nu} [\tilde{W}^{\dagger}]
   )
+8 \Re[
\tilde{V}_{\nu} \tilde{W}^{\dagger}_{\mu}
]
\Big)
 \right]
    , \label{eq:dJ_1p1_final} 
    \end{align}
    \begin{align}
     \i \tilde{d}_{\mu}  j^{5,\mu} =
 \frac{1 }{ 4 \pi  \sqrt{\det[B]}}&
       \left[
 2
\Big( 
-\i\tilde{d}^{\mu} ( \tilde{W}^{\dagger}_{\mu} ) 
-\i  \tilde{d}^{\mu \dagger} ( \tilde{W}_{\mu} )
- 
( \tilde{V}^{\mu} \tilde{W}^{\dagger}_{\mu}
- \tilde{W}^{\mu} \tilde{V}^{\dagger}_{\mu} )
+ ( \tilde{V}^{\mu \dagger} \tilde{W}_{\mu}
-  \tilde{W}^{\mu \dagger} \tilde{V}_{\mu} )
\Big)
-\varepsilon^{\mu \nu}
\Big(
(
  \i \tilde{F}_{\mu \nu}[ \tilde{V}^{\dagger}] 
         -
  \i \tilde{F}^{\dagger}_{\mu \nu}[ \tilde{V}] 
  )
 \Big)
 \right]
    , \label{eq:dJ5_1p1_final}
\end{align}
where $B$ is defined in Eq.~\eqref{eq:constA}.

To facilitate further comparison between these three systems, we refer to the divergence of chiral currents in Hermitianized, anti-Hermitianized, and non-Hermitian systems by $[\tilde{d}_{\mu} j^{5,\mu}]_{\rm h}$, $[\tilde{d}_{\mu} j^{5,\mu}]_{\rm ah}$ and $[\tilde{d}_{\mu} j^{5,\mu}]_{\rm nh}$, respectively. The non-Hermitian anomaly $[\tilde{d}_{\mu} j^{5,\mu}]_{\rm nh}$ is different from the naive summation of $[\tilde{d}_{\mu} j^{5,\mu}]_{\rm h}+[\tilde{d}_{\mu} j^{5,\mu}]_{\rm ah}$, as can be seen by noticing, for example, that the $det[B]$ is different for all three cases.

\subsubsection{$4$ dimensions} \label{App:Fujikawa3p1}

\paragraph{\bf Chiral Transformation-.}

To evaluate the change of the path-integral measure under an infinitesimal chiral rotation, we first expand the spinor fields in the eigenbases of Laplacian operators as
\begin{align}
\Psi_{\rm rot} &= e^{-\i \gamma^{5} \beta} \Psi(x)=\left(1-\i \beta \gamma^{5} \right) \sum_{m}a_{m} \xi_{m}(x) = \sum\limits_{m}  \xi_{m} a^{\rm rot}_{m},\\
\bar{\Psi}_{\rm rot} &= \bar{\Psi}(x) e^{-\i \gamma^{5} \beta}= \sum_{m}\bar{b}_{m} \eta^{\dagger}_{m}(x)  \left(1-\i \beta \gamma^{5} \right)= \sum_{m}\bar{b}^{\rm rot}_{m} \eta^{\dagger}_{m}(x). 
\end{align}
Here the orthogonality of eigenfunctions imposes that
\begin{align}
a_{n}^{\rm rot} = \left(\delta_{nm} + C_{nm} \right)a_{m} =
\left[ \delta_{nm} - \i \beta \int {\rm d}^4 x \xi^{\dagger}_{n}(x) \gamma^{5} \xi_{m}(x) \right]a_{m}
,\\
 \bar{b}_{n}^{\rm rot} = \bar{b}_{m} \left(\delta_{nm} + D_{nm} \right) = \bar{b}_{m}
\left[ \delta_{nm} - \i \beta \int {\rm d}^4 x \eta^{\dagger}_{m}(x) \gamma^{5} \eta_{n}(x) \right].
\end{align}
The path-integral measure for the rotated spinors, and up to the first-order in $C$ and $D$, then yields
\begin{align}
{\cal D} \Psi_{\rm rot} {\cal D} \bar{\Psi}_{\rm rot} &=
\left[
{\rm det}(1+C) {\rm det}(1+D)
\right]^{-1}
{\cal D} \Psi {\cal D} \bar{\Psi} = e^{-\tr [D+C]} {\cal D} \Psi {\cal D} \bar{\Psi} ,\\
&=
e^{
\i \int {\rm d}^{4} x 
\beta(x)
\left(
\sum\limits_{n} \xi^{\dagger}_{n}(x) \gamma^{5} \xi_{n}(x)
+
 \sum\limits_{n} \eta^{\dagger}_{n}(x) \gamma^{5} \eta_{n}(x)
\right)
}
{\cal D} \Psi {\cal D} \bar{\Psi} 
= e^{\i \int {\rm d}^{4} x  \beta(x) {\cal A}_{5}(x)} {\cal D} \Psi {\cal D} \bar{\Psi} = e^{{\cal S}_{5}[\beta]}{\cal D} \Psi {\cal D} \bar{\Psi}. \label{eq:Sbeta_3p1}
\end{align}
Here ${\cal S}_{5}[\beta]$ is the Jacobian contribution on the action.
${\cal A}_{5}$ is ill-defined as using the completeness relation in Eq.~\eqref{eq:norm_comp_1p1} results in ${\cal A}_{5} \propto \tr[\gamma^{5}.\delta(0)]$.
To evaluate ${\cal A}_{5}$, we use a heat-kernel regulator~\cite{Nakahara1990}, with a Gaussian cut-off $m \rightarrow \infty$, such that
\begin{align}
{\cal A}_{5} (x) &= \lim_{m \rightarrow \infty} \sum\limits_{n} 
e^{\frac{-\lambda_{n}^{2}}{m^2}}
 \left[
  \xi^{\dagger}_{n}(x) \gamma^{5} \xi_{n}(x)
  +
    \eta^{\dagger}_{n}(x)
     \gamma^{5} \eta_{n}(x)
 \right],\\
 &=
  \lim_{m \rightarrow \infty} \sum\limits_{n} 
 \left[
  \xi^{\dagger}_{n}(x)   
 \gamma^{5}   
 e^{\frac{-\tilde{\slashed{\zc D}}^{\dagger} \tilde{\slashed{\zc D}} }{m^2}}
 \xi_{n}(x)
  +
    \eta^{\dagger}_{n}(x)
     \gamma^{5}
  e^{\frac{-\tilde{\slashed{\zc D}}  \tilde{\slashed{\zc D}}^{\dagger} }{m^2}}     
      \eta_{n}(x)
 \right]
 ,\\
 &=
   \lim_{m \rightarrow \infty} \sum\limits_{n} 
   \int \frac{{\rm d}^4 k}{(2\pi)^4}
   e^{-\i k x}
   \tr \left[
 \gamma^{5}   
 e^{\frac{-\tilde{\slashed{\zc D}}^{\dagger} \tilde{\slashed{\zc D}} }{m^2}}
  +
     \gamma^{5}
  e^{\frac{-\tilde{\slashed{\zc D}}  \tilde{\slashed{\zc D}}^{\dagger} }{m^2}}     
 \right]
 e^{\i k x},\label{eq:A5_3p1}
\end{align}
where we have substituted the Fourier transforms of the eigenbases and used the completeness relation of the eigenvectors in the last line.

Inserting Eqs.~(\ref{eq:Dddag},~\ref{eq:DdagD}) into Eq.~\eqref{eq:A5_3p1} reproduces Eqs.~(\ref{eq:smpddagd2}, \ref{eq:smpdddag2}). Following similar steps as we presented in Sec.~\ref{sec:1p1app}, we again encounter a Gaussian integral which reads
\begin{eqnarray}
\int \frac{{\rm d}^{4} p }{(2 \pi)^{4}}  e^{p_{\alpha} f_{\mu}^{*\alpha} \gamma^{\mu} \gamma^{\nu} f_{\nu}^{\beta} p_{\beta}}
&=& \int \frac{{\rm d}^{4} p }{(2 \pi)^{4}} 
 e^{p_{\alpha} f_{\mu}^{*\alpha}
  \big(
 \frac{1}{2} \{\gamma^{\mu} , \gamma^{\nu} \}
 +
 \frac{1}{2} [ \gamma^{\mu},  \gamma^{\nu}]
 \big)
  f_{\nu}^{\beta} p_{\beta}}
  ,\\
  &=& 
 \int \frac{{\rm d}^{4} p }{(2 \pi)^{4}} 
 e^{p_{\alpha} f_{\mu}^{*\alpha}
  \big(
g^{\mu \nu}
 +
 \frac{1}{2} [ \gamma^{\mu},  \gamma^{\nu}]
 \big)
  f_{\nu}^{\beta} p_{\beta}}
  ,\\
  \label{eq:Bwithsym}
  &=& 
   \int \frac{{\rm d}^{4} p }{(2 \pi)^{4}} 
 e^{p_{\alpha} f_{\mu}^{*\alpha}
  \big(
-\delta^{\mu \nu}
 +
 \frac{1}{2} [ \gamma^{\mu},  \gamma^{\nu}]
 \big)
  f_{\nu}^{\beta} p_{\beta}}
  ,\\
  \label{eq:Bnosym}
   &=& 
     \int \frac{{\rm d}^{4} p }{(2 \pi)^{4}} 
 e^{-p_{\alpha} 
 \Big(
 \delta^{\mu \nu}
 f_{\mu}^{*\alpha} f_{\nu}^{\beta}
 -
  \frac{1}{2} [ \gamma^{\mu},  \gamma^{\nu}]
\frac{f_{\mu}^{*\alpha} f_{\nu}^{\beta} 
- f_{\mu}^{\alpha} f_{\nu}^{*\beta}   
}{2}
 \Big)
 p_{\beta}}=\frac{\pi^2}{\sqrt{{\rm det}[B]}} ,\label{eq:G0_3p1}
\end{eqnarray}
where the matrix elements of $B$ given in Eq.~\eqref{eq:constA}.

Using for example $f=\mathrm(v e^{\i \phi},v,v,1)$, the determinant of $B$ reads
\begin{align}
\det[B]= v^6[3\cos^2(\phi)-2]
\label{eq:detconstA3d},
\end{align}

We next expand the exponential functions in Eqs.~(\ref{eq:smpddagd2}, \ref{eq:smpdddag2}) as power series, take the limit $m \rightarrow \infty$ and only keep terms which are are proportional to $1/m^2$ and with a nonzero trace of gamma matrices. After performing momentum integration, we finally obtain
\begin{align}
{\cal S}_{5}[\beta] &= \i \int {\rm d}^{4} x  \beta(x) {\cal A}_{5}(x), \label{eq:Sbeta_3p1_2}\\
{\cal A}_{5}
& =
   \lim_{m \rightarrow \infty} \sum\limits_{n} 
   \int \frac{{\rm d}^4 k}{(2\pi)^4}
   e^{-\i k x}
   \tr \left[ \gamma^{5} [
     e^{-\frac{\slashed{\zc D}  \slashed{\zc D}^{\dagger} }{m^2}}  
 +
 e^{-\frac{\slashed{\zc D}^{\dagger} \slashed{\zc D} }{m^2}}
 ]
 \right]
 e^{\i k x},\\
 =&
   \frac{1}{32 \pi^2 \sqrt{{\rm det}[B]}} 
     \left[
  4
\Big(
+\i \tilde{d}^{\mu} ( \tilde{V}^{\dagger}_{\mu} ) 
+\tilde{V}^{\mu} \tilde{V}^{\dagger}_{\mu}
-\tilde{W}^{\mu} \tilde{W}^{\dagger}_{\mu}
\Big)
\Big(
-\i \tilde{d}^{\nu} ( \tilde{W}^{\dagger}_{\nu} ) 
-  \tilde{V}^{\nu} \tilde{W}^{\dagger}_{\nu}
+ \tilde{W}^{\nu} \tilde{V}^{\dagger}_{\nu}
\Big)
  \right.
  \nonumber \\
  & \qquad \qquad 
+4
 \Big(
-\i \tilde{d}^{\mu \dagger} ( \tilde{V}_{\mu} ) 
+\tilde{V}^{\mu \dagger} \tilde{V}_{\mu}
-\tilde{W}^{\mu \dagger} \tilde{W}_{\mu}
\Big)\Big(
-\i \tilde{d}^{\nu \dagger} ( \tilde{W}_{\nu} )
+ ( \tilde{V}^{\nu \dagger} \tilde{W}_{\nu}
-  \tilde{W}^{\nu \dagger} \tilde{V}_{\nu} )
\Big)
  \nonumber \\
  & 
  - \varepsilon^{\mu \nu \eta \zeta}
\Big(
   +
  \i \tilde{F}_{\mu \nu}[ \tilde{V}^{\dagger}] 
  +\big( \tilde{V}_{\mu} \tilde{V}^{\dagger}_{\nu} - \tilde{V}_{\nu} \tilde{V}^{\dagger}_{\mu}  \big)   
-\big( \tilde{W}_{\mu} \tilde{W}^{\dagger}_{\nu} - \tilde{W}_{\nu} \tilde{W}^{\dagger}_{\mu}  \big)   
  \Big)
\Big(
   +
  \i \tilde{F}_{\eta \zeta}[ \tilde{V}^{\dagger}] 
  +\big( \tilde{V}_{\eta} \tilde{V}^{\dagger}_{\zeta} 
  - \tilde{V}_{\zeta} \tilde{V}^{\dagger}_{\eta}  \big)   
-\big( \tilde{W}_{\eta} \tilde{W}^{\dagger}_{\zeta} 
- \tilde{W}_{\zeta} \tilde{W}^{\dagger}_{\eta}  \big)   
  \Big)
    \nonumber \\
  & 
    -  \varepsilon^{\mu \nu \eta \zeta}
  \Big(
       -
  \i \tilde{F}^{\dagger}_{\mu \nu}[ \tilde{V}] 
  +\big( \tilde{V}^{\dagger}_{\mu} \tilde{V}_{\nu} - \tilde{V}^{\dagger}_{\nu} \tilde{V}_{\mu}  \big)   
-\big( \tilde{W}^{\dagger}_{\mu} \tilde{W}_{\nu} - \tilde{W}^{\dagger}_{\nu} \tilde{W}_{\mu}  \big)   
\Big) 
\Big(
       -
  \i \tilde{F}^{\dagger}_{\eta \zeta}[ \tilde{V}] 
  +\big( \tilde{V}^{\dagger}_{\eta} \tilde{V}_{\zeta}
   - \tilde{V}^{\dagger}_{\zeta} \tilde{V}_{\eta}  \big)   
-\big( \tilde{W}^{\dagger}_{\eta} \tilde{W}_{\zeta}
 - \tilde{W}^{\dagger}_{\zeta} \tilde{W}_{\eta}  \big)   
\Big) 
    \nonumber \\
  &
    - \varepsilon^{\mu \nu \eta \zeta}
 \Big(
   -\i  \tilde{F}^{\dagger}_{\mu \nu} [\tilde{W}]
+ \big( 
\tilde{V}^{\dagger}_{\mu} \tilde{W}_{\nu}
-\tilde{V}^{\dagger}_{\nu} \tilde{W}_{\mu}
-\tilde{W}^{\dagger}_{\mu} \tilde{V}_{\nu} 
+ \tilde{W}^{\dagger}_{\nu} \tilde{V}_{\mu}
 \big)
\Big) 
\Big(
   -\i   \tilde{F}^{\dagger}_{\eta\zeta} [\tilde{W}]
+ \big( 
\tilde{V}^{\dagger}_{\eta} \tilde{W}_{\zeta}
-\tilde{V}^{\dagger}_{\zeta} \tilde{W}_{\eta}
-\tilde{W}^{\dagger}_{\eta} \tilde{V}_{\nu} 
+ \tilde{W}^{\dagger}_{\zeta} \tilde{V}_{\eta}
 \big)
\Big) 
    \nonumber \\
  & \left.
    - \varepsilon^{\mu \nu \eta \zeta}
\Big(
  -\i   \tilde{F}_{\mu \nu} [\tilde{W}^{\dagger}]
+ \big( 
-\tilde{V}_{\mu} \tilde{W}^{\dagger}_{\nu}
+\tilde{V}_{\nu} \tilde{W}^{\dagger}_{\mu}
+\tilde{W}_{\mu} \tilde{V}_{\nu}^{\dagger} 
- \tilde{W}_{\nu} \tilde{V}_{\mu}^{\dagger}
 \big)
  \Big)
\Big(
  -\i  \tilde{F}_{\eta \zeta} [\tilde{W}^{\dagger}]
+  \big( 
-\tilde{V}_{\eta} \tilde{W}^{\dagger}_{\zeta}
+\tilde{V}_{\zeta} \tilde{W}^{\dagger}_{\eta}
+\tilde{W}_{\eta} \tilde{V}_{\zeta}^{\dagger} 
- \tilde{W}_{\zeta} \tilde{V}_{\eta}^{\dagger}
 \big)
  \Big)
 \right]
.\label{eq:A5_3p1_final}
\end{align}
Here, the matrix elements of $B$ are defined in Eq.~\eqref{eq:constA} with $f$ given in Table~\ref{tabapp:convert} for Hermitianized, anti-Hermitianized, and non-Hermitian systems.

Note that by setting $M=\id_{4 \times 4}$ and inserting $V=\Re[V]$ and $W=0$, we retrieve the Hermitian result ${\cal A}_{5}= \varepsilon_{\mu \nu \rho \sigma} F^{\mu \nu} F^{\rho \sigma}/16 \pi^2 $~\cite{Bertlmann1996}.

\paragraph{\bf Non-Chiral Transformation-.}

Under an infinitesimal non-chiral rotation, we can again expand the spinor fields in the eigenbases of Laplacian operators as
\begin{align}
\Psi_{\rm rot} &= e^{\i  \kappa} \Psi(x)=\left(1+\i \kappa \right) \sum_{m}a_{m} \xi_{m}(x) = \sum\limits_{m} a^{\rm rot}_{m} \xi_{m},\\
\bar{\Psi}_{\rm rot} &= \bar{\Psi}(x) e^{-\i  \kappa}= \sum_{m}\bar{b}_{m} \eta^{\dagger}_{m}(x)  \left(1-\i \kappa \right)= \sum_{m}\bar{b}^{\rm rot}_{m} \eta^{\dagger}_{m}(x). 
\end{align}
Here the orthogonality of eigenvectors imposes that
\begin{align}
a_{n}^{\rm rot} = \left(\delta_{nm} + C_{nm} \right)a_{m} =
\left[ \delta_{nm} + \i \kappa \int {\rm d}^4 x \xi^{\dagger}_{n}(x)  \xi_{m}(x) \right],\\ \bar{b}_{n}^{\rm rot} = \left(\delta_{nm} + D_{nm} \right)a_{m} =
\left[ \delta_{nm} - \i \kappa \int {\rm d}^4 x \eta^{\dagger}_{n}(x) \eta_{m}(x) \right].
\end{align}
The path-integral measure for the rotated spinors, and up to the first-order in $C$ and $D$, then yields
\begin{align}
{\cal D} \Psi_{\rm rot} {\cal D} \bar{\Psi}_{\rm rot} &=
\left[
{\rm det}(1+C) {\rm det}(1+D)
\right]^{-1}
{\cal D} \Psi {\cal D} \bar{\Psi}
=
e^{-\tr[D+C]}
{\cal D} \Psi {\cal D} \bar{\Psi}
,\\
&=
e^{
\i \int {\rm d}^{4} x 
\left(
-\kappa(x) \sum\limits_{n} \xi^{\dagger}_{n}(x)  \xi_{n}(x)
+
\kappa(x) \sum\limits_{n} \eta^{\dagger}_{n}(x)\eta_{n}(x)
\right)
}
{\cal D} \Psi {\cal D} \bar{\Psi} = e^{\i \int {\rm d}^{4} x  \kappa(x) {\cal A}(x)} {\cal D} \Psi {\cal D} \bar{\Psi}= e^{{\cal S}[\kappa]}{\cal D} \Psi {\cal D} \bar{\Psi}. \label{eq:Skappa_3p1}
\end{align}
Here ${\cal S}[\kappa]$ is the Jacobian contribution on the action.
To evaluate ${\cal A}$, we employ the Gaussian regulator~\cite{Nakahara1990}, with $m \rightarrow \infty$, such that
\begin{align}
{\cal A} (x) &= \lim_{m \rightarrow \infty} \sum\limits_{n} 
e^{\frac{-\lambda_{n}^{2}}{m^2}}
 \left[
 - \xi^{\dagger}_{n}(x) \xi_{n}(x)
  +
    \eta^{\dagger}_{n}(x)
      \eta_{n}(x)
 \right],\\
 &=
  \lim_{m \rightarrow \infty} \sum\limits_{n} 
 \left[
 - \xi^{\dagger}_{n}(x)   
 e^{\frac{-\tilde{\slashed{\zc D}}^{\dagger} \tilde{\slashed{\zc D}} }{m^2}}
 \xi_{n}(x)
  +
    \eta^{\dagger}_{n}(x)
  e^{\frac{-\tilde{\slashed{\zc D}}  \tilde{\slashed{\zc D}}^{\dagger} }{m^2}}     
      \eta_{n}(x)
 \right]
 ,\\
 &=
   \lim_{m \rightarrow \infty} \sum\limits_{n} 
   \int \frac{{\rm d}^4 k}{(2\pi)^4}
   e^{-\i k x}
   \tr \left[
 -
 e^{\frac{-\tilde{\slashed{\zc D}}^{\dagger} \tilde{\slashed{\zc D}} }{m^2}}
  +
  e^{\frac{-\tilde{\slashed{\zc D}}  \tilde{\slashed{\zc D}}^{\dagger} }{m^2}}     
 \right]
 e^{\i k x},\label{eq:A_3p1}
\end{align}
where we have substituted the Fourier transforms of the eigenbases and used the completeness relation of eigenvectors in the last line.
Here $\tilde{\slashed{\zc D}}  \tilde{\slashed{\zc D}}^{\dagger}$ and $\tilde{\slashed{\zc D}}^{\dagger}  \tilde{\slashed{\zc D}}$ are given in Eqs. (\ref{eq:Dddag},~\ref{eq:DdagD}).
To finally evaluate~\eqref{eq:A_3p1}, we apply a change of variables $k_{\mu} \rightarrow p_{\mu}/m$, and in the limit $m \rightarrow \infty$, we only keep terms proportional to $1/m^4$ and with nonzero trace of gamma matrices. After performing the momentum integration, we obtain
\begin{align}
{\cal S}[\kappa] &= \i \int {\rm d}^{4} x  \kappa(x) {\cal A}(x), \label{eq:Skappa_3p1_2}\\
{\cal A} 
 &=
   \lim_{m \rightarrow \infty} \sum\limits_{n} 
   \int \frac{{\rm d}^4 k}{(2\pi)^4}
   e^{-\i k x}
   \tr \left[
     e^{\frac{\slashed{\zc D}  \slashed{\zc D}^{\dagger} }{m^2}}  
 -
 e^{\frac{\slashed{\zc D}^{\dagger} \slashed{\zc D} }{m^2}}
 \right]
 e^{\i k x},\\
 &= 
  \frac{1}{32 \pi^2 \sqrt{{\rm det}[B]}} 
   \left[ 4
\Big[ 
+\i \tilde{d}^{\mu} ( \tilde{V}^{\dagger}_{\mu} ) 
\Big(
+\tilde{V}^{\mu} \tilde{V}^{\dagger}_{\mu}
-\tilde{W}^{\mu} \tilde{W}^{\dagger}_{\mu}
\Big)
-\i\gamma^{5} \tilde{d}^{\mu} ( \tilde{W}^{\dagger}_{\mu} ) 
\Big(
- \gamma^{5} \tilde{V}^{\mu} \tilde{W}^{\dagger}_{\mu}
+ \gamma^{5} \tilde{W}^{\mu} \tilde{V}^{\dagger}_{\mu}
\Big)
\Big]
  \right.
\nonumber \\
& \qquad \qquad \qquad \quad 
-
   4
 \Big[ 
-\i \tilde{d}^{\mu \dagger} ( \tilde{V}_{\mu} )
\Big(
+\tilde{V}^{\mu \dagger} \tilde{V}_{\mu}
-\tilde{W}^{\mu \dagger} \tilde{W}_{\mu}
\Big)
-\i \gamma^{5} \tilde{d}^{\mu \dagger} ( \tilde{W}_{\mu} )
\Big(
+ \gamma^{5} ( \tilde{V}^{\mu \dagger} \tilde{W}_{\mu}
-  \tilde{W}^{\mu \dagger} \tilde{V}_{\mu} )
\Big)
\Big]
  \nonumber \\
  & 
  - \varepsilon^{\mu \nu \eta \zeta} 
\Big(
  \i \tilde{F}_{\mu \nu}[ \tilde{V}^{\dagger}] 
  +\big( \tilde{V}_{\mu} \tilde{V}^{\dagger}_{\nu} 
  - \tilde{V}_{\nu} \tilde{V}^{\dagger}_{\mu}  \big)   
-\big( \tilde{W}_{\mu} \tilde{W}^{\dagger}_{\nu} 
- \tilde{W}_{\nu} \tilde{W}^{\dagger}_{\mu}  \big)   
\Big)
  \Big(
   -\i  \tilde{F}_{\eta \zeta} [\tilde{W}^{\dagger}]
+  \big( 
\tilde{V}_{\zeta} \tilde{W}^{\dagger}_{\eta}
-\tilde{V}_{\eta} \tilde{W}^{\dagger}_{\zeta}
+\tilde{W}_{\eta} \tilde{V}_{\zeta}^{\dagger} 
- \tilde{W}_{\zeta} \tilde{V}_{\eta}^{\dagger}
 \big)
\Big)
\nonumber \\
&\left.
+  \varepsilon^{\mu \nu \eta \zeta}
\Big(
       -
  \i \tilde{F}^{\dagger}_{\mu \nu}[ \tilde{V}] 
+\big( \tilde{V}^{\dagger}_{\mu} \tilde{V}_{\nu} 
- \tilde{V}^{\dagger}_{\nu} \tilde{V}_{\mu}  \big)   
-\big( \tilde{W}^{\dagger}_{\mu} \tilde{W}_{\nu} 
- \tilde{W}^{\dagger}_{\nu} \tilde{W}_{\mu}  \big)   
\Big)
\Big(
   -\i   \tilde{F}^{\dagger}_{\eta \zeta} [\tilde{W}] 
 +  \big( 
\tilde{V}^{\dagger}_{\eta} \tilde{W}_{\zeta}
-\tilde{V}^{\dagger}_{\zeta} \tilde{W}_{\eta}
-\tilde{W}^{\dagger}_{\eta} \tilde{V}_{\zeta} 
+ \tilde{W}^{\dagger}_{\zeta} \tilde{V}_{\eta}
 \big)
 \Big)
 \right],\label{eq:A_3p1_final}
 \end{align}
where $B$ is given in Eq.~\eqref{eq:constA}.
Note that by setting $M=\id_{4\times 4}$ and inserting $V=\Re[V]$ and $W=0$, we retrieve the topological Hermitian result of ${\cal A}=0 $~\cite{Bertlmann1996}.

\paragraph{\bf Anomalous chiral and vector currents-.}

Under the chiral and vector transformations in Eqs.~\eqref{eq:barpsi_psirot}, the rotated action in Eq.~\eqref{eq:action_1p1} shifts such that
\begin{align}
    \tilde{\cal S}_{\rm rot} -\tilde{\cal S} =& -  \int {\rm d}^{2} x \Big[\beta(x) \tilde{d}_{\nu}  j^{5,\mu}
    - \kappa(x) 
    \tilde{d}_{\nu}  j^{\mu} \Big],
\end{align}
where the chiral and vector currents are $j^{5,\mu}=\bar{\Psi} \gamma^{\mu} \gamma^{5} \Psi$ and $j^{\mu} = \bar{\Psi} \gamma^{\mu} \Psi$, respectively. Finally, the variation of the partition function with respect to $\beta$ and $\kappa$ enforces the invariance of $\tilde{\cal Z}$ under the chiral and vector rotations by satisfying ${\cal A}_{5}=  \i \tilde{d}_{\mu}  j^{5,\mu} $, and ${\cal A}= -\i \tilde{d}_{\mu}  j^{\mu}$. As a result we get
\begin{align}
 - \i \tilde{d}_{\mu}  j^{\mu}&= 
    \frac{1}{32 \pi^2 \sqrt{{\rm det}[B]}} 
   \left[ 4
\Big[ 
+\i \tilde{d}^{\mu} ( \tilde{V}^{\dagger}_{\mu} ) 
\Big(
+\tilde{V}^{\mu} \tilde{V}^{\dagger}_{\mu}
-\tilde{W}^{\mu} \tilde{W}^{\dagger}_{\mu}
\Big)
-\i\gamma^{5} \tilde{d}^{\mu} ( \tilde{W}^{\dagger}_{\mu} ) 
\Big(
- \gamma^{5} \tilde{V}^{\mu} \tilde{W}^{\dagger}_{\mu}
+ \gamma^{5} \tilde{W}^{\mu} \tilde{V}^{\dagger}_{\mu}
\Big)
\Big]
  \right.
\nonumber \\
& \qquad \qquad \qquad \quad 
-
   4
 \Big[ 
-\i \tilde{d}^{\mu \dagger} ( \tilde{V}_{\mu} )
\Big(
+\tilde{V}^{\mu \dagger} \tilde{V}_{\mu}
-\tilde{W}^{\mu \dagger} \tilde{W}_{\mu}
\Big)
-\i \gamma^{5} \tilde{d}^{\mu \dagger} ( \tilde{W}_{\mu} )
\Big(
+ \gamma^{5} ( \tilde{V}^{\mu \dagger} \tilde{W}_{\mu}
-  \tilde{W}^{\mu \dagger} \tilde{V}_{\mu} )
\Big)
\Big]
  \nonumber \\
  & 
  - \varepsilon^{\mu \nu \eta \zeta} 
\Big(
  \i \tilde{F}_{\mu \nu}[ \tilde{V}^{\dagger}] 
  +\big( \tilde{V}_{\mu} \tilde{V}^{\dagger}_{\nu} 
  - \tilde{V}_{\nu} \tilde{V}^{\dagger}_{\mu}  \big)   
-\big( \tilde{W}_{\mu} \tilde{W}^{\dagger}_{\nu} 
- \tilde{W}_{\nu} \tilde{W}^{\dagger}_{\mu}  \big)   
\Big)
  \Big(
   -\i  \tilde{F}_{\eta \zeta} [\tilde{W}^{\dagger}]
+  \big( 
-\tilde{V}_{\eta} \tilde{W}^{\dagger}_{\zeta}
+\tilde{V}_{\zeta} \tilde{W}^{\dagger}_{\eta}
+\tilde{W}_{\eta} \tilde{V}_{\zeta}^{\dagger} 
- \tilde{W}_{\zeta} \tilde{V}_{\eta}^{\dagger}
 \big)
\Big)
\nonumber \\
&\left.
+  \varepsilon^{\mu \nu \eta \zeta}
\Big(
       -
  \i \tilde{F}^{\dagger}_{\mu \nu}[ \tilde{V}] 
+\big( \tilde{V}^{\dagger}_{\mu} \tilde{V}_{\nu} - \tilde{V}^{\dagger}_{\nu} \tilde{V}_{\mu}  \big)   
-\big( \tilde{W}^{\dagger}_{\mu} \tilde{W}_{\nu} - \tilde{W}^{\dagger}_{\nu} \tilde{W}_{\mu}  \big)   
\Big)
\Big(
   -\i   \tilde{F}^{\dagger}_{\eta \zeta} [\tilde{W}] 
 +  \big( 
\tilde{V}^{\dagger}_{\eta} \tilde{W}_{\zeta}
-\tilde{V}^{\dagger}_{\zeta} \tilde{W}_{\eta}
-\tilde{W}^{\dagger}_{\eta} \tilde{V}_{\zeta} 
+ \tilde{W}^{\dagger}_{\zeta} \tilde{V}_{\eta}
 \big)
 \Big)
 \right]
    , \label{eq:dJ_3p1_final} 
    \end{align}
    \begin{align}
  \i \tilde{d}_{\mu}  j^{5,\mu} &=
    \frac{1}{32 \pi^2 \sqrt{{\rm det}[B]}} 
     \left[
  4
\Big(
+\i \tilde{d}^{\mu} ( \tilde{V}^{\dagger}_{\mu} ) 
+\tilde{V}^{\mu} \tilde{V}^{\dagger}_{\mu}
-\tilde{W}^{\mu} \tilde{W}^{\dagger}_{\mu}
\Big)
\Big(
-\i \tilde{d}^{\nu} ( \tilde{W}^{\dagger}_{\nu} ) 
-  \tilde{V}^{\nu} \tilde{W}^{\dagger}_{\nu}
+ \tilde{W}^{\nu} \tilde{V}^{\dagger}_{\nu}
\Big)
  \right.
  \nonumber \\
  & \qquad \qquad 
+4
 \Big(
-\i \tilde{d}^{\mu \dagger} ( \tilde{V}_{\mu} ) 
+\tilde{V}^{\mu \dagger} \tilde{V}_{\mu}
-\tilde{W}^{\mu \dagger} \tilde{W}_{\mu}
\Big)\Big(
-\i \tilde{d}^{\nu \dagger} ( \tilde{W}_{\nu} )
+ ( \tilde{V}^{\nu \dagger} \tilde{W}_{\nu}
-  \tilde{W}^{\nu \dagger} \tilde{V}_{\nu} )
\Big)
  \nonumber \\
  & 
  - \varepsilon^{\mu \nu \eta \zeta}
\Big(
  \i \tilde{F}_{\mu \nu}[ \tilde{V}^{\dagger}] 
  +\big( \tilde{V}_{\mu} \tilde{V}^{\dagger}_{\nu} - \tilde{V}_{\nu} \tilde{V}^{\dagger}_{\mu}  \big)   
-\big( \tilde{W}_{\mu} \tilde{W}^{\dagger}_{\nu} - \tilde{W}_{\nu} \tilde{W}^{\dagger}_{\mu}  \big)   
  \Big)
\Big(
  \i \tilde{F}_{\eta \zeta}[ \tilde{V}^{\dagger}] 
  +\big( \tilde{V}_{\eta} \tilde{V}^{\dagger}_{\zeta} 
  - \tilde{V}_{\zeta} \tilde{V}^{\dagger}_{\eta}  \big)   
-\big( \tilde{W}_{\eta} \tilde{W}^{\dagger}_{\zeta} 
- \tilde{W}_{\zeta} \tilde{W}^{\dagger}_{\eta}  \big)   
  \Big)
    \nonumber \\
  & 
    -  \varepsilon^{\mu \nu \eta \zeta}
  \Big(
       -
  \i \tilde{F}^{\dagger}_{\mu \nu}[ \tilde{V}] 
  +\big( \tilde{V}^{\dagger}_{\mu} \tilde{V}_{\nu} - \tilde{V}^{\dagger}_{\nu} \tilde{V}_{\mu}  \big)   
-\big( \tilde{W}^{\dagger}_{\mu} \tilde{W}_{\nu} - \tilde{W}^{\dagger}_{\nu} \tilde{W}_{\mu}  \big)   
\Big) 
\Big(
       -
  \i \tilde{F}^{\dagger}_{\eta \zeta}[ \tilde{V}] 
  +\big( \tilde{V}^{\dagger}_{\eta} \tilde{V}_{\zeta}
   - \tilde{V}^{\dagger}_{\zeta} \tilde{V}_{\eta}  \big)   
-\big( \tilde{W}^{\dagger}_{\eta} \tilde{W}_{\zeta}
 - \tilde{W}^{\dagger}_{\zeta} \tilde{W}_{\eta}  \big)   
\Big) 
    \nonumber \\
  &
    - \varepsilon^{\mu \nu \eta \zeta}
 \Big(
   -\i  \tilde{F}^{\dagger}_{\mu \nu} [\tilde{W}]
+ \big( 
\tilde{V}^{\dagger}_{\mu} \tilde{W}_{\nu}
-\tilde{V}^{\dagger}_{\nu} \tilde{W}_{\mu}
-\tilde{W}^{\dagger}_{\mu} \tilde{V}_{\nu} 
+ \tilde{W}^{\dagger}_{\nu} \tilde{V}_{\mu}
 \big)
\Big) 
\Big(
   -\i   \tilde{F}^{\dagger}_{\eta\zeta} [\tilde{W}]
+ \big( 
\tilde{V}^{\dagger}_{\eta} \tilde{W}_{\zeta}
-\tilde{V}^{\dagger}_{\zeta} \tilde{W}_{\eta}
-\tilde{W}^{\dagger}_{\eta} \tilde{V}_{\nu} 
+ \tilde{W}^{\dagger}_{\zeta} \tilde{V}_{\eta}
 \big)
\Big) 
    \nonumber \\
  & \left.
    - \varepsilon^{\mu \nu \eta \zeta}
\Big(
  -\i   \tilde{F}_{\mu \nu} [\tilde{W}^{\dagger}]
+ \big( 
\tilde{V}_{\nu} \tilde{W}^{\dagger}_{\mu}
-\tilde{V}_{\mu} \tilde{W}^{\dagger}_{\nu}
+\tilde{W}_{\mu} \tilde{V}_{\nu}^{\dagger} 
- \tilde{W}_{\nu} \tilde{V}_{\mu}^{\dagger}
 \big)
  \Big)
\Big(
  -\i  \tilde{F}_{\eta \zeta} [\tilde{W}^{\dagger}]
+  \big( 
\tilde{V}_{\zeta} \tilde{W}^{\dagger}_{\eta}
-\tilde{V}_{\eta} \tilde{W}^{\dagger}_{\zeta}
+\tilde{W}_{\eta} \tilde{V}_{\zeta}^{\dagger} 
- \tilde{W}_{\zeta} \tilde{V}_{\eta}^{\dagger}
 \big)
  \Big)
 \right]
    . \label{eq:dJ5_3p1_final}
\end{align}
Here the matrix $B$ is given in Eq.~\eqref{eq:constA}.

To facilitate further comparison between these three systems, we refer to the divergence of chiral currents in Hermitianized, anti-Hermitianized, and non-Hermitian systems by $[\tilde{d}_{\mu} j^{5,\mu}]_{\rm h}$, $[\tilde{d}_{\mu} j^{5,\mu}]_{\rm ah}$ and $[\tilde{d}_{\mu} j^{5,\mu}]_{\rm nh}$, respectively.
Our results for $4$ dimensions in Eq.~\eqref{eq:dJ5_3p1_final} reveal that $[\tilde{d}_{\mu} j^{5,\mu}]_{\rm nh} \neq [\tilde{d}_{\mu} j^{5,\mu}]_{\rm h} + [\tilde{d}_{\mu} j^{5,\mu}]_{\rm ah} $. Examples of terms in $[\tilde{d}_{\mu} j^{5,\mu}]_{\rm nh}$ which are not present neither in  $[\tilde{d}_{\mu} j^{5,\mu}]_{\rm h} $ nor in $ [\tilde{d}_{\mu} j^{5,\mu}]_{\rm ah} $ are $ \varepsilon^{\mu \nu  \eta \zeta} 
B_{\mu \nu  \eta \zeta}^{\iota \alpha \tau \upsilon  }
\partial_{\iota}\Im[V]_{\alpha}
 \partial_{\tau}\Re[V]_{\upsilon}
$
 and
 $ \varepsilon^{\mu \nu  \eta \zeta} 
C_{\mu \nu  \eta \zeta}^{\iota \alpha \tau \upsilon  }
\partial_{\iota}\Im[V]_{\alpha}
 \partial_{\tau}\Re[V]_{\upsilon}
$, where 
\begin{align}
    B_{\mu \nu  \eta \zeta}^{\iota \alpha \tau \upsilon  }
=\Re[M_{\mu}^{\iota}]  \Im[M_{\nu}^{\alpha}]
\Re[M_{\eta}^{\tau}]  \Re[M_{\zeta}^{\upsilon}]
,\\
C_{\mu \nu  \eta \zeta}^{\iota \alpha \tau \upsilon  }=
  \i \Im[M_{\mu}^{\iota}]  \Im[M_{\nu}^{\alpha}]
\Re[M_{\eta}^{\tau}]  \Re[M_{\zeta}^{\upsilon}].
\end{align}

\section{Bosonization in $2$ dimensions}

In this section, we present the bosonized form of the Minkowski partition function in $2$ dimensions under chiral and vector rotations introduced in Eq.~\eqref{eq:barpsi_psirot}.

To bosonize the generating function associated with the action in Eq.~\eqref{eq:action_1p1}, we use the bosnonization relation $\Psi = e^{i \varsigma}$ with $\varsigma$ being the bosonic field. We also employ the free-fermion action in the language of bosonic gauge field which reads~\cite{Fujikawa2004, Rao2001}

\begin{align}
  \int {\cal D} \bar{\Psi}  {\cal D} \Psi e^{ - \int {\rm d}^{2} x \bar{\Psi} \big( \gamma^{\mu} 
   f_{\mu}^{\nu} 
  \partial_{\nu} \big) \Psi }
  = 
  \int {\cal D} \varsigma  e^{ \frac{-1 }{{\cal F}_{2}}
  \int {\rm d}^{2} x 
   f_{\mu}^{\nu}  f_{\eta}^{*\mu}
  \big[
    \partial^{\eta} \varsigma 
  \partial_{\nu} \varsigma
  \big]  } ,
   \label{eq:bzfree}
\end{align}
where ${\cal F}_{2} = 2 \pi \sqrt{{\rm det}[B]}$ with $B$ given in Eq.~\eqref{eq:constA}.
Performing chiral and vector rotations in Eq.~\eqref{eq:barpsi_psirot} brings the action in Eq.~\eqref{eq:action_1p1} into a free-fermionic action if the rotating angles satisfy Eq.~\eqref{eq:VW_kb}. These transformations, as we discussed in Sec.~\ref{App:Fujikawa1p1}, introduce additional anomalous contributions, in the form of the Jacobian of the path-integral measures, in the generating function such that
\begin{align}
    \tilde{\cal Z} \propto \int {\cal D} \Psi_{\rm rot} {\cal D} \bar{\Psi}_{\rm rot}
    e^{\i \int {\rm d}^{2}x \bar{\Psi}_{\rm rot} \tilde{\slashed{\zc D}} \Psi_{\rm rot}}
    =\int {\cal D} \Psi {\cal D} \bar{\Psi}
    e^{{\cal S}[\delta \kappa]} e^{{\cal S}_{5}[\delta \beta]}
    e^{\i \int {\rm d}^{2}x \bar{\Psi} ( \i \tilde{\slashed{d}})\Psi},
    \label{eq:Zbz}
\end{align}
where ${\cal S}_{5}[\delta \beta]$ and ${\cal S}[\delta \kappa]$ are given in Eqs.~(\ref{eq:Sbeta_1p1_2}, \ref{eq:Skappa_1p1_2}), respectively. To emphasize the infinitesimal transformations, we have represented the infinitesimal rotating angles by $\delta \kappa$ and $\delta \beta$. Using the parameter $s \in [0,1]$, we express these angles as $\delta \kappa = {\rm d} s \kappa$ and $\delta \beta= {\rm d}s \beta$. Eq.~\eqref{eq:VW_kb} suggests that $V$ and $W$, in ${\cal S}_{5}[\delta \beta]$ and ${\cal S}[\delta \kappa]$, can be written in terms of rotating angles as
\begin{align}
    \tilde{V}_{\mu} = f_{\mu}^{\nu} V_{\nu} = f_{\mu}^{\nu} (1-s) \partial_{\nu} \kappa, \qquad
    \tilde{W}_{\mu} =f_{\mu}^{\nu} W_{\nu}  = -f_{\mu}^{\nu} (1-s) \partial_{\nu} \beta.
    \label{eq:bz-VW}
\end{align} 
Note that from now on we consider $f$ as a diagonal matrix with diagonal elements $f_{\mu}^{\mu}=u_{\mu}$.
The parameter $s$ of infinitesimal transformations is added to impose Eq.~\eqref{eq:VW_kb} at $s=0$, and zero rotation angles, i.e., $s=1$ and ${\rm d}s =0$, when $V$ and $W$ fields are zero. Substituting Eq.~\eqref{eq:bz-VW} in Eqs.~(\ref{eq:Sbeta_1p1_2}, \ref{eq:Skappa_1p1_2}), and using anti-symmetric properties of the Levi-Civita symbol~($\varepsilon^{\mu \nu}$) then yields
\begin{align}
{\cal S}[ \delta \kappa] = \frac{-1 }{{\cal F}_{2}}  \int {\rm d}^{2} x
 &
\kappa(x)  
     \left[ 
 \tilde{d}^{\mu} ( \tilde{V}^{\dagger}_{\mu} ) 
+ \tilde{d}^{\mu \dagger} ( \tilde{V}_{\mu} ) 
 \right]
 ,\\
   = \frac{-1 }{{\cal F}_{2}}  \int {\rm d}^{2} x
 &
 \int_{0}^{1} {\rm d} s  (1-s)
 u_{\mu} u^{*}_{\mu}
     \left[ 
 \partial^{\mu} \kappa(x)   \partial_{\mu} \kappa^{*} 
+ \partial^{\mu} \kappa(x)  \partial_{\mu} \kappa
 \right]
    \label{eq:skbz}
,
\end{align}
\begin{align}
{\cal S}_{5}[\delta \beta] = \frac{1 }{{\cal F}_{2}}  \int {\rm d}^{2} x 
&
\beta(x)
       \left[
\tilde{d}^{\mu} ( \tilde{W}^{\dagger}_{\mu} ) 
+  \tilde{d}^{\mu \dagger} ( \tilde{W}_{\mu} )
- \i
( \tilde{V}^{\mu} \tilde{W}^{\dagger}_{\mu}
- \tilde{W}^{\mu} \tilde{V}^{\dagger}_{\mu} )
+\i ( \tilde{V}^{\mu \dagger} \tilde{W}_{\mu}
-  \tilde{W}^{\mu \dagger} \tilde{V}_{\mu} )
\right]
,\\
= \frac{-1 }{{\cal F}_{2}}  \int {\rm d}^{2} x
&
 \int_{0}^{1} {\rm d} s  (1-s)
u_{\mu} u^{*}_{\mu}
       \left[
\partial^{\mu} \beta(x) \partial_{\mu} \beta^{*}
+ \partial^{\mu}\beta(x)  \partial_{\mu} \beta 
+\i \beta(x)
\big(
 \partial^{\mu} \kappa  \partial_{\mu} \beta^{*}
- \partial^{\mu} \beta \partial_{\mu} \kappa^{*}
\big)
\right]
,\label{eq:s5bbz}
\end{align}
where in writing the last lines of Eqs.~(\ref{eq:skbz}, \ref{eq:s5bbz}), we applied integration by parts and assume $u_{\mu}$ to be constant.
Using Eqs.~(\ref{eq:bzfree}, \ref{eq:skbz}, \ref{eq:s5bbz}), we express $\tilde{\cal Z}$ in Eq.~\eqref{eq:Zbz} as
\begin{align}
    \tilde{\cal Z} \propto & \int {\cal D} \varsigma  e^{ \frac{-1 }{{\cal F}_{2}}
  \int {\rm d}^{2} x {\cal W}[\varsigma, \beta, \kappa]},\\
  {\cal W}[\varsigma, \beta, \kappa] = 
 u_{\mu} u^{*}_{\mu}
  \Big[ &
  \frac{1}{2}  \partial^{\mu} \varsigma 
  \partial_{\mu} \varsigma
-\frac{1}{2}
     \left[ 
 \partial^{\mu} \kappa(x)   \partial_{\mu} \kappa^{*} 
+ \partial^{\mu} \kappa(x)  \partial_{\mu} \kappa
 \right]
  \Big] 
  \nonumber \\
  & -\frac{1}{2}
      \left[
\partial^{\mu} \beta(x) \partial_{\mu} \beta^{*}
+ \partial^{\mu}\beta(x)  \partial_{\mu} \beta 
+\i \beta(x)
\big(
 \partial^{\mu} \kappa  \partial_{\mu} \beta^{*}
- \partial^{\mu} \beta \partial_{\mu} \kappa^{*}
\big)
\right]
  \Big] .
\end{align}
To simplify $\cal W$, we first shift the bosonic field by $\varsigma \rightarrow \varsigma +  \kappa +  \beta$ and note that the path-integral measure remains unchanged under this translation shift, i.e., ${\cal D} \varsigma ={\cal D} (\varsigma +  \kappa +  \beta) $. As a result, $\cal W$ reads
\begin{align}
{\cal W}[\varsigma, \beta, \kappa] = &
 u_{\mu} u^{*}_{\mu}
  \Big[ 
     \frac{1}{2}  \partial^{\mu} \varsigma 
  \partial_{\mu} \varsigma
  +
    \frac{1}{2} 
   \partial^{\mu} \varsigma 
   \big(
   \partial_{\mu} \kappa +\partial_{\mu} \beta
   \big)
   +
     \frac{1}{2} 
   \big(
   \partial^{\mu} \kappa +\partial^{\mu} \beta
   \big)
   \partial_{\mu} \varsigma 
   \nonumber \\
   &
+   \frac{1}{2}     \partial^{\mu} \kappa
  \partial_{\mu} \kappa
  +    \frac{1}{2}    \partial^{\mu} \kappa
  \partial_{\mu} \beta
  +    \frac{1}{2}    \partial^{\mu} \beta
  \partial_{\mu} \kappa
-\frac{1}{2}
     \left[ 
 \partial^{\mu} \kappa(x)   \partial_{\mu} \kappa^{*} 
+ \partial^{\mu} \kappa(x)  \partial_{\mu} \kappa
 \right]
  \Big] 
  \nonumber \\
  & 
  +    \frac{1}{2}   \partial^{\mu} \beta
  \partial_{\mu} \beta
  -\frac{1}{2}
      \left[
\partial^{\mu} \beta(x) \partial_{\mu} \beta^{*}
+ \partial^{\mu}\beta(x)  \partial_{\mu} \beta 
+\i \beta(x)
\big(
 \partial^{\mu} \kappa  \partial_{\mu} \beta^{*}
- \partial^{\mu} \beta \partial_{\mu} \kappa^{*}
\big)
\right]
  \Big] . \label{eq:W3}
\end{align}
Next we introduce a local source function~($v$) such that $\slashed{D} =\i \slashed{d} + M_{\mu}^{\nu} \gamma^{\mu} v_{\nu} $ and $v_{\mu} = {V}_{\mu} + \gamma^{5} {W}_{\mu}$. Using Eq.~\eqref{eq:VW_kb}, the source function reads
\begin{align}
    &v_{\mu} =   \partial_{\mu} \kappa + \varepsilon_{\mu \nu} \partial^{\nu} \beta, \\
    &\partial^{\mu} v_{\mu} + \varepsilon^{\nu \mu} \partial_{\nu} v_{\mu}  = \partial^{\mu} \partial_{\mu} \kappa + \partial^{\mu} \partial_{\mu} \beta.\label{eq:v_betakappa}
\end{align}
We next perform integration by parts on the linear terms proportional to $\partial^{\mu} \varsigma$, and use Eq.~\eqref{eq:v_betakappa} to express various terms in Eq.~\eqref{eq:W3} in terms of the source function. The resulting action is
\begin{align}
{\cal W}[\varsigma, \beta, \kappa] = &
     u_{\mu} u^{*}_{\mu}
  \Big[ 
     \frac{1}{2}  \partial^{\mu} \varsigma 
  \partial_{\mu} \varsigma
  + 
   \big(
   v_{\mu}  \partial^{\mu}    \varsigma 
   + v_{\mu}  \varepsilon^{\nu \mu} \partial_{\nu}   \varsigma 
   \big)
   \nonumber \\
   & \qquad 
  +    \frac{1}{2} 
    \partial^{\mu} \kappa
  \partial_{\mu} \beta
  +    \frac{1}{2} 
    \partial^{\mu} \beta
  \partial_{\mu} \kappa
-\frac{1}{2}
 \partial^{\mu} \kappa(x)   \partial_{\mu} \kappa^{*} 
  \nonumber \\
  & \qquad 
  -\frac{1}{2}
      \left[
\partial^{\mu} \beta(x) \partial_{\mu} \beta^{*}
+\i \beta(x)
\big(
 \partial^{\mu} \kappa  \partial_{\mu} \beta^{*}
- \partial^{\mu} \beta \partial_{\mu} \kappa^{*}
\big)
\right]
  \Big],
  \label{eq:W3f}
  \\
  = & {\cal W}_{0}[\varsigma] + {\cal W}_{1} [\kappa, \beta].
\end{align}
Here ${\cal W}_{0}[\varsigma]$ is 
\begin{align}
    {\cal W}_{0}[\varsigma] = &
     u_{\mu} u^{*}_{\mu}
  \Big[ 
     \frac{1}{2}  \partial^{\mu} \varsigma 
  \partial_{\mu} \varsigma
  + 
   \big(
   v_{\mu}  \partial^{\mu}    \varsigma 
   + v_{\mu}  \varepsilon^{\nu \mu} \partial_{\nu}   \varsigma 
   \big)
  \Big] ,
\end{align}
and ${\cal W}_{1} [\kappa, \beta]$ reads
\begin{align}
    {\cal W}_{1}[\beta, \kappa] = &
      \frac{  u_{\mu} u^{*}_{\mu} }{2}
  \Big[ 
    \partial^{\mu} \kappa
  \partial_{\mu} \beta
  +   
    \partial^{\mu} \beta
  \partial_{\mu} \kappa
-
 \partial^{\mu} \kappa(x)   \partial_{\mu} \kappa^{*} 
  -
      \left[
\partial^{\mu} \beta(x) \partial_{\mu} \beta^{*}
+\i \beta(x)
\big(
 \partial^{\mu} \kappa  \partial_{\mu} \beta^{*}
- \partial^{\mu} \beta \partial_{\mu} \kappa^{*}
\big)
\right]
  \Big].
\end{align}
Finally, we can express the bosonized partition function $\tilde{\cal Z}$ as
\begin{align}
   \tilde{{\cal Z}} \propto & \int {\cal D} \varsigma  e^{ \frac{-1 }{{\cal F}_{2}}
  \int {\rm d}^{2} x {\cal W}_{1}[\beta, \kappa]}
  e^{ \frac{-1 }{ {\cal F}_{2}}
  \int {\rm d}^{2} x   
     u_{\mu} u^{*}_{\mu}
  \Big[ 
     \frac{1}{2}  \partial^{\mu} \varsigma 
  \partial_{\mu} \varsigma
  + 
   \big(
   v_{\mu}  \partial^{\mu}    \varsigma 
   + v_{\mu}  \varepsilon^{\nu \mu} \partial_{\nu}   \varsigma 
   \big)
  \Big]\label{eq:Zbos}
  }.
\end{align}

Eq.~\eqref{eq:Zbos} reproduces the well-known bosonized partition function for Lorentz preserving Hermitian systems with $u_{\mu}=1$ if we set $\kappa =0$ and eliminate the term proportional to $v_{\mu}  \partial^{\mu}    \varsigma $ from the action~\cite{Fujikawa2004}.
This is because the vector current is zero in these systems, i.e.,
in Lorentz invariant Hermitian systems, ${\cal S}[\kappa]$ has no contribution in the bosonized action and $v_{\mu} = \gamma^{5} W_{\mu}$. For our generic non-Hermitian system, the bosonized action is rescaled by the factor $u_{\mu} u^{*}_{\mu}$, and higher-order contributions from rotating angels ($\kappa$, $\beta$), collected in ${\cal W}_{1}$, are generated in the partition function.

\section{The non-Hermitian chiral anomaly: The diagrammatic method}
 \label{App:oneloop}
In this section, we calculate the chiral anomaly for the Hermitianized, anti-Hermitianized, and non-Hermitian systems through the diagrammatic method up to one loop.
We formulate an effective theory of the external non-Hermitian fields $V_\mu$ and $W_\mu$ for each system by integrating out the fermionic degrees of freedom from the underlying action and expanding the remaining functional determinant to second order in the external fields.
The vector and chiral current is defined as the sum of the functional derivative of the effective action with respect to $V_\mu$ and $V^\dagger_\mu$ and $W_\mu$, $W_\mu^\dagger$ respectively.
We discuss how the functional determinant is treated in the three cases, especially when the underlying action is not Hermitian.
The calculations are performed in Minkowski space, with matrix signature $(+,-)$. The gamma matrices are given by $\gamma^{0}=\sigma^{2}$, $\gamma^{1}=\i \sigma^{1}$ and $\gamma^{5}=\gamma^0\gamma^1= \sigma^{3}$, where $\sigma$ stands for the Pauli matrices and our gamma matrices satisfy $\tr[\gamma^{5} \gamma^{\mu} \gamma^{\nu}]=-2 \varepsilon^{\mu \nu}$ with $\varepsilon^{01}=-1$. The diagonal velocity matrix, $M_\mu^\nu$, is a complex valued constant.

\subsection{Defining the functional determinant of a non-Hermitian action} 

The effective action for a Hermitian Dirac action with a Dirac operator, $\zc D_\mu$  and Lagrangian density $\mathcal{L}_{\rm{h}}=\Psi^\dagger(\i \gamma^0\slashed{\zc D})\psi$ can be expressed as 
\begin{align}
\Gamma&=-\i\ln[\det(\i\gamma^0\slashed{\zc D})]\\
&=-\frac{\i}{2}\ln[\det[(\i \gamma^0\slashed{\zc D})^2]],
\end{align}
where the argument of the determinant is Hermitian.
Using the product rule of determinants and logarithms and discarding the constant terms, the above reduces to
\begin{align}
\Gamma&=-\i\ln[\det(\i \slashed{\zc D})]\\
&=-\frac{\i}{2}\ln[\det[(\i \slashed{\zc D})^2]].
\end{align}
For a non-Hermitian Dirac action, evaluating the above functional determinant is not straight forward as we should use the biorthogonal basis of a non-Hermitian system.
To avoid these complexities, we instead use the Hermitian form constructed from $ \slashed{\zc D} \slashed{\zc D}_{\mathcal{L}}^\dagger$ rather than the operator $(\i \slashed{\zc D})^2$ itself.
Here $\slashed{\zc D}_{\mathcal{L}}^\dagger\equiv\gamma^\mu\zc D^\dagger$, which stems from $(\gamma^0\slashed{\zc D})^\dagger=\gamma^0\gamma^\mu\zc D^\dagger$ of the underlying Lagrangian density.

To write an effective theory, starting from a non-Hermitian action, we reverse the background manifold so that $\slashed{\zc D}$ goes to $\slashed{\zc D}_\mathcal{L}^\dagger$. 
We use the notation $\Gamma_\uparrow$ to define the effective action corresponding to the operator $\slashed{\zc D}$, and $\Gamma_\downarrow$ to define the  effective action corresponding to the operator $\slashed{\zc D}_\mathcal{L}^\dagger$, such that
\begin{align}
\Gamma[V,W]&=\Gamma_\uparrow+\Gamma_\downarrow\nonumber\\
& = -\frac{\i}{2}\ln\det(\i \gamma^0\slashed{\zc D})-\frac{\i}{2}\ln\det(-\i\gamma^0 \slashed{\zc D}_\mathcal{L}^\dagger)\\
& = -\frac{\i}{2}\ln\det(\i \slashed{\zc D})-\frac{\i}{2}\ln\det(-\i\slashed{\zc D}_\mathcal{L}^\dagger).
\end{align}
Defining the variables $a+b=1$ and manipulating the above, yields
\begin{align}
(a+b)\Gamma_\uparrow+(a+b)\Gamma_\downarrow & =\frac{a}{2}[-i\ln\det(\i \slashed{\zc D})-\i\ln\det(-\i \slashed{\zc D}_\mathcal{L}^\dagger)]+\frac{b}{2}[-\i\ln\det(-\i \slashed{\zc D}_\mathcal{L}^\dagger)-\i\ln\det(\i \slashed{\zc D})]\nonumber\\
&=\frac{a}{2}\big(-\i\ln[\det(\i\slashed{\zc D})\det(-\i \slashed{\zc D}_\mathcal{L}^\dagger)]\big)+\frac{b}{2}\big(-\i\ln[\det(-\i\slashed{\zc D}_\mathcal{L}^\dagger)\det(\i\slashed{\zc D})]\big)\nonumber\\
&=\frac{a}{2}\big(-\i\ln[\det(\slashed{\zc D}\slashed{\zc D}_\mathcal{L}^\dagger)]\big)+\frac{b}{2}\big(-\i\ln[\det(\slashed{\zc D}_\mathcal{L}^\dagger\slashed{\zc D})]\big).
\end{align}
We choose $a=b=1/2$, and consider the effective action to be
\begin{align}
\Gamma[V,W]&=-\frac{\i}{4}\ln[\det(\slashed{\zc D}\slashed{\zc D}_\mathcal{L}^\dagger)]-\frac{\i}{4}\ln[\det(\slashed{\zc D}_\mathcal{L}^\dagger\slashed{\zc D})]\\
&=-\frac{\i}{2}\ln[\det(\slashed{\zc D}\slashed{\zc D}_\mathcal{L}^\dagger)]\label{eq:app_nH_effaction}
\end{align}
which is Hermitian and subsequently well-defined.
\subsection{The chiral anomaly from the Hermitianized action $\cal S_{\rm{h}}$}
 \label{App:oneloop_S_H}
In this subsection, we calculate the Hermitian chiral anomaly where the starting point is the Hermitianized action,  
 \begin{align}
 \label{eqapp:Z_h_loop}
  {\cal Z}_{\rm{h}} &\propto \int {\cal D} \Psi {\cal D} \bar{\Psi} 
e^{ \i {\cal S}_{\rm{h}}},\\
{\cal S}_{\rm{h}} &= \frac{\i}{2} \int {\rm d}^{d} x 
\Big[
\bar{\Psi} \gamma^{\mu} ({\zc D}_{\rm{nh}, \mu} \Psi)
-\overline{({\zc D}_{\rm{nh}, \mu} \Psi)} \gamma^{\mu} \Psi
\Big]
= \i 
\int {\rm d}^{2} x 
\bar{\Psi} \gamma^{\mu}
{\zc D}_{\rm{h}, \mu} \Psi, \\
{\zc D}_{\rm{h}, \mu}
 &=
  \Big[
\Re[M_{\mu}^{\nu}] \partial_{\nu}
 -\i \Re[M_{\mu}^{\nu}  V_{\nu} ]
 -\i \gamma^{5}  \Re[M_{\mu}^{\nu}  W_{\nu}]
 \Big] ,
\end{align}
where the Dirac operator, ${\zc D}_{\rm{nh},\mu}$, is 
\begin{equation}
\slashed{\zc D}_{\rm{nh}}  =\gamma^{\mu} {\zc D}_{\rm{nh}, \mu} = \gamma^{\mu}  M_{\mu}^{\nu}
\partial_{\nu}
-\i \gamma^{\mu} M_{\mu}^{\nu}
\left(
V_{\nu}
+\gamma^{5} W_{\nu}
\right),\label{eq:diracminkowski}
\end{equation}
Using the unified notation described in Table.~\ref{tabapp:convert}, the Hermitian Dirac operator, ${\zc D}_{\rm{h}, \mu}$,  is written as
\begin{align}
{\zc D}_{\rm{h}, \mu}
 &=
 \tilde{d}_{\mu} -\i \tilde{V}_{\mu} - \i \gamma^{5} \tilde{W}_{\mu}.
\end{align}
${\zc D}_{\rm{h}}$ has the same form as the associated Dirac operator for the well-known Schwinger model with external vector- and chiral fields.
By integrating out the fermionic degrees of freedom from the action, the effective theory of the gauge fields reads
\begin{align}
\label{eq:app_gamma_herm}
\Gamma[\tilde{V},\tilde{W}]=-\i\ln[\det(\i\slashed{\zc D}_{\rm{h}})].
\end{align}

We expand the effective action to second order in the fields, to extract the linear response terms:
\begin{align}
\Gamma^{(2)}[\tilde{V}, \tilde{W}]&=\frac{\i}{2}\Big[\text{Tr}[G(x)\slashed{\tilde{V}}G(y) \slashed{\tilde{V}}]+\text{Tr}[G(x)\gamma^5\slashed{\tilde{W}}G(y) \gamma^5\slashed{\tilde{W}}]-\text{Tr}[G(x)\slashed{\tilde{V}} G(y) \gamma^5\slashed{\tilde{W}}]-\text{Tr}[G(x)\gamma^5\slashed{\tilde{W}}G(y) \slashed{\tilde{V}}]\Big],
\end{align}
where the Green's function is 
\begin{align}
\label{Greens_+}
G(x)=\frac{1}{\i\tilde{\slashed{d}}},
\end{align}
and where the trace is considered over both  position and spin space.
After calculating the trace and integrating over the loop momentum, the effective action becomes
\begin{align}
\Gamma^{(2)}[\tilde{V}, \tilde{W}]&=\frac{1}{2\det(\Re[M])^2}\int \frac{{\rm d}^{2} \tilde{k} }{(2 \pi)^{2}} \big[\tilde{V}_{\mu}(\tilde{k})\tilde{V}_{\nu}(-\tilde{k})\Pi_{\rm{h}}^{\mu\nu}+\tilde{W}_{\mu}(\tilde{k})\tilde{W}_{\nu}(-\tilde{k})\Pi_{\rm{h}}^{\mu\nu}+2\tilde{V}_{\mu}(\tilde{k})\tilde{W}_{\nu}(-\tilde{k})\Pi_{5,\rm{h}}^{\mu\nu}\big],
\label{eq:tmpG2}
\end{align}
where the vacuum polarization tensors are
\begin{align}
\label{Pi_plus}
\Pi_{\rm{h}}^{\mu\nu}
&=-\frac{1}{\pi}\frac{\tilde{k}^\mu \tilde{k}^\nu-\tilde{k}^2g^{\mu\nu}}{ \tilde{k}^2},\\
\label{Pi5_plus}
\Pi_{5,\rm{h}}^{\mu\nu}
&=-\frac{1}{\pi}\frac{\varepsilon^{\nu\beta}( \tilde{k}^\mu \tilde{k}_\beta-\delta_\beta^\mu\tilde{k}^2)}{ \tilde{k}^2}.
\end{align}
The factor $\det(\Re[M])^2$ is the product of the Jacobians obtained from the variable transforms of both the loop momentum~($p$) as  $\tilde{p}_\mu=\Re[M^\nu_\mu]p_\nu$, and the external momentum~($k$) such that $\tilde{k}_\mu=\Re[M^\nu_\mu]k_\nu$.
By inserting Eqs.~(\ref{Pi_plus}, \ref{Pi5_plus}) in  Eq.~\eqref{eq:tmpG2}, the effective action in terms of the bare momenta and fields reads 
\begin{align}
\label{eq:GammaH}
\Gamma^{(2)}[V,W]&=\frac{1}{2\det(\Re[M])}\int \frac{{\rm d}^{2} k }{(2 \pi)^{2}} \big[\nonumber\\
&+\frac{1}{4}(M_\mu^\alpha M_\nu^\beta V_\alpha V_\beta+M_\mu^{*\alpha} M_\nu^{*\beta} V^\dagger_\alpha V^\dagger_\beta+M_\mu^{\alpha} M_\nu^{*\beta} V_\alpha V^\dagger_\beta+M_\mu^{*\alpha} M_\nu^{\beta} V^\dagger_\alpha V_\beta)\Pi_{\rm{h}}^{\mu\nu}\nonumber\\
&+\frac{1}{4}(M_\mu^\alpha M_\nu^\beta W_\alpha W_\beta+M_\mu^{*\alpha} M_\nu^{*\beta} W^\dagger_\alpha W^\dagger_\beta+M_\mu^{\alpha} M_\nu^{*\beta} W_\alpha W^\dagger_\beta+M_\mu^{*\alpha} M_\nu^{\beta} W^\dagger_\alpha W_\beta)\Pi_{\rm{h}}^{\mu\nu}\nonumber\\
&+\frac{1}{2}(M_\mu^\alpha M_\nu^\beta V_\alpha W_\beta+M_\mu^{*\alpha} M_\nu^{*\beta} V^\dagger_\alpha W^\dagger_\beta+M_\mu^{\alpha} M_\nu^{*\beta} V_\alpha W^\dagger_\beta+M_\mu^{*\alpha} M_\nu^{\beta} V^\dagger_\alpha W_\beta)\Pi_{5,\rm{h}}^{\mu\nu}\big].
\end{align}

The corresponding rescaled vector current is given by the functional derivative of the effective action with respect to the gauge field and its Hermitian conjugate,
\begin{align}
\Re[M_\mu^\sigma]  j^\mu&=\frac{\delta \Gamma^{(2)}}{\delta V_\sigma}+\frac{\delta \Gamma^{(2)}}{\delta V^\dagger_\sigma}\nonumber\\
&= \frac{1}{2}\Re[M_\mu^\sigma](M_\nu^\beta V_\beta+M_\nu^{*\beta} V^\dagger_\beta)\frac{\Pi_{\rm{h}}^{\mu\nu}}{\det(\Re[M])}+\frac{1}{2}\Re[M^\sigma_\mu] (M_\nu^\beta W_\beta+M_\nu^{*\beta} W^\dagger_\beta)\frac{\Pi^{\mu\nu}_{5,\rm{h}}}{\det(\Re[M])}.
\end{align}
Similarly, the rescaled chiral current is obtained by summing the functional derivatives of the effective action with respect to the chiral gauge field and its Hermitian conjugate:
\begin{align}
\Re[M^\sigma_\mu] j^{5,\mu} &=\frac{\delta \Gamma^{(2)}}{\delta W_\sigma}+\frac{\delta \Gamma^{(2)}}{\delta W^\dagger_\sigma}\nonumber\\
&= \frac{1}{2}\Re[M^\sigma_\mu](M_\nu^\beta W_\beta+M_\nu^{*\beta} W^\dagger_\beta)\frac{\Pi^{\mu\nu}_{\rm{h}}}{\det(\Re[M])}+\frac{1}{2}\Re[M^\sigma_\nu] (M_\mu^\beta V_\beta+M_\mu^{*\beta} V^\dagger_\beta)\frac{\Pi^{\mu\nu}_{5,\rm{h}}}{\det(\Re[M])}.
\end{align}
The divergence of the rescaled vector and chiral currents is
\begin{align}
&\Re[M_\mu^\alpha] \partial_\alpha j^\mu=0,\label{eq:app:H_dj}\\
&\Re[M_\mu^\alpha] \partial_\alpha j^{5,\mu} 
=\frac{1}{\det(\Re[M])\pi}\varepsilon^{\mu\nu}\Re[M_\mu^\alpha] \partial_\alpha \Re[M_\nu^\beta V_\beta].\label{eq:app:H_dj5}
\end{align}

\subsection{The chiral anomaly from the anti-Hermitianized action $\cal S_{\rm{ah}}$}

In this subsection, we calculate the chiral anomaly from the anti-Hermitian action Eq.~\eqref{eq:action_ah}, defined in terms of a complex velocity matrix and complex gauge fields as

 \begin{align}
  \label{eqapp:Z_ah_loop}
  {\cal Z}_{\rm{ah}} &\propto \int {\cal D} \Psi {\cal D} \bar{\Psi} 
e^{ \i {\cal S}_{\rm{ah}}},\\
{\cal S}_{\rm{ah}} &= \frac{\i}{2} \int {\rm d}^{d} x 
\Big[
\bar{\Psi} \gamma^{\mu} ({\zc D}_{\rm{nh},\mu} \Psi)
+\overline{({\zc D}_{\rm{nh},\mu} \Psi)} \gamma^{\mu} \Psi
\Big]
= \i 
\int {\rm d}^{2} x 
\bar{\Psi} \gamma^{\mu}
{\zc D}_{\rm{ah}, \mu} \Psi, \\
{\zc D}_{\rm{ah}, \mu}
 &=
  \Big[
\i \Im[M_{\mu}^{\nu}] \partial_{\nu}
 + \Im[M_{\mu}^{\nu}  V_{\nu} ]
 + \gamma^{5}  \Im[M_{\mu}^{\nu}  W_{\nu}]
 \Big] ,
\end{align}
where the operator, ${\zc D}_{\rm{nh},\mu}$, is defined in Eq.~(\ref{eq:diracminkowski}).
Using the unified notation described in Table.~\ref{tabapp:convert}, the anti-Hermitian Dirac operator, ${\zc D}_{\rm{ah}, \mu}$,  is written as
\begin{align}
{\zc D}_{\rm{ah}, \mu}
 &=
 \tilde{d}_{\mu} -\i \tilde{V}_{\mu} - \i \gamma^{5} \tilde{W}_{\mu}.
\end{align}
The operator $\gamma^0\i\slashed{\zc D}_{\rm{ah}}$ is anti-Hermitian and therefore diagonalizable with a complete set of orthogonal eigenvectors. Hence, the functional determinant of $\gamma^0\i\slashed{\zc D}_{\rm{ah}}$ is well-defined and the subsequent effective action is
\begin{align}
\Gamma[V,W]=-\frac{\i}{2}\ln[\det(\i \slashed{\zc D}_{\rm{ah}})].
\end{align}
$\Gamma[V,W]$ is expanded analogously to the effective theory of the  Hermitianized action in~Sec \ref{App:oneloop_S_H} and results in the effective action, up to second order in the fields
\begin{align}
\label{eq:GammaAH}
\Gamma^{(2)}[V,W]&=\frac{1}{2\det(\Im[M])}\int \frac{{\rm d}^{2} k }{(2 \pi)^{2}} \big[\nonumber\\
&+\frac{1}{4}(M_\mu^\alpha M_\nu^\beta V_\alpha V_\beta+M_\mu^{*\alpha} M_\nu^{*\beta} V^\dagger_\alpha V^\dagger_\beta-M_\mu^{\alpha} M_\nu^{*\beta} V_\alpha V^\dagger_\beta-M_\mu^{*\alpha} M_\nu^{\beta} V^\dagger_\alpha V_\beta)\Pi^{\mu\nu}\nonumber\\
&+\frac{1}{4}(M_\mu^\alpha M_\nu^\beta W_\alpha W_\beta+M_\mu^{*\alpha} M_\nu^{*\beta} W^\dagger_\alpha W^\dagger_\beta-M_\mu^{\alpha} M_\nu^{*\beta} W_\alpha W^\dagger_\beta-M_\mu^{*\alpha} M_\nu^{\beta} W^\dagger_\alpha W_\beta)\Pi^{\mu\nu}\nonumber\\
&+\frac{1}{2}(M_\mu^\alpha M_\nu^\beta V_\alpha W_\beta+M_\mu^{*\alpha} M_\nu^{*\beta} V^\dagger_\alpha W^\dagger_\beta-M_\mu^{\alpha} M_\nu^{*\beta} V_\alpha W^\dagger_\beta-M_\mu^{*\alpha} M_\nu^{\beta} V^\dagger_\alpha W_\beta)\Pi_{5}^{\mu\nu}\big],
\end{align}
\begin{align}
\Pi_{\rm{ah}}^{\mu\nu}
&=\frac{1}{\pi}\frac{\tilde{M}_\alpha^\mu \tilde{M}_\beta^\nu k^\alpha k^\beta-\bar{M}_\alpha^\sigma \tilde{M}_\sigma^\beta k^\alpha k_\beta g^{\mu\nu}}{ \bar{M}_\kappa^\rho \tilde{M}_\rho^\delta k^\kappa k_\delta},
\end{align}
\begin{align}
\Pi_{\rm{ah},5}^{\mu\nu}
&=\frac{1}{\pi}\frac{\varepsilon^{\nu\beta}( \tilde{M}_\alpha^\mu \tilde{M}_\beta^\sigma k^\alpha k_\sigma-\delta_\beta^\mu \tilde{M}_\alpha^\sigma \tilde{M}_\sigma^\lambda k^\alpha k_\lambda)}{ \tilde{M}_\kappa^\rho \tilde{M}_\rho^\delta k^\kappa k_\delta},
\end{align}
where $\tilde{M}_\nu^\mu=\Im[M_\nu^\mu]$.
The associated vector current is defined as the sum of the functional derivatives of the effective action with respect to both $V$ and $V^\dagger$
\begin{align}
\i\Im[M^\sigma_\mu] j^\mu&=\frac{\delta \Gamma^{(2)}}{\delta V_\sigma}+\frac{\delta \Gamma^{(2)}}{\delta V^\dagger_\sigma}\nonumber\\
&=\frac{\i}{2}\Im[M^\sigma_\mu] (M_\nu^\beta V_\beta-M_\nu^{*\beta} V^\dagger_\beta)\frac{\Pi^{\mu\nu}}{\det(\Im[M])}+\frac{\i}{2}\Im[M^\sigma_\mu](M_\nu^\beta W_\beta-M_\nu^{*\beta} W^\dagger_\beta)\frac{\Pi^{\mu\nu}_5}{\det(\Im[M])}\\
\i\Im[M^\sigma_\mu]j^{5,\mu} &=\frac{\delta \Gamma^{(2)}}{\delta W_\sigma}+\frac{\delta \Gamma^{(2)}}{\delta W^\dagger_\sigma}\nonumber\\
&= \frac{\i}{2}\Im[M^\sigma_\mu] (M_\nu^\beta W_\beta-M_\nu^{*\beta} W^\dagger_\beta)\frac{\Pi^{\mu\nu}}{\det(\Im[M])}+\frac{\i}{2}\Im[M^\sigma_\nu] (M_\mu^\beta V_\beta-M_\mu^{*\beta} V^\dagger_\beta)\frac{\Pi^{\mu\nu}_5}{\det(\Im[M])}.
\end{align}
The divergence of both currents are
\begin{align}
&\i\Im[M_\mu^\alpha] \partial_\alpha j^\mu=0,\label{eq:app:aH_dj}\\
&\i\Im[M_\mu^\alpha] \partial_\alpha j^{5,\mu}
= \frac{1}{\det(\Im[M])\pi}\varepsilon^{\mu\nu}\Im[M_\mu^\alpha]  \partial_\alpha\Im[M_\nu^\beta V_\beta].\label{eq:app:aH_dj5}
\end{align}

\subsection{The chiral anomaly from the non-Hermitian action $\cal S_{\rm{nh}}$}

The generic non-Hermitian action introduced in Eq.~\ref{eqapp:action_nh} is in Minkowski space given by
\begin{align}
 \label{eqapp:Z_nh_loop}
  {\cal Z}_{\rm{nh}} &\propto \int {\cal D} \Psi {\cal D} \bar{\Psi} 
e^{ \i {\cal S}_{\rm{nh}}},\\
{\cal S}_{\rm nh} &= \i \int {\rm d}^{2} x 
\Big[
\bar{\Psi} \gamma^{\mu} ({\zc D}_{{\rm nh}, \mu} \Psi)
\Big]
, \\
\slashed{\zc D}_{\rm nh} &=\gamma^{\mu} {\zc D}_{{\rm nh}, \mu} = \gamma^{\mu}  M_{\mu}^{\nu}
\partial_{\nu}
-\i \gamma^{\mu} M_{\mu}^{\nu}
\left(
V_{\nu}
+\gamma^{5} W_{\nu}
\right),
\end{align}
where we have kept the gauge fields non-Hermitian, but considered a real Fermi velocity matrix~($M$), due to the technical challenges arising from complex momentum integrals. 
For notational convenience, we define $\tilde{V}_\mu=M_\nu^\nu V_\nu$, $\tilde{W}_\mu=M_\nu^\nu W_\nu$, and $\tilde{d}_\mu=M_\mu^\nu \partial_\nu$, such that the non Hermitian Dirac operator reads
\begin{align}
\zc D_{{\rm nh},\mu} &={\zc D}_\mu= \tilde{d}_\mu-\i \tilde{V}_\mu-\i \tilde{W}_\mu.
\end{align}
The effective action of the fields, obtained by integrating out the fermionic degrees of freedom takes the form of Eq.~\eqref{eq:app_nH_effaction} as
\begin{align}
 \label{eq:eff_action_nH}
 \Gamma[V,W]=-\frac{\i}{2}\ln[\det( \slashed{\zc D} \slashed{\zc D}_{\mathcal{L}} ^\dagger)],
 \end{align}
 since the underlying action is non-Hermitian.
 Here
 \begin{align}
 \i\slashed{\zc D} &=i\tilde{\slashed{d}}+ \slashed{\tilde{V} }+\varepsilon^{\mu\nu}\gamma_\nu \tilde{W} _\mu\\
 (\i\slashed{\zc D}_{\mathcal{L}} )^\dagger&=i\tilde{\slashed{d}}+ \slashed{\tilde{V} }^\dagger+\varepsilon^{\mu\nu}\gamma_\nu \tilde{W} ^\dagger_\mu,
 \end{align}
where $\varepsilon^{\mu\nu}\gamma_\nu=\gamma^\mu\gamma^5$. 
The product between $\slashed{\zc D}$ and $\slashed{\zc D}_{\mathcal{L}}^\dagger$ is 
 \begin{align}
 \label{eq:DDdagger_nH}
\slashed{\zc D}\slashed{\zc D}_{\mathcal{L}}^\dagger
 &=-\tilde{d}^2+\tilde{V} _\mu \tilde{V} ^{\dagger\mu}+\i \tilde{d}_\mu \tilde{V} ^{\dagger\mu}+\i \tilde{V} ^{\dagger\mu}\tilde{d}_\mu +\i \tilde{V} ^{\mu}\tilde{d}_\mu+\frac{1}{2}[\gamma^\mu,\gamma^\nu](\i \tilde{d} \tilde{V} _\nu^\dagger+\i \tilde{V} ^\dagger_\nu d\tilde{d}_\mu+\i \tilde{V} _\mu \tilde{d}_\nu)\nonumber\\
 &+\gamma^\mu\gamma^\alpha(\i \tilde{d}_\mu+ \tilde{V} _\mu)\varepsilon^{\nu\beta}g_{\beta\alpha}\gamma^\alpha \tilde{W} ^\dagger_\nu+\gamma^\sigma\gamma^\nu\varepsilon^{\mu\rho}g_{\rho\sigma}\gamma^\sigma \tilde{W} _\mu
 (\i \tilde{d}_\nu +\tilde{V} ^\dagger_\nu)+\gamma^\sigma\gamma^\alpha \varepsilon^{\mu\rho}g_{\rho\sigma} \tilde{W} _\mu \varepsilon^{\nu\beta}g_{\beta\alpha}\tilde{W} ^\dagger_\nu,
 \end{align}
where, since we are only interested in the terms contributing to the linear response, all terms that are zero in the expansion to the second-order in the fields are omitted; an example is the term proportional to $[\gamma^\mu,\gamma^\nu]\tilde{V} _\mu \tilde{V} ^\dagger_\nu$ which is zero when the trace over the spin is taken.
By simplifying Eq.~\eqref{eq:DDdagger_nH} further, and only keeping the relevant terms, it is rewritten as
 \begin{align}
\slashed{\zc D}\slashed{\zc D}_{\mathcal{L}} ^\dagger=-\tilde{d}^2+\Delta^{(1)}+\Delta^{(2)}+\Delta^{(s)},
 \end{align}
 where
 \begin{align}
 \Delta^{(1)}&=\i \tilde{d}_\mu \tilde{V} ^{\dagger\mu}+\i \tilde{V} ^{\dagger\mu}\tilde{d}_\mu +\i \tilde{V} ^{\mu}\tilde{d}_\mu+\varepsilon^{\mu\nu}(\i \tilde{d}_\nu \tilde{W} ^\dagger_\mu+\i  \tilde{W} ^\dagger_\mu \tilde{d}_\nu+\i \tilde{W} _\mu \tilde{d}_\nu )\\
 \Delta^{(2)}&=g^{\mu\nu}(\tilde{V} _\mu \tilde{V} ^{\dagger}_\nu-\tilde{W} _\mu \tilde{W} ^\dagger_\nu)+\varepsilon^{\mu\nu}(\tilde{V} _\nu \tilde{W} ^\dagger_\mu +\tilde{W} _\mu \tilde{V} ^\dagger_\nu) \\
 \Delta^{(s)}&=\frac{1}{2}[\gamma^\mu,\gamma^\nu](\i \tilde{d}_\mu \tilde{V} _\nu^\dagger+\i \tilde{V} ^\dagger_\nu \tilde{d}_\mu+\i \tilde{V} _\mu \tilde{d}_\nu)+\frac{1}{2}[\gamma^\nu,\gamma^\alpha]\varepsilon^{\mu\beta}g_{\beta\alpha}(\i \tilde{d}_\nu \tilde{W} ^\dagger_\mu+\i  \tilde{W} ^\dagger_\mu \tilde{d}_\nu- \i \tilde{W} _\mu \tilde{d}_\nu),
 \end{align}
Expanding the effective action in Eq.\eqref{eq:eff_action_nH} up to the second-order casts
 \begin{align}
 \label{eq:eff_action_nH1}
 \Gamma^{(2)}[V,W ]&=-\frac{\i}{2}\ln\det(\tilde{\slashed{\zc D}} \tilde{\slashed{\zc D}} _{\mathcal{L}}^\dagger)\nonumber\\
 &=-\frac{\i}{2}\big(\tr[(-\tilde{d}^2)^{-1}\Delta^{(2)}]-\frac{1}{2}\tr[(-\tilde{d}^2)^{-1}\Delta^{(1)}(-\tilde{d}^2)^{-1}\Delta^{(1)}]-\frac{1}{2}\tr[(-\tilde{d}^2)^{-1}\Delta^{(s)}(-\tilde{d}^2)^{-1}\Delta^{(s)}]\big).
 \end{align}
By defining the internal and external momenta in terms of the velocities as $\tilde{p}_\mu=M_\mu^\nu p$ and $\tilde{k}_\mu=M_\mu^\nu k$, the traces in  Eq.~\eqref{eq:eff_action_nH1} yield
\begin{align}
\tr[(-\tilde{d}^2)^{-1}\Delta^{(2)}]&=\frac{2}{\det[M]^2}\int \frac{{\rm d}^{2} \tilde{k} }{(2 \pi)^{2}} \int \frac{{\rm d}^{2} \tilde{p} }{(2 \pi)^{2}}\frac{g^{\mu\nu}(\tilde{V}_\mu \tilde{V}^{\dagger}_\nu-\tilde{W}_\mu \tilde{W}^\dagger_\nu)+\varepsilon^{\mu\nu}(\tilde{V}_\nu \tilde{W}^\dagger_\mu +\tilde{W}_\mu \tilde{V}^\dagger_\nu) }{\tilde{p}^2(\tilde{p}+\tilde{k})^2}(\tilde{p}+\tilde{k})^2,\\
\tr[(-\tilde{d}^2)^{-1}\Delta^{(1)}(-\tilde{d}^2)^{-1}\Delta^{(1)}]&=\frac{2}{\det[M]^2}\int \frac{{\rm d}^{2} \tilde{k} }{(2 \pi)^{2}} \int \frac{{\rm d}^{2} \tilde{p} }{(2 \pi)^{2}}\frac{1}{\tilde{p}^2(\tilde{p}+\tilde{k})^2}\nonumber\\
&\times\Big[\big[\tilde{V}^\dagger_\mu \tilde{V}^\dagger_\nu+\tilde{V}_\mu \tilde{V}_\nu\big]\big[\tilde{p}^\mu(\tilde{p}+\tilde{k})^\nu\big]+\tilde{V}^\dagger_\mu \tilde{V}_\nu(\tilde{k}^\mu\tilde{k}^\nu +2 \tilde{p}^\mu \tilde{p}^\nu+ \tilde{p}^\mu \tilde{k}^\nu+ \tilde{k}^\mu\tilde{p}^\nu)\nonumber\\
&+\big[\tilde{W}^\dagger_\mu \tilde{W}^\dagger_\nu+\tilde{W}_\mu \tilde{W}_\nu\big]\big[-(\tilde{p}^\beta(\tilde{p}+\tilde{k})_\beta)g^{\mu\nu}+\tilde{p}^\nu(\tilde{p}+\tilde{k})^\mu\big]\nonumber\\
&+\tilde{W}^\dagger_\mu \tilde{W}_\nu\big[-(\tilde{k}^2+2\tilde{p}^2+2\tilde{p}^\alpha \tilde{k}_\alpha)g^{\mu\nu}+\tilde{k}^\mu \tilde{k}^\nu +2 \tilde{p}^\mu \tilde{p}^\nu+ \tilde{p}^\mu \tilde{k}^\nu+ \tilde{k}^\mu \tilde{p}^\nu\big]\nonumber\\
&+\big[\tilde{V}^{\dagger}_\alpha \tilde{W}^\dagger_\mu +\tilde{V}_\alpha \tilde{W}^\dagger_\mu\big]\big[\varepsilon^{\mu\nu}(2 \tilde{p}_\nu \tilde{p}^\alpha+\tilde{p}_\nu \tilde{k}^\alpha+\tilde{k}_\nu \tilde{p}^\alpha)\big]\nonumber\\
&+\big[\tilde{V}^{\dagger}_\alpha \tilde{W}_\mu +\tilde{V}_\alpha \tilde{W}^\dagger_\mu \big]\big[\varepsilon^{\mu\nu}(2 \tilde{p}_\nu \tilde{p}^\alpha+\tilde{p}_\nu \tilde{k}^\alpha+\tilde{k}_\nu \tilde{p}^\alpha+ \tilde{k}_\nu \tilde{k}^\alpha)
\big]
\Big],
\end{align}
and
\begin{align}
\tr[(-\tilde{d}^2)^{-1}\Delta^{(s)}(-\tilde{d}^2)^{-1}\Delta^{(s)}]&=\frac{2}{\det[M]^2}\int \frac{{\rm d}^{2} \tilde{k} }{(2 \pi)^{2}} \int \frac{{\rm d}^{2} \tilde{p} }{(2 \pi)^{2}}\frac{1}{\tilde{p}^2(\tilde{p}+\tilde{k})^2}\nonumber\\
&\times\Big[\big[\tilde{V}^\dagger_\nu \tilde{V}^\dagger_\mu+\tilde{V}_\nu \tilde{V}_\mu\big]\big[-\tilde{p}^\alpha(\tilde{p}+\tilde{k})_\alpha g^{\mu\nu}+\tilde{p}^\mu(\tilde{p}+\tilde{k})^\nu\big]\nonumber\\
&+\tilde{V}^\dagger_\nu \tilde{V}_\mu\big[(\tilde{k}^2+2\tilde{p}^2+2\tilde{k}_\mu \tilde{p}^\mu)g^{\nu\mu}-(\tilde{k}^\nu \tilde{k}^\mu+2\tilde{p}^\nu \tilde{p}^\mu+\tilde{p}^\nu \tilde{k}^\mu+\tilde{k}^\nu \tilde{p}^\mu)\big]\nonumber\\
&+\big[\tilde{W}^\dagger_\mu \tilde{W}^\dagger_\nu+\tilde{W}_\nu \tilde{W}_\mu\big]\big[\tilde{p}^\mu(\tilde{p}+k)^\nu\big]+\tilde{W}^\dagger_\mu \tilde{W}_\nu\big[-(\tilde{k}^\mu \tilde{k}^\nu+2\tilde{p}^\mu \tilde{p}^\nu)+\tilde{p}^\mu \tilde{k}^\nu+\tilde{k}^\mu \tilde{p}^\nu\big]\nonumber\\
&+\big[\tilde{V}^{\dagger}_\nu \tilde{W}^\dagger_\mu +\tilde{V}_\nu \tilde{W}_\mu \big]\big[-\varepsilon^{\mu\nu}2\tilde{p}^\alpha(\tilde{p}+\tilde{k})_\alpha+\varepsilon^{\mu\alpha}2\tilde{p}_\alpha(\tilde{p}+\tilde{k})^\nu\big]\nonumber\\
&+\big[\tilde{V}^{\dagger}_\nu \tilde{W}_\mu +\tilde{V}_\nu \tilde{W}^{\dagger}_\mu \big]\big[\varepsilon^{\mu\nu}(2\tilde{p}^2+\tilde{k}^2+2\tilde{p}_\alpha \tilde{k}^\alpha)-\varepsilon^{\mu\alpha}(\tilde{k}_\alpha \tilde{k}^\nu+2\tilde{p}_\alpha \tilde{p}^\nu+2\tilde{k}_\alpha \tilde{p}^\nu)\big]
\Big].
\end{align}
Here each factor of $\det[M]^{-1}$ is the Jacobian of the transformation of the two momentum variables.
Collecting the expressions for the traces results in
\begin{align}
\Gamma^{(2)}[V,W]
&=-\frac{\i}{\det[M]^2}\int \frac{{\rm d}^{2} \tilde{k} }{(2 \pi)^{2}} \int \frac{{\rm d}^{2} \tilde{p} }{(2 \pi)^{2}}\frac{1 }{\tilde{p}^2(\tilde{p}+\tilde{k})^2}\nonumber\\
&\times\Big[\big[\tilde{V}_\mu \tilde{V}_\nu^\dagger-\tilde{W}_\mu \tilde{W}_\nu^\dagger)\big]g^{\mu\nu}\big[(\tilde{p}+\tilde{k})^2-\frac{1}{2}(\tilde{k}^2+2\tilde{p}^2+2\tilde{p}\tilde{k})\big]\nonumber\\
&+\big[\tilde{W}_\mu \tilde{V}_\nu^\dagger+\tilde{V}_\nu \tilde{W}_\mu^\dagger q\big]\varepsilon^{\mu\nu}\big[(\tilde{p}+\tilde{k})^2-\frac{1}{2}(2\tilde{p}^2+\tilde{k}^2+2\tilde{p}\tilde{k})\big]\nonumber\\
&+\big[\tilde{V}_\mu^\dagger \tilde{V}_\nu^\dagger+\tilde{V}_\mu \tilde{V}_\nu+\tilde{W}_\mu \tilde{W}_\nu+\tilde{W}_\mu^\dagger \tilde{W}_\nu^\dagger\big]
\big[-p^\mu(\tilde{p}+\tilde{k})^\nu+\frac{1}{2}(\tilde{p}^2+\tilde{p}\tilde{k})g^{\mu\nu}\big]\nonumber\\
&+\big[\tilde{W}_\mu^\dagger \tilde{V}_\nu^\dagger+\tilde{W}_\mu \tilde{V}_\nu\big] \big[-\varepsilon^{\mu\alpha}(2\tilde{p}_\alpha \tilde{p}^\nu+\tilde{k}_\alpha \tilde{p}^\nu+\tilde{p}_\alpha \tilde{k}^\nu)+\varepsilon^{\mu\nu}(\tilde{p}^2+\tilde{p}\tilde{k})\big]\Big].
\end{align}
By introducing a Feynman parameter $x$ and changing variables $\tilde{p}_\mu=\ell_\mu-x\tilde{k}_\mu$, and only keeping terms linear in $\ell$ reduces the effective action to
\begin{align}
\Gamma^{(2)}[V,W]&=-\frac{\i}{\det[M]^2}\int \frac{{\rm d}^{2} \tilde{k} }{(2 \pi)^{2}} \int_0^1 {\rm{d}}  x\int \frac{{\rm d}^{2} \ell }{(2 \pi)^{2}}\frac{1 }{[\ell^2-\Delta\tilde{k}^2]^2}\nonumber\\
&\times \Big[\big[g^{\mu\nu}(\tilde{V}_\mu \tilde{V}_\nu^\dagger-\tilde{W}_\mu \tilde{W}_\nu^\dagger)+\varepsilon^{\mu\nu}(\tilde{W}_\mu \tilde{V}_\nu^\dagger+\tilde{V}_\nu \tilde{W}_\mu^\dagger )\big]\big[\frac{1}{2}(1-2x)\tilde{k}^2\big]\nonumber\\
&-\frac{1}{2}\big[\tilde{V}_\mu^\dagger \tilde{V}_\nu^\dagger+\tilde{V}_\mu \tilde{V}_\nu+\tilde{W}_\mu \tilde{W}_\nu+\tilde{W}_\mu^\dagger \tilde{W}_\nu^\dagger\big]
\big[2\ell^\mu\ell^\nu-\ell^2g^{\mu\nu}+2\Delta \tilde{k}^\mu \tilde{k}^\nu-\Delta \tilde{k}^2 g^{\mu\nu}\big]\nonumber\\
&+\big[\tilde{W}_\mu^\dagger \tilde{V}_\nu^\dagger +\tilde{W}_\mu \tilde{V}_\nu \big]\big[-\varepsilon^{\mu\alpha}(2\ell_\alpha\ell^\nu+2\Delta \tilde{k}_\alpha \tilde{k}^\nu)+\varepsilon^{\mu\nu}(\ell^2-\Delta\tilde{k}^2)\big]\Big],
\end{align}
where $\Delta=-x(1-x)$.
The remaining integrals over $x$ and $\ell$ are standard~\cite{Peskin1995}, and the final expression for $\Gamma^{(2)}[V,W]$ is
\begin{align}
\label{eq:nH_Gamma2_1_final}
\Gamma^{(2)}[\tilde{V},\tilde{W}]&=
\frac{1}{\det[M]^2}\int \frac{{\rm d}^{2}\tilde{k} }{(2 \pi)^{2}}\left[
\Big[\tilde{V}_\mu^\dagger \tilde{V}_\nu^\dagger+\tilde{V}_\mu \tilde{V}_\nu+\tilde{W}_\mu \tilde{W}_\nu+\tilde{W}_\mu^\dagger \tilde{W}_\nu^\dagger\Big]
\Big[-\frac{1}{4\pi}\left(\frac{\tilde{k}^\mu\tilde{k}^\nu-\tilde{k}^2 g^{\mu\nu}}{\tilde{k}^2}\right)\Big]
\right.
\nonumber\\
&\left.
+\Big[\tilde{W}_\mu^\dagger \tilde{V}_\nu^\dagger +\tilde{W}_\mu \tilde{V}_\nu \Big]\Big[-\frac{1}{2\pi}\left(\frac{\varepsilon^{\mu\alpha}(\tilde{k}_\alpha \tilde{k}^\nu-\delta_\alpha^\nu \tilde{k}^2)}{\tilde{k}^2}\right)\Big]
\right],
\end{align}
which in terms of the bare fields is:
\begin{align}
\Gamma^{(2)}[VW]=\frac{1}{2}\int \frac{{\rm d}^{2} k }{(2 \pi)^{2}}\Big[&\left[M_\mu^{\alpha} M_\nu^{\beta} V^\dagger_\alpha V^\dagger_\beta+M_\mu^{\alpha} M_\nu^{\beta} V_\alpha V_\beta+M_\mu^{\alpha} M_\nu^{\beta} W^\dagger_\alpha W^\dagger_\beta+M_\mu^{\alpha} M_\nu^{\beta} W_\alpha W_\beta\right]\nonumber\\
&\times\Big[-\frac{1}{2\pi \det[M]}\left(\frac{ M_\rho^\mu  M^\nu_\sigma k^\rho k^\sigma-M_\eta^\rho M_\sigma^\eta k_\rho k^\sigma g^{\mu\nu}}{ M_\kappa^\lambda M_\delta^\kappa k_\lambda k^\delta}\right)\nonumber\\
&+(M_\mu^{\alpha} M_\nu^{\beta}  W_\alpha^\dagger V_\beta^\dagger+M_\mu^{\alpha} M_\nu^{\beta}  W_\alpha V_\beta )[-\frac{1}{\pi \det[M]}\left(\frac{\varepsilon^{\mu\gamma}(M_\gamma^\rho M_\epsilon^\nu k_\rho k^\epsilon-\delta_\gamma^\nu  M_\eta^\rho M_\epsilon^\eta k_\rho k^\epsilon)}{M_\kappa^\lambda M_\delta^\kappa k_\lambda k^\delta }\right)\Big].
\end{align}
The corresponding vector current is defined as  the functional derivative of the effective action with respect to both $V_\mu$ and $V_\mu^\dagger$, such that
\begin{align}
M_\nu^\sigma j^\nu&=\frac{\delta \Gamma^{(2)}}{\delta V_\sigma}+\frac{\delta \Gamma^{(2)}}{\delta V^\dagger_\sigma}\nonumber\\
&=2M_\nu^\sigma M_\mu^\alpha \Re[V_\alpha]\left[-\frac{1}{2\pi \det[M]}\left(\frac{ M_\rho^\nu  M^\mu_\gamma k^\rho k^\gamma-M_\eta^\rho M_\gamma^\eta k_\rho k^\gamma g^{\mu\nu}}{ M_\kappa^\lambda M_\delta^\kappa k_\lambda k^\delta}\right)\right]\nonumber\\
&+ M_\nu^\sigma M_\mu^\alpha \Re[ W_\alpha]\left[-\frac{1}{\pi \det[M]}\left(\frac{\varepsilon^{\mu\gamma}(M_\gamma^\rho M_\epsilon^\nu k_\rho k^\epsilon-\delta_\gamma^\nu  M_\eta^\rho M_\epsilon^\eta k_\rho k^\epsilon)}{M_\kappa^\lambda M_\delta^\kappa k_\lambda k^\delta }\right)\right],
\end{align}
and the chiral current is similarly
\begin{align}
M_\mu^\sigma j^{5,\mu} &=\frac{\delta \Gamma^{(2)}}{\delta W_\sigma}+\frac{\delta \Gamma^{(2)}}{\delta W^\dagger_\sigma}\nonumber\\
&=2M_\mu^\sigma M_\nu^\alpha \Re[W_\alpha]\left[-\frac{1}{2\pi \det[M]}\left(\frac{ M_\rho^\mu  M^\nu_\gamma k^\rho k^\gamma-M_\eta^\rho M_\gamma^\eta k_\rho k^\gamma g^{\mu\nu}}{ M_\kappa^\lambda M_\delta^\kappa k_\lambda k^\delta}\right)\right]\nonumber\\
&+M_\mu^\sigma M_\nu^\alpha\Re[ V_\alpha]\left[-\frac{1}{\pi \det[M]}\left(\frac{\varepsilon^{\mu\gamma}(M_\gamma^\rho M_\epsilon^\nu k_\rho k^\epsilon-\delta_\gamma^\nu  M_\eta^\rho M_\epsilon^\eta k_\rho k^\epsilon)}{M_\kappa^\lambda M_\delta^\kappa k_\lambda k^\delta }\right)\right].
\end{align}
Finally, the divergences of the currents are
\begin{align}
& M_\mu^\alpha \partial_\alpha j^\mu =0,\label{eq:app:nH_dj}\\
&M_\mu^\alpha \partial_\alpha j^{5,\mu} =\frac{1}{\pi \det[M]}\varepsilon^{\mu\nu} M_\mu^\alpha  \partial_\alpha \Re[M_\nu^\beta V_\beta].\label{eq:app:nH_dj5}
\end{align}

The expressions for the divergence of the currents for the Hermitianized action, Eqs.~(\ref{eq:app:H_dj}, \ref{eq:app:H_dj5}), the Anti Hermitianized action, Eqs.~(\ref{eq:app:aH_dj}, \ref{eq:app:aH_dj5}), and the non-Hermitian action,  Eqs.~(\ref{eq:app:nH_dj}, \ref{eq:app:nH_dj5}), can be expressed in the unified notation, (Table~\ref{tabapp:convert}), as
\begin{align}
&  \tilde{d}_\mu j^\mu = 0,\label{eq:app:nH_dj_uni}\\
& \tilde{d}_\mu j^{5,\mu} =\frac{1}{\pi \det[f]}\varepsilon^{\mu\nu}  \tilde{d}_\mu \tilde{V}_\nu,\label{eq:app:nH_dj5_uni}
\end{align}
where, to further simplify the notation, we have re-defined $\tilde{V}_\mu=M_\mu^\nu \Re[V_\nu]$ and $\tilde{W}_\mu=M_\mu^\nu \Re[W_\nu]$ for the non-Hermitian case. 

The anomalies in Eqs.~(\ref{eq:app:nH_dj_uni}, \ref{eq:app:nH_dj5_uni}) are the consistent chiral anomalies, contrary to the results in Eqs.~(\ref{eq:dJ_1p1_final}, \ref{eq:dJ5_1p1_final}), which are given in terms of the covariant anomaly~\cite{Bardeen1984}.
Note that the vector current in Eqs.~(\ref{eq:app:nH_dj_uni}) is conserved, which is due to the use of a gauge invariant regulator in the calculation of the loop integrals.
By choosing a different regulator the anomaly equations, Eqs.~(\ref{eq:app:nH_dj_uni}, \ref{eq:app:nH_dj5_uni}) are equivalently expressed as
\begin{align}
&  \tilde{d}_\mu j^\mu = \frac{1}{2\pi \det[f]}\varepsilon^{\mu\nu} \tilde{d}_\mu \tilde{W}_\nu,\label{eq:app:nH_dj_mixed}\\
& \tilde{d}_\mu j^{5,\mu} =\frac{1}{2\pi \det[f]}\varepsilon^{\mu\nu}  \tilde{d}_\mu \tilde{V}_\nu\label{eq:app:nH_dj5_mixed},
\end{align}
where the anomaly is spread over both the vector and axial currents.
The choice of a specific regulator amounts to adding Bardeen counter terms~\cite{Bardeen1969,Landsteiner2016} on the level of the effective action; the counter terms guarantee that the consistent anomaly always can be expressed in a manner which preserves the vector current.
By adding a Bardeen counter term of the form $\Gamma_{\rm{B}}[\tilde{V},\tilde{W}]=1/(2\pi \det[f])\int\frac{\diff^2 p}{(2\pi)^2}\varepsilon^{\mu\nu}\tilde{W}_\mu \tilde{V}_\nu$ to the effective action underlying Eqs.~(\ref{eq:app:nH_dj_mixed}, \ref{eq:app:nH_dj5_mixed}), we get back Eqs.~(\ref{eq:app:nH_dj_uni}, \ref{eq:app:nH_dj5_uni}).

The results in Eqs.~(\ref{eq:app:nH_dj_mixed}, \ref{eq:app:nH_dj5_mixed}), recover the results~\footnote{The difference of sign of the axial current between the results is due to the chosen convention of the sign of the angle of the chiral transformation in the Fujikawa method.}, Eqs.~(\ref{eq:dJ_1p1_final}, \ref{eq:dJ5_1p1_final}), where the covariant form of the anomaly is a factor of two larger than the consistent form~\cite{Bardeen1984}.

\section{Witten effect and unquantized monopole charge}

In the main text we derived the non-hermitian polarization
\begin{align}
\label{eqApp:Pol1}
M_{\nu}^{\alpha} P^{\nu}
=&\frac{ \varepsilon^{0 \nu \eta \zeta}  }{4 \pi \sqrt{\det[B]} }
\Re[M^{\alpha *}_{\nu} M^{\iota}_{\eta} M^{\rho}_{\zeta} B^{\dagger}_{\iota \rho}],
\end{align}
Let us choose the velocity matrix as $M = \mathrm{diag}(v e^{i\phi},v,v,1)$ and the non-hermitian magnetic field $B^z$ in the $z$ direction given by $B^{z} = \frac{1}{2}\epsilon^{z\mu\nu} B_{\mu\nu}$ with
\begin{equation}
    B_{xy}= \exp[2 \i \phi] \partial_{y} V_{x} \equiv \exp[2 \i \phi]B^{z}.
\end{equation}
%
%
Introducing these definitions in the polarization \eqref{eqApp:Pol1}, the polarization reads
\begin{align}
v P^{z}
=&\frac{v^3\cos(\phi)}{2 \pi \sqrt{\det[B]} }\Re[B^z],
\end{align}
Using Maxwell's equations modified by the presence of $M^\mu_\nu$ we can follow \cite{Rosenberg2010} to derive that a unit magnetic monopole will induce a charge
\begin{align}
Q = \dfrac{\Delta \beta}{2\pi}\frac{v^3\cos(\phi)}{\sqrt{\det[B]}}e,
\end{align}
where, in our notation, the change in the $\theta$-angle of Ref.\cite{Rosenberg2010} is labelled as $\Delta \beta$. This is called the Witten effect.
If $\Delta \beta = 0$, as in trivial insulators, $Q=0$ \cite{Rosenberg2010}.
If $\Delta \beta = \pi$, as in topological insulators,
\begin{align}
\label{eq:chargenHTI}
Q =  \left(\frac{\cos(\phi)}{\sqrt{3\cos^2\phi-2}}\right)\dfrac{e}{2},
\end{align}
where we used that $\sqrt{\det[B]} =v^3\sqrt{3\cos^2\phi-2}$ (see \eqref{eq:detconstA3d}). In the Hermitian case, where $\phi=0$, the induced charge is $e/2$, as in Ref.~\cite{Rosenberg2010}. In the non-Hermitian case, the induced charge is also finite, but now $Q$ can take any value, given by \eqref{eq:chargenHTI}.

\end{document}